\documentclass[11pt]{article}

\usepackage{cite}
\usepackage{amsmath,amssymb,amsfonts}
\usepackage{algorithmic}
\usepackage{graphicx}
\usepackage{textcomp}
\usepackage{hyperref}

\usepackage{graphics} 
\usepackage{bm}
\usepackage{epsfig} 
\usepackage{mathtools}

\usepackage{color}

\usepackage{color}
\usepackage{diagbox}

\usepackage{enumerate}   

\usepackage{tabu}

\usepackage[mathscr]{euscript}

\usepackage{tikz}

\usepackage{pgfplots}

\usepackage{framed}
\usepackage[framed]{ntheorem}

\usetikzlibrary{arrows,shapes,snakes,automata,backgrounds,petri}
\usetikzlibrary{graphs}

\usetikzlibrary{decorations.markings}
\tikzset{
    set arrow inside/.code={\pgfqkeys{/tikz/arrow inside}{#1}},
    set arrow inside={end/.initial=>, opt/.initial=},
    /pgf/decoration/Mark/.style={
        mark/.expanded=at position #1 with
        {
            \noexpand\arrow[\pgfkeysvalueof{/tikz/arrow inside/opt}]{\pgfkeysvalueof{/tikz/arrow inside/end}}
        }
    },
    arrow inside/.style 2 args={
        set arrow inside={#1},
        postaction={
            decorate,decoration={
                markings,Mark/.list={#2}
            }
        }
    },
}

\newtheorem{thm}{Theorem}
 
\newtheorem*{prop_n}{Proposition}

\newtheorem{dfn}{Definition}

\newtheorem{prop}{Proposition}

\newtheorem{crl}{Corollary}

\newtheorem{rem}{Remark}

\newtheorem{lem}{Lemma}

\newtheorem{expl}{Example}

\newcommand{\RR}{{\mathbb R}}

\newcommand{\dotex}{{\frac{d}{dt}}}

\newcommand{\axel}[1]{#1}
\newcommand{\axelout}[1]{}

\newcommand{\dimEmbedG}{d}
\newcommand{\dimG}{d}
\newcommand{\dimFixV}{q}
\newcommand{\dimBodyV}{r}

\newcommand{\nbVectFix}{N_1}
\newcommand{\nbVectBody}{N_2}
\newcommand{\Ad}[1]{\Matrix{Ad}_{#1}}

\newcommand{\Matrix}{\pmb}

\newcommand{\fixX}{x}
\newcommand{\bodyX}{\text{\sc x}}
\newcommand{\body}[1]{\text{\sc #1}}
\newcommand{\fixV}{V}
\newcommand{\bodyV}{B}
\newcommand{\bodyb}{\text{\sc b}}
\newcommand{\fixb}{b}
\newcommand{\bodyh}{\text{\sc h}}
\newcommand{\fixh}{h}
\newcommand{\bodyY}{\text{\sc Y}}
\newcommand{\fixY}{y}
\newcommand{\bodyE}{\text{\sc e}}
\newcommand{\fixE}{e}
\newcommand{\bodyZ}{\text{\sc z}}
\newcommand{\fixZ}{z}

\newcommand{\generic}[1]{{#1}}
\newcommand{\fixF}{\Matrix{F}}
\newcommand{\fixd}{d}
\newcommand{\fixC}{\Matrix{C}}
\newcommand{\fixu}{u}
\newcommand{\bodyF}{\Matrix{\Phi}}
\newcommand{\bodyd}{\body{d}}
\newcommand{\bodyC}{\Matrix{\Gamma}}
\newcommand{\bodyu}{\body{u}}

\newcommand{\dNoiseFix}{G^\fixX}
\newcommand{\dNoiseBody}{G^\bodyX}

\newcommand{\bodyXi}{\xi^\bodyE}
\newcommand{\fixXi}{\xi^\fixE}

\newcommand{\outSpace}{\mathcal{Y}}
\newcommand{\elem}[1]{#1}
\newcommand{\action}{*}

\newcommand{\rep}[2]{ \Matrix{ #1 }{\action}}

\newcommand{\repu}[2]{\Matrix{(\elem{#1})}_{\action}}

\newcommand{\dg}[1]{\Matrix{(#1)}_{\action}}

\newcommand{\grouplaw}{ } 

\newcommand{\biggrouplaw}{\bullet}

\newcommand{\tFrames}{G^+_{\fixV, \bodyV}}

\newcommand{\Id}[1]{\Matrix{I}_{#1}}
\newcommand{\0}[1]{\Matrix{0}_{#1}}

\newcommand{\target}{\beta}

\newcommand{\Rot}[1]{\Matrix{\rho} \left( #1 \right)}
\newcommand{\nuMat}[1]{\Matrix{\nu_2} \left( #1 \right)}

\newcommand{\expm}[2]{\Matrix{\exp_{#1} \left( #2 \right)}}

\begin{document}
\title{The Geometry of Navigation Problems}
\author{Axel Barrau, Silv\` ere Bonnabel 
\thanks{A. Barrau is with  SAFRAN TECH, Groupe Safran, Rue des Jeunes
Bois - Ch\^ateaufort, 78772 Magny Les Hameaux CEDEX, France {\tt\small axel.barrau@safran.fr}. S. Bonnabel is with MINES
ParisTech, PSL Research University, Centre for
robotics, 60 Bd St Michel 75006 Paris, France and with  {Université de la Nouvelle-Calédonie, Institut de Sciences Exactes et Appliquées,   98851 Nouméa Cedex}. {\tt\small silvere.bonnabel@mines-paristech.fr}          }}
\date{}

The following document is the preprint for:

 \vspace{.2 cm}
 
 \textbf{The Geometry of Navigation Problems}, Axel Barrau and Silv\`ere Bonnabel, \emph{IEEE Transactions on Automatic Control}, published 21 January 2022. DOI: 10.1109/TAC.2022.3144328.
 
 \vspace{.2 cm}
 
 At the end it includes supplementary material to the above-mentioned article.

\tableofcontents
\clearpage

\maketitle


\begin{abstract}
While many works exploiting an existing Lie group structure have been proposed for state estimation, in particular the Invariant Extended Kalman Filter (IEKF), few papers   address the construction of a group structure that  allows casting a given system into the  framework of invariant filtering.  In this paper we introduce a large class of systems encompassing most problems involving a navigating vehicle encountered in practice. For those systems we introduce  a novel methodology that systematically provides a group structure for the state space, including vectors of the body frame such as biases. We use  it to derive observers  having properties akin to those of linear observers or filters. The proposed unifying and versatile framework encompasses all systems where IEKF has   proved successful, improves state-of-the art ``imperfect" IEKF for inertial navigation with sensor biases,   and allows addressing novel examples, like GNSS antenna  lever arm   estimation.    \end{abstract}

\section{Introduction}\label{intro:pop}

The Kalman filter  introduced in 1960 was immediately applied   to the  localization of the manned space capsule   going to the Moon and back.   Though sixty years have  since passed, state estimation for vehicles that navigate still offers challenges. This is because estimating the state of a navigating rigid body inevitably implies estimating the operator  that is needed to move the object from a reference placement to its current placement, that is, an operator mapping the fixed reference frame to a frame being attached to the body. This includes a rotation, and rotations do not form a vector space, making the problem inherently nonlinear.

In the 2000s, with the advent of the aerial robotics field, an  approach   to the   problem of estimating attitude revolving around symmetries and equivariance  emerged, see  \cite{mahony-et-al-IEEE,Vasconcelos,batista2014attitude,hua2010attitude} to cite a few, and   \cite{mahony2020equivariant} for a more recent exposition,    see also the theory of symmetry-preserving or invariant observer design \cite{bonnabel2008symmetry,arxiv-08}. This body of work  dedicated to  providing  constant-gain  observers with convergence properties, and especially the complementary filter of \cite{mahony-et-al-IEEE},  underpinned the first generation of small unmanned aerial vehicles (UAVs) or drones. 

Invariant Kalman filtering, namely the invariant extended Kalman filter (IEKF)   early introduced in \cite{bonnabel2007left,bonnabel2009left} and whose modern form is introduced in \cite{barrau2017invariant,barrau2018invariant}, which targets Jacobians that do not depend on the state, has  proved successful in various   
applications. The main theoretical properties of the IEKF having been brought to light so far may be summarized as follows: 1- IEKF possesses convergence properties when used as an observer for  systems with group affine dynamics and a specific form of outputs. For these systems error equation is state trajectory independent, and its propagation is actually governed by a linear equation \cite{barrau2017invariant},    a property called error \emph{log-linearity} implying no linearization error is made by IEKF at propagation. The discovery that  dynamics associated with (unbiased) inertial measurement units (IMU) are group affine in  \cite{barrau2017invariant} has led  to various recent experimental and theoretical successes, e.g. \cite{Hartley-RSS-18,hartley2019contact,wang2020hybrid,cohen2020navigation,allakconsistent,van2020invariant}. 2- IEKF possesses consistency properties   when the system is not fully observable as in the problem of Simultaneous Localization and Mapping (SLAM)   \cite{barrau2015non,barrau2015ekf,brossard2018exploiting}, as exploited  in  \cite{wu_invariant-ekf_2017, heo_consistent_2018,brossard2017unscented, caruso_2018, heo2018consistent,zhang2017convergence}. 3- When  the actual  state is physically restricted within or near a subspace of the state space, the IEKF's estimate  reflects this information  \cite{  barrau2019extended,barrau2015non}, contrary to the EKF, as   experimentally confirmed  in  \cite{hartley2019contact}. 4- Group affine dynamics possess the  preintegration property \cite{barrau2019linear,bonnabel2020mathematical} that plays key role in modern robotics \cite{forster2017manifold}. The theory is remarkable in that those four  properties  are characteristic of the linear case, and are  usually lost in the nonlinear case. 
\begin{figure}[h]\centering
\includegraphics[width=0.65\columnwidth]{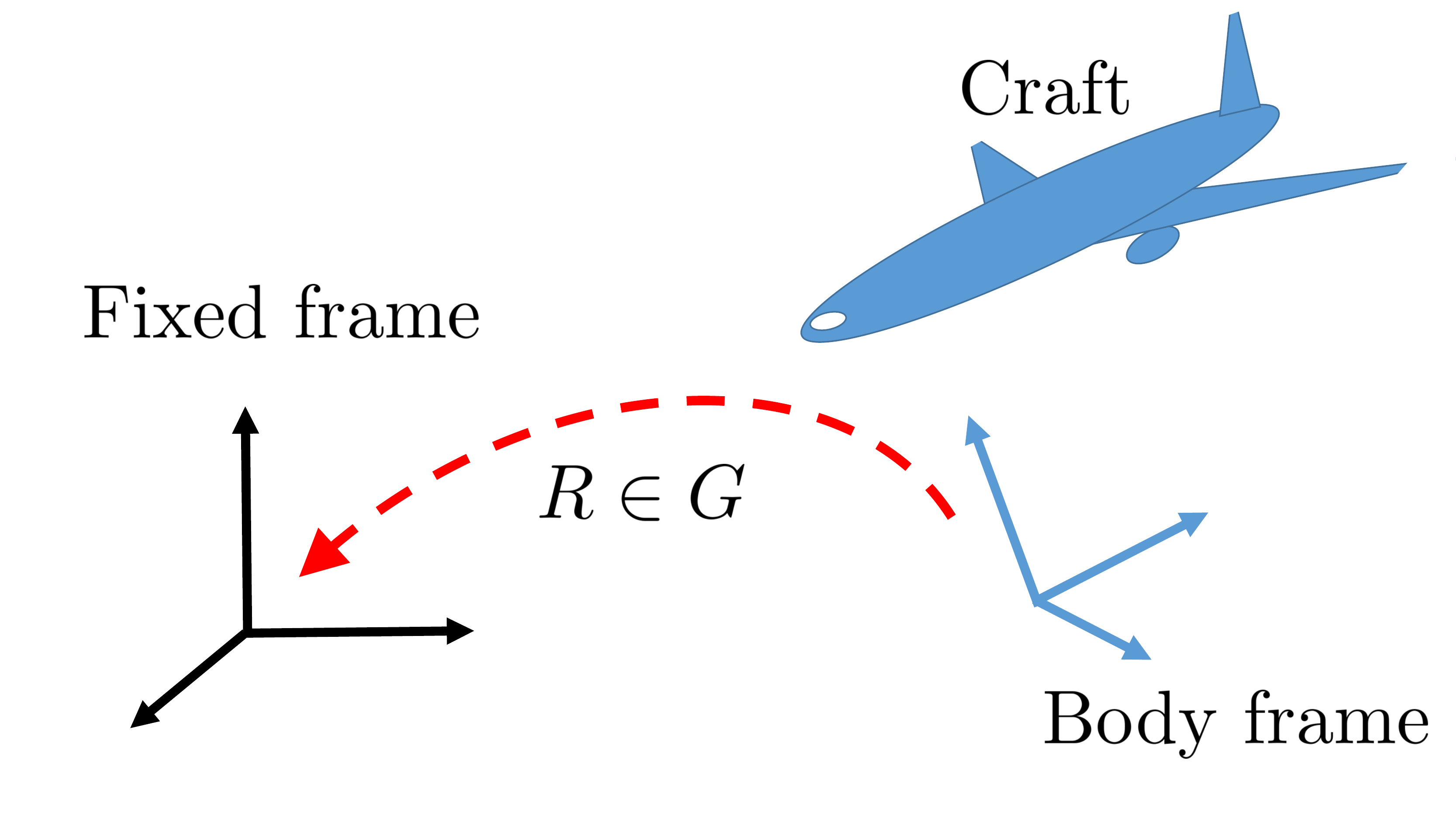}
\caption{The state of a two-frames system, to be estimated,  consists of a frame transformation operator $R$,    \emph{along with vectors}, e.g., the  position, velocity, or sensor biases, stacked  in  $\fixX$ (and  $\bodyX$) when expressed in the fixed (resp. body) frame.}\label{aircraft}
\end{figure} 

The big question when it comes to invariant observers/filtering is:  \emph{how do we find a  group structure for  the state space    that comes with theoretical results,   given a system}? The only generic approaches to date revolve around the search for symmetries of the original system, but recent successes of the IEKF turn out not to be  a direct application of this idea. The present paper makes a leap forward in this respect. We introduce  a   class of systems modeling rigid bodies in space,  called ``two-frames systems'', see Fig. \ref{aircraft}, covering  most practical navigation problems. We use a transformation group encoding the orientation of the vehicle as a building block for a larger general group called the two-frames group (TFG). This second group endows the state space with a group structure, allowing for the application of invariant filtering theory \cite{barrau2017invariant, barrau2018invariant,barrau2019linear}. This leads to a class of observers having striking properties, and to an IEKF  based on the TFG. The practitioner only needs  to check the system is a two-frames system, and then the methodology is constructive and systematic.  

If we consider for example the (bias-free) inertial navigation example of  \cite{barrau2017invariant} the TFG boils down to   $SE_2(3)$ - a group introduced in the latter paper for navigation  and successfully used since in many articles - and we immediately recover the IEKF of  \cite{barrau2017invariant}. Similarly, if we consider the SLAM problem,   the TFG then  boils down to      $SE_k(d)$  - a group introduced for SLAM in    \cite{bonnabel2012symmetries,barrau2015ekf} and successfully used since   - and we necessarily recover the   IEKF of  \cite{barrau2015ekf,barrau2018invariant}.  Besides, novel  examples   to which invariant filtering theory apparently did not apply      before are presented, and addressed by strictly following the proposed methodology. A   wheeled robot with an unknown  GNSS antenna lever arm being estimated online is proved to benefit from  invariant filtering   theory thanks to the TFG. So does the 3D  SLAM problem with moving objects tracking (SLAMMOT).  The TFG structure  also  improves state-of-the-art IEKF-based inertial navigation with IMU biases  as used in the recent papers \cite{Hartley-RSS-18,hartley2019contact,wu_invariant-ekf_2017,heo_consistent_2018,heo2018consistent, barrau2018invariant,brossard2017unscented, cohen2020navigation}. Those practically relevant examples are   contributions in themselves.

In Section \ref{2frame:sec} \axel{we introduce new concepts capturing the specificity of navigation problems. Two-frames state spaces (Def. \ref{def::2-frame-state}) are proposed as generic navigation states. Natural outputs (Def. \ref{def::2frame_obs}) and natural dynamics (Def. \ref{def::vector_discrete_dynamics} and \ref{def::frame-shift}), that encapsulate the ``intrinsicness'' of the system's equations, are then introduced as building blocks for our main concept of natural two-frames system (Def. \ref{def::two-frames-system}).} 
In Section \ref{TFG:sec}  we define the TFG structure \axel{(Def. \ref{def::2frame_law}),} a novel and non-trivial group law, and prove it makes two-frames systems fall within the invariant filtering framework \axel{(Thm. \ref{thm::output} for outputs, Thm. \ref{thm::group-affine} and Thm. \ref{thm::unbiased} for dynamics, and Thm. \ref{thm:aff} for the full system).} This opens the door to a new class of invariant observers  (including invariant Kalman filters) for two-frames systems that come with strong properties, see  Section \ref{sect::design} \axel{(Thm. \ref{thm:error}, \ref{thm::log-linear} and Prop. \ref{prop:con}). }  By transposing the group-theoretic results into the original state variables, Section \ref{explicit:sec} provides   ``turn-key'' concrete formulas for observer implementation and convergence analysis.  Finally, the remainder of Section \ref{sect::expl} exhibits new systems to which invariant filtering can be applied thanks to the TFG structure.  An appendix provides the technical methods and  ingredients for  IEKF implementation on the TFG. A supplementary material  document reproduced in Section \ref{supp:sec} provides extended proofs and  complements. The reader is referred to \cite{barrau2018invariant,barrau2019linear} for an introduction to the topic.

%

\section{Natural two-frames  systems}\label{2frame:sec}
\label{sect::2frames}

This section   introduces the novel notion of  two-frames systems, which brings new  insights into many estimation problems related to navigation.  

\subsection{Two frames state space}
\label{sect::def2frames1}
To describe the motion of a rigid mobile body such as a robot or an aircraft we define two frames: one is attached to the body, and another one is considered as ``fixed'' with respect to the ``world''. The orientation of the body at time step $n$ is generally described by a rotation matrix $R_n$ that maps vectors expressed in the body frame to vectors expressed in the world frame, see Fig. \ref{aircraft}. The state of the mobile body may then be described by the rotation $R_n$ which encodes its orientation, a collection of vectors whose expressions are given in the world frame, such as  its position and its velocity, and a collection of vectors  whose expressions are given in the body frame, like sensor biases or lever arms, to be estimated as well.  
\begin{dfn}[Two-frames state space]
\label{def::2-frame-state}
A two-frames state space is a product space $ G \times \fixV \times \bodyV$, with $ \fixV, \bodyV$ two vector spaces of dimensions respectively $q$, $r$ and $G$ a $\dimG$-dimensional Lie group. Elements $\chi$ of this space will be indifferently denoted   in lines or columns, with $ R\in G, \fixX \in \fixV, \bodyX \in \bodyV$:
$$
\chi=\begin{pmatrix}
R \\ \fixX \\ \bodyX
\end{pmatrix}
=
(R, \fixX, \bodyX).
$$
\end{dfn}
$G$ should be viewed as the group of frame changes, $\fixV$ denotes a space of vectors regrouped in the variable $\fixX_n$ written in ``fixed-frame'' coordinates and $\bodyV$  denotes a space of vectors regrouped in the variable $\bodyX_n$ written in ``body-frame'' coordinates. Most often $G$ denotes a group of rotations acting on  multi-vectors of $\fixV$ and $\bodyV$.
We thus see that Def. \ref{def::2-frame-state} is actually the basic setting for any attitude,   navigation or SLAM  estimation problem: an attitude $R$, a set of variables written in the fixed frame (the vector $\fixX$) and a set of variables written in the body frame (the vector $\bodyX$).  

 \underline{Notation}: Throughout the paper,   {lowercase indicates   vectors expressed in the fixed frame, and small uppercase   vectors expressed in the body frame. }


\subsection{The class of two-frames natural systems}
\label{sect::def2frames}
The point of the present paper is to show many navigation problems share common properties, which suggests addressing them through a unifying framework.   To this end, we are going to define a class of two-frames systems we consider as ``natural'' in the sense that they show invariance/equivariance with respect to frame changes induced by $G$. 
\begin{dfn}[Group action]
\label{def::action}
A (left) group action of $G$ on a set $S$ is a map $(G,S)\to S$ that we denote  as $(\elem{R},s)\mapsto\elem{R} \action s$, and which verifies the following conditions:
\begin{itemize}
\item $\elem{Id}\action s=s$ where $\elem{Id}$ denotes the identity element of $G$,
\item $(\elem{R}_1\grouplaw \elem{R}_2) \action s = \elem{R}_1 \action (\elem{R}_2 \action s)$ for all $\elem{R}_1,\elem{R}_2 \in G$, $s \in S$.
\end{itemize}Throughout the paper we assume the mapping $x \mapsto \elem{R} \action x$ is linear (in group theory this is called a ``representation'' \cite{Hall}).  
\end{dfn}
The main benefit of Def. \ref{def::action} is allowing the same element of a group to define different transformations over different sets. From that point onwards, we   assume $G$ acts both on $V$ and $B$ {and shall denote by $\action$ both of these actions, although different}. The following example   should be meaningful:
\begin{expl}[Term-by-term   rotation]
\label{expl::term-by-term}
Assume that $G=SO(d)$ is a group of rotations  with typically $d=2$ or $d= 3$ and consider a vector space  $W =  \RR^{Nd} $   whose elements are $N$-uples of vectors of $\RR^d$. Then the following operator $\action$ defines an action of $G$ on $W$:
\begin{equation}
 {R} \action \left( x^{1}, \cdots, x^{N} \right) := \left( \Matrix{R}x^{1}, \cdots,\Matrix{R}x^{N} \right)
\label{tbt:eq}
\end{equation}
\end{expl}

\begin{dfn}[Commuting actions]
\label{def::comute}
Let $G$ be a group acting on vector spaces $W_1$ and $W_2$, and let $\Matrix{H}:W_1\to W_2$ be a    linear operator (a matrix).  The symbol $\action$  denotes as previously the action on  {both} spaces $W_1$, $W_2$. \color{black}We say $\Matrix{H}$  commutes with the action of $G$ if we have for any element $R\in G$ and $x\in W_1$:
$$
R\action(\Matrix{H}x) = \Matrix{H}(R\action x)
$$
\end{dfn}
The following  result will cover most examples.
\begin{prop}\label{easy:prop}
Let us consider  the term-by-term action \eqref{tbt:eq} of Example \ref{expl::term-by-term} on both spaces $W_1 =  \RR^{Nd} $ and $W_2 =  \RR^{Md}  $, and a block matrix $\Matrix{H}$ of the form:
\begin{equation}
\Matrix{H} = \begin{pmatrix}
\alpha_{11} \Matrix{I}_d & \dots & \alpha_{1N} \Matrix{I}_d
\\\vdots&  &\vdots\\
\alpha_{M1} \Matrix{I}_d & \dots & \alpha_{MN} \Matrix{I}_d
\end{pmatrix}\label{block:eq}
\end{equation}
with $\alpha_{ij}$'s  real numbers. Then  $\Matrix{H}$   commutes with the action of $G$, as can be easily verified.
\end{prop}

\underline{Notation}: Throughout the paper,    {bold mathematical symbols} are reserved for linear  {operators, or more prosaically  matrices}, like $\Matrix{H}$ in the definition above. Besides, elements of matrix Lie groups, e.g., $R\in SO(d)$,  will be written in \textbf{bold} when their matrix nature is to be emphasized, in particular when they act on vectors through matrix-vector multiplication.

\vspace{.1 cm}

  Let us introduce the output space, and a family of output maps that  represent observations, i.e.,  measurements:
\begin{dfn}[Natural two-frames output]
\label{def::2frame_obs}
Let the output space $\mathcal{Y}$ be a vector space  \emph{on which $G$ acts} through an action denoted by $\action$.  
We call a natural output  in the fixed frame (or the body frame)  a map $G\times V\times B\mapsto \outSpace$ defined by:
\begin{align} \Aboxed{
\text{\normalfont{fixed-frame:}} ~\fixh (R,\fixX, \bodyX) & =  \Matrix{H}^\fixX \fixX + R \action [\Matrix{H}^\bodyX \bodyX + \bodyb] \label{eq::output_fixed} }\\\Aboxed{
\text{\normalfont{body-frame:}} ~\bodyh(R, \fixX, \bodyX) & =  R^{-1} \action [\fixb-\Matrix{H}^\fixX \fixX] - \Matrix{H}^\bodyX \bodyX 
\label{eq::output_body} }
\end{align}
with $\Matrix{H}^\fixX: \fixV \mapsto \outSpace$ and $\Matrix{H}^\bodyX: \bodyV \mapsto \outSpace$ two linear maps that \emph{commute with the action of $G$}, see Def. \ref{def::comute}, and $\fixb,\bodyb\in\mathcal{Y}$.
\end{dfn}
Note the minus signs in Eq. \eqref{eq::output_body}  make computations to follow clearer, but are to some extent arbitrary. 
%
The purpose of elements $R$ and $R^{-1}$ appearing in Def. \ref{def::2frame_obs} is to bring  variables either from body to fixed coordinates, or the opposite.  
%
\begin{dfn}[Natural vector dynamics]
\label{def::vector_discrete_dynamics}
Given a two-frames state space (Def. \ref{def::2-frame-state}) $  G\times \fixV\times \bodyV $ we define discrete-time    {natural  vector dynamics} as  $\chi_n= f_n(\chi_{n-1})$ where:
\begin{equation}\boxed{
f_n \begin{pmatrix}
R \\ \fixX \\ \bodyX
\end{pmatrix} = 
\begin{pmatrix}
R \\
[\fixF_n \fixX + \fixd_n] + R \action [\fixC_n \bodyX + \bodyu_n] \\
[\bodyF_n \bodyX + \bodyd_n] + R^{-1}\action  [\bodyC_n \fixX + \fixu_n] 
\end{pmatrix},}
\label{eq::vector_dynamics}
\end{equation}
and where matrices $\fixF_n: \fixV \mapsto \fixV, \quad \bodyF_n: \bodyV \mapsto \bodyV, \quad \fixC_n: \bodyV \mapsto \fixV, \quad \bodyC_n: \fixV \mapsto \bodyV $ \emph{all commute with the action of $G$},  and $\fixd_n, \bodyu_n$ (resp. $\bodyd_n$, $\fixu_n$) are vectors of $V$ (resp. $B$).
%
\end{dfn}
This family of dynamics is ubiquitous in robotics and navigation problems, as soon as some  dynamics $\generic{s}_n$ allows   transformation $R$ (that encodes a change of frame) to evolve as well. A ``natural'' class of such frame dynamics is now introduced, also based on a commutation hypothesis:

\begin{dfn}[Natural frame dynamics]
\label{def::frame-shift}
We define  natural frames dynamics  as $\chi_n= s_n(\chi_{n-1})$ where:
\begin{equation}
\label{eq::frame-shift}\boxed{
s_n\begin{pmatrix}
R \\
\fixX \\
\bodyX
\end{pmatrix}
=
\begin{pmatrix}\generic{s}_n^R(\chi) 
 \\
\fixX \\
\bodyX
\end{pmatrix}=
\begin{pmatrix}O_n\grouplaw R \grouplaw \Omega_n
 \\
\fixX \\
\bodyX
\end{pmatrix}}
\end{equation}
 with $O_n, \Omega_n \in G$   known elements of $G$ (inputs).  Moreover it is required that the maps $\fixX \mapsto O_n * \fixX$  and $\bodyX \mapsto \Omega_n * \bodyX$  \emph{commute  with the action} of $G$ on $\fixV$ and  $\bodyV$ respectively.
\end{dfn}

Each of the preceding notions brings properties. When all   are combined, we obtain a two-frames system we call natural.
\begin{dfn}[Natural two-frames observed system]
\label{def::two-frames-system}
A natural two-frames observed system is defined on a two-frames state space $(G, \fixV, \bodyV)$ as follows 
\begin{equation}\boxed{
\begin{aligned}
\chi_{n}  = & s_n \circ f_n \left( \chi_{n-1} \right),  \\
\fixY_n  = \fixh_n(\chi_n)& \quad \text{\emph{or}} \quad \bodyY_n = \bodyh_n(\chi_n),\end{aligned}}\quad\text{where:}\label{eq::two-frame-dynamics}
\end{equation}
\begin{itemize}\item $\fixh_n$ and $\bodyh_n$ are the natural outputs of Definition \ref{def::2frame_obs},\item $f_n$ is a natural vector dynamics of Definition \ref{def::vector_discrete_dynamics}, \item $s_n$ is a natural frame dynamics of Definition \ref{def::frame-shift}.\end{itemize}
\end{dfn}

In practice, some systems of interest are not completely ``natural'', so that we will also need the following definition. 
\begin{dfn}[Generic frame dynamics]
\label{biased:def}
When $\generic{s}_n^R(\chi)$ in Eq. \eqref{eq::frame-shift} is generic, that is, either it does not satisfy the commutation requirements of Def. \ref{def::frame-shift} or it is \emph{not} even of the form $
O_n\grouplaw R \grouplaw \Omega_n$ and  possibly depends   on state variables $\fixX$ and $\bodyX$, we speak of generic frame dynamics. 
\end{dfn} 
 \subsection{A prototypical example: navigation on flat earth}
 \label{expl::nav_IMU}
Consider a mobile body  equipped with an inertial measurement unit (IMU) providing gyroscope and accelerometer measurements, and a GNSS receiver providing position measurements $Y_n$. A simple discretization of the continuous equations       \cite{forster2017manifold} yields the  discrete-time dynamics:
 \begin{equation}\label{eq::nav-flat}
\left\lbrace \begin{aligned}
 \Matrix{R}_n & = \Matrix{R}_{n-1} \expm{m}{\Delta t{ \Matrix{[    \omega_n+b^\omega_{n-1}]_\times}}} \\
v_n & = v_{n-1}+\Delta t~(g + \Matrix{R}_{n-1} \left( a_n + b^a_{n-1} \right))\\
p_n &= p_{n-1} + \Delta t ~v_{n-1}
\\
b^\omega_n  &= b^\omega_{n-1},\quad 
b^a_n   = b^a_{n-1}
\end{aligned} \right. ,
 \end{equation} 
with observation $
Y_n = p_n$. In the above $\Delta t$ is a time step, $\Matrix{R}_n\in G=SO(3)$ denotes the transformation at time step $n$ that maps the frame attached to the IMU (body) to the earth-fixed  frame, $p_n \in \RR^3$ denotes the position of the body in space, $v_n \in\RR^3$  denotes \color{black}its velocity, $g$ is the earth gravity vector, $a_n,\omega_n \in \RR^3$ the accelerometer and gyroscope signals, $b^a_n$ the accelerometer bias, $b^\omega_n$ the gyroscope bias, $\expm{m}{}$ denotes the matrix exponential,   and     for any vector $\beta\in\RR^3$, the quantity \color{black} $(\beta)_\times$ denotes the skew-symmetric matrix such that $(\beta)_\times\gamma=\beta\times\gamma$ for any $\gamma\in\RR^3$. Let us cast Eq. \eqref{eq::nav-flat} into the framework of two-frames systems.  We can define a two-frames state space   where the group $G=SO(3)$ encodes the orientation $R_n$, the vector space $\fixV =  \RR^3 \times \RR^3$ encodes $\fixX_n=(p_n,v_n)$ and  $\bodyV =\RR^3 \times \RR^3$ encodes $\bodyX_n=(b^\omega_n , b^a_n) $. The group $G=SO(3)$ acts through the term-by-term action \eqref{tbt:eq} as: $\Matrix{R} \action \fixX=\Matrix{R} \action (v,p) = (\Matrix{R} v, \Matrix{R} p)$ and  $\Matrix{R} \action \bodyX = \Matrix{R} \action (b^\omega,b^a) = (\Matrix{R} b^\omega , \Matrix{R} b^a)$. The system matches Eq. \eqref{eq::two-frame-dynamics} where  components of $\generic{f}_n$ are:
\begin{equation}
\begin{aligned}
&\fixF_n = 
\begin{pmatrix}
\Id{3} & \0{3} \\
\Delta t \Id{3} &\Id{3}
\end{pmatrix},\quad 
\fixC_n =
\begin{pmatrix}
\0{3} & \Delta t \Id{3} \\
\0{3} & \0{3}
\end{pmatrix},~~
\fixd_n = \begin{pmatrix} \Delta t ~ g \\ \0{3,1} \end{pmatrix} \\
&\bodyu_n =\begin{pmatrix} \Delta t ~ a_n \\ \0{3,1} \end{pmatrix},~
\bodyF_n = 
\begin{pmatrix}
\Id{3} & \0{3} \\
\0{3} &\Id{3}
\end{pmatrix}, ~
\bodyC_n =
\0{6},~
\bodyd_n = 
\fixu_n = \0{6,1} 
\end{aligned}\label{TFS:inert}
\end{equation}and where $s_n^R$  is easily retrieved from the first line of \eqref{eq::nav-flat}. The position observation $
Y_n = p_n$ matches Eq. \eqref{eq::output_fixed} with:
\begin{equation*}
\Matrix{H}^\fixX =
\begin{pmatrix}
\0{3} & \Id{3}  
\end{pmatrix}, \qquad
\Matrix{H}^\bodyX = \0{3, 6}.
\end{equation*}
We let $G$ act  on the output through the term-by-term action \eqref{tbt:eq} as $\Matrix{R} \action Y_n=\Matrix{R} Y_n$. 
It can then be immediately  checked from Prop. \ref{easy:prop} that matrices $\fixF_n, \fixC_n, \bodyF_n, \bodyC_n, \Matrix{H}^\fixX, \Matrix{H}^\bodyX$ commute with the action of $SO(3)$, i.e., with rotations. This proves that we are here dealing with natural outputs and natural vector dynamics indeed. 
 However, 
we see in this example that the system fails to be entirely ``natural'', as  frame dynamics $\generic{s}_n$ is generic (see Def. \ref{biased:def}), owing to $\Omega_n= \expm{m}{\Delta t{ \Matrix{[    \omega_n+b^\omega_{n-1}]_\times}}}$ containing state element $b^\omega$.  In the present paper, we show that many systems of interest are   natural two-frames systems, and even for the present example where the frame dynamics fails to be natural, we show in Sect. \ref{simuss:sec} the proposed theory  brings  sensible improvement over existing methods, which makes it a general tool for systems defined on a two-frames state space.

%

 \section{The two-frames group structure}
 \label{TFG:sec}

The core of this paper is the introduction of a novel group structure on the class of two-frames  systems that allows casting them into the invariant filtering framework. In the case of natural two-frames system, it provides the practitioner with a systematic method to design observers  or filters that come with theoretical properties akin to linear observers and filters.

 \subsection{A group structure for two-frames systems}
 \label{sect::group-structure}
We start by introducing a relevant group structure.

\begin{dfn}[Two-frames group structure]
\label{def::2frame_law}
The $\biggrouplaw$ operation below defines a group structure on the two-frames state space $ G \times \fixV\times  \bodyV $ of Definition \ref{def::2-frame-state}:
\begin{align}\label{group:comp}
\begin{pmatrix} R_1 \\ \fixX_1 \\ \bodyX_1 \end{pmatrix} \biggrouplaw
\begin{pmatrix} R_2 \\ \fixX_2 \\ \bodyX_2 \end{pmatrix} =
\begin{pmatrix} R_1 \grouplaw R_2 \\ \fixX_1 + R_1 \action \fixX_2 \\ \bodyX_2 + R_2^{-1} \action \bodyX_1 \end{pmatrix}
\end{align}
We call it the two-frames group (TFG) law. Its identity element is $(Id, 0, 0)$ and the inverse is defined as:
\begin{equation}\label{eq::inverse}
(R, \fixX,\bodyX)^{-1}=(R^{-1}, -R^{-1} \action \fixX, -R \action \bodyX).
\end{equation}
$G \times \fixV\times  \bodyV  $ endowed with this structure is denoted by $\tFrames$. We also write it as $SO(\dimEmbedG)^+_{N_1,N_2}$ when $G=SO(\dimEmbedG)$ and   $\fixV=   \RR^{N_1\dimG}  $ and $\bodyV=  \RR^{N_2\dimG}$ are multi-vector spaces of Example \ref{expl::term-by-term}.
 \end{dfn}

The definition encompasses recently introduced Lie groups: $SO(\dimEmbedG)^+_{k,0}$ coincides with $SE_k(d)$ introduced for SLAM in \cite{bonnabel2012symmetries},  named $SE_k(d)$ in \cite{barrau2015ekf}, and exploited in \cite{wu_invariant-ekf_2017, heo_consistent_2018,brossard2017unscented, caruso_2018, heo2018consistent}, see also \cite{mahony2017geometric}. Similarly $SO(3)^+_{2,0}$ coincides with $SE_2(3)$, introduced in  \cite{barrau2017invariant,barrau2015non} for navigation and  exploited in \cite{Hartley-RSS-18,wang2020hybrid,hartley2019contact,brossard2017unscented,cohen2020navigation}.
In various implementations related to the IEKF, researchers - including the authors  -  have proposed to treat IMU biases in the state linearly, that is, complete $SE_2(3)$ group law with  $\bodyX_1 +  \bodyX_2$  regarding body variables (biases). This is advocated notably in \cite{Hartley-RSS-18,hartley2019contact,wu_invariant-ekf_2017,heo_consistent_2018,heo2018consistent, barrau2018invariant,brossard2017unscented,van2020invariant,cohen2020navigation} and   the term ``imperfect IEKF'' that refers to it, as well as the technique itself, was coined in \cite{barrau2015non}. We demonstrate in this paper that  the proposed group law \eqref{group:comp} is a far better option,  both theoretically and in practice.

Our purpose  is to \emph{cast natural two-frames systems into the invariant filtering framework}, i.e., make them fit into the framework of  \cite{barrau2019linear}, and beyond \cite{barrau2018invariant,barrau2017invariant},  using the TFG law. Invariant filtering builds upon two ingredients: output maps defined as group actions, and group affine dynamics.

\subsection{Natural two-frames outputs as TFG  actions}
\label{sect::2frames-results} 
We may now use the action $\action$ of $G$ on the output space $\mathcal{Y}$  as a building block to build a  group action of the TFG on $\mathcal{Y}$.
\begin{lem}[Two-frames group action]
\label{prop::action}
Let $\Matrix{H}^\fixX, \Matrix{H}^\bodyX$ be two matrices defining linear maps from, respectively, $\fixV$ and $\bodyV$ to the output space $\mathcal{Y}$.   $G$ acts on $\mathcal{Y}$  and we assume    \emph{its  action commutes with} $\Matrix{H}^\fixX$ and $\Matrix{H}^\bodyX$. Then the following  operation, denoted by  $\star$, defines a  group action on the output space:
\begin{align}
\tFrames \times \mathcal{Y} & \mapsto \mathcal{Y} \nonumber \\
(R, \fixX, \bodyX) \star \target & = \Matrix{H}^\fixX \fixX + R \action [\Matrix{H}^\bodyX \bodyX + \target] \label{star}
\end{align}
\end{lem}
\paragraph{proof}
We consider $\chi_1=(R_1, \fixX_1, \bodyX_1)$ and $\chi_2=(R_2, \fixX_2, \bodyX_2)$ and prove the identity $ [\chi_1 \biggrouplaw \chi_2]\star \target = \chi_1 \star [ \chi_2 \star \target ] $. As $R$ commutes with $\Matrix{H}^\fixX$ and $\Matrix{H}^\bodyX$ we have on the one hand:
\begin{align*}
[\chi_1 \biggrouplaw & \chi_2]\star \target =  (R_1 \grouplaw R_2, \fixX_1+R_1 \action \fixX_2, \bodyX_2 + R_2^{-1} \action \bodyX_1) \star \target \\
 & = \Matrix{H}^\fixX (\fixX_1 + R_1 \action \fixX_2) + R_1 \grouplaw R_2  \action [ \Matrix{H}^\bodyX (\bodyX_2 + R_2^{-1} \action \bodyX_1) + \target] \\
 & =  \Matrix{H}^\fixX   \fixX_1 + \Matrix{H}^\fixX R_1 \action \fixX_2 + \Matrix{H}^\bodyX R_1 \grouplaw R_2 \action  \bodyX_2+ \Matrix{H}^\bodyX R_1 \action \bodyX_1 + R_1 \grouplaw R_2 \action \target, 
\end{align*}
and on the other hand:
\begin{align*}
\chi_1 \star & [ \chi_2 \star \target ] = 
  (R_1, \fixX_1, \bodyX_1) \star [\Matrix{H}^\fixX \fixX_2 + R_2 \action (\Matrix{H}^\bodyX \bodyX_2 + \target)] \\
 & = \Matrix{H}^\fixX \fixX_1 + R_1 \action (\Matrix{H}^\bodyX \bodyX_1 + \Matrix{H}^\fixX \fixX_2 + R_2  \action (\Matrix{H}^\bodyX \bodyX_2 + \target)) \\
 & = \Matrix{H}^\fixX \fixX_1 + \Matrix{H}^\bodyX R_1  \action \bodyX_1 + \Matrix{H}^\fixX R_1 \action \fixX_2 + \Matrix{H}^\bodyX  R_1 \grouplaw R_2 \action \bodyX_2 + R_1 \grouplaw R_2 \action \target.
\end{align*}
Both expressions match, hence $\star$ is a group action of the TFG. $\blacksquare$

\vspace{.2cm}

Lemma \ref{prop::action} allows for our first major result, that is as follows:
\begin{thm}
\label{thm::output}
Consider a two-frames state space $\tFrames$, an output space $\mathcal{Y}$  endowed with an action of $G$, and   let   $\star$ be the operation defined by  \eqref{star}.  Then the  natural output map $\fixh$ of Eq. \eqref{eq::output_fixed}  merely writes: 
 \begin{align}\label{output11}
\fixY=\fixh(\chi)=\chi\star \bodyb.
\end{align}
Similarly,  the natural output map $\bodyh$ of Eq. \eqref{eq::output_body} writes:
\begin{align}\label{output12}
\bodyY=\bodyh(\chi)=\chi^{-1} \star \fixb.
\end{align}where in both cases $\star$ is a group action, from Lemma \ref{prop::action}.
\end{thm}\paragraph{proof}Proving \eqref{output11} immediately follows from   \eqref{star}. Regarding   body frame output  \eqref{output12},   $
\chi^{-1} \star \fixb   = (R^{-1}, -R^{-1} \action  \fixX, -R \action \bodyX) \star \fixb = -\Matrix{H}^\fixX R^{-1} \action   \fixX + R^{-1} \action  (\fixb-\Matrix{H}^\bodyX R \action  \bodyX) = R^{-1} \action  [\fixb-\Matrix{H}^\fixX  \fixX] - \Matrix{H}^\bodyX \bodyX$, which we recognize as \eqref{eq::output_body}. $\blacksquare$

\vspace{.2cm}

Using our novel group law \eqref{group:comp} we have defined a family of group actions of the TFG on the output such that all natural outputs write as group actions. This is a major property of two-frames systems, and one of our main contributions. 

\subsection{Natural vector dynamics as group affine dynamics}\label{GAP:exc}
Our second step is to prove that vector dynamics possess a key property of invariant filtering, namely the group affine property, with respect to the TFG. Let us first recall what this property is. Consider  general discrete-time dynamics on  the two-frames state space $\tFrames$ endowed with the TFG structure:
\begin{equation}
\chi_{n}   =   \phi_n(\chi_{n-1}). \label{genedyna}
\end{equation}
\begin{dfn}[Group affine dynamics \cite{barrau2017invariant,barrau2018invariant}]Dynamics \eqref{genedyna} is called group affine if $\phi_n: \tFrames \rightarrow \tFrames$ satisfies the ``group affine property'': for all $ \chi_1, \chi_2 \in\tFrames,$ we have \begin{equation} \phi_n(\chi_1 \biggrouplaw \chi_2)=\phi_n(\chi_1) \biggrouplaw \phi_n(Id)^{-1} \biggrouplaw \phi_n(\chi_2).\label{groupaffine:prop}\end{equation}
It may be checked the composition of two group affine  maps, i.e., maps that satisfy  \eqref{groupaffine:prop}, is also group affine.\label{groupaffine:def}\end{dfn}

Group affine dynamics have proved key to generalize properties of invariant observers related to autonomous estimation error equations.  Indeed, the first article wholly dedicated to error  autonomy is  \cite{arxiv-08}. This article and others on symmetry-preserving observers, see e.g.,  \cite{bonnabel2008symmetry,Vasconcelos,phogat2020discrete,khosravian2016state}
 focus on maps  of the form $\phi_n(\chi)=\chi\biggrouplaw\omega_n$ (left-invariance) or $\phi_n(\chi)=\nu_n\biggrouplaw\chi\biggrouplaw\omega_n$ (mixed-invariance, as in \cite{khosravian2016state}). Note both are particular instances of  maps satisfying  \eqref{groupaffine:prop} indeed.


{
\begin{rem}The   property \eqref{groupaffine:prop} is equivalent to   having $
 \phi_n(\chi)=g_n(\chi) \biggrouplaw \phi_n(Id)$ where  $g_n$ satisfies the automorphism property
\begin{equation}
\label{eq::morphism}
g_n(\chi_1 \biggrouplaw \chi_2) = g_n(\chi_1) \biggrouplaw g_n(\chi_2).
\end{equation}
This is shown by   multiplying both sides of \eqref{groupaffine:prop}  on the right by a factor $\phi_n(Id)^{-1}$, see  \cite{barrau2019linear} for more details. \end{rem}
}

To date,  a major shortcoming of the invariant filtering theory is that the group affine property or more restrictively  left or right invariance have been studied on a case-by-case basis. We  show in this section that the framework of two-frames systems provides a wide range of group affine dynamics, as well as a systematic approach to this property, thanks to the TFG law \eqref{group:comp}. Our second major result  is indeed as follows.
\begin{thm}
\label{thm::group-affine}
Natural vector dynamics  of Def.  \ref{def::vector_discrete_dynamics}, satisfies the group affine property \eqref{groupaffine:prop} with respect to the TFG law \eqref{group:comp}.
\end{thm}

\paragraph{proof}
We let $\phi_n=f_n$ as defined in \eqref{eq::vector_dynamics} and check it is group affine. {Letting $ g_n(\chi) = \phi_n(\chi)\biggrouplaw  \phi_n(Id)^{-1} $}, all we have to prove is \eqref{eq::morphism}. Denoting by $g_n^R(\chi), g_n^\fixX(\chi), g_n^\bodyX(\chi)$ the  components of $g_n(\chi)$ and applying the TFG law formula \eqref{group:comp}, desired identity \eqref{eq::morphism} becomes the following set of identities:
\begin{align}
g_n^R(\chi_1 \biggrouplaw \chi_2) & = g_n^R(\chi_1) g_n^R(\chi_2) \label{eq::morphism_R} \\
g_n^\fixX(\chi_1 \biggrouplaw \chi_2) & = g_n^\fixX(\chi_1) + g_n^R(\chi_1) * g_n^\fixX(\chi_2) \label{eq::morphism_fix}\\
g_n^\bodyX(\chi_1 \biggrouplaw \chi_2) & = g_n^\bodyX(\chi_2) + g_n^R(\chi_2)^{-1} * g_n^\bodyX(\chi_1)\label{eq::morphism_body}
\end{align}
First, let us write down explicit formulas for $g_n^R, g_n^\fixX, g_n^\bodyX$ using the definition $g_n(\chi)=\phi_n(\chi) \biggrouplaw \phi_n(Id)^{-1}$. Setting $R=Id, \fixX=0, \bodyX=0$ in \eqref{eq::vector_dynamics} we write $\phi_n(Id)$, then $\phi_n(Id)^{-1}$ using \eqref{eq::inverse}:
\begin{equation*}
\phi_n(Id) = 
\begin{pmatrix}
Id \\
\fixd_n + \bodyu_n \\
\bodyd_n + \fixu_n
\end{pmatrix}, \quad \phi_n(Id)^{-1} = 
\begin{pmatrix}
Id \\
-\fixd_n - \bodyu_n \\
-\bodyd_n - \fixu_n
\end{pmatrix}.
\end{equation*}
The product $g_n(\chi) = \phi_n(\chi) \biggrouplaw \phi_n(Id)^{-1}$ is then computed combining this expression for $\phi_n(Id)^{-1}$ with \eqref{eq::vector_dynamics} and \eqref{group:comp} 
which immediately yields for $g_n(
R , \fixX ,\bodyX)$ the components:
\begin{align}
g_n^R(\chi) & = R \label{eq::gR}\\
g_n^\fixX(\chi) & = \fixF_n \fixX + R \action \fixC_n \bodyX + \left( Id-R \right) \action \fixd_n \label{eq::gfix}\\
g_n^\bodyX(\chi) & = \bodyF_n \bodyX + R^{-1}\action  \bodyC_n \fixX + \left( R^{-1}-Id \right) \action \fixu_n.\label{eq::gbody}
\end{align}
Now, let us use those to check \eqref{eq::morphism_R}, \eqref{eq::morphism_fix}, \eqref{eq::morphism_body} hold. Checking \eqref{eq::morphism_R} is easy as both sides are simply $R_1 R_2$, see above. Left-hand side of \eqref{eq::morphism_fix} can be computed using \eqref{group:comp} then \eqref{eq::gfix}:
\begin{multline*}
g_n(\chi_1 \biggrouplaw \chi_2) =
\fixF_n (\fixX_1 + R_1 \fixX_2) + 
R_1 R_2 \fixC_n (\bodyX_2 + R_2^{-1}\bodyX_1) + \fixd_n - R_1 R_2 \fixd_n,
\end{multline*}
where we omit symbol $\action$ to alleviate notation. Expanding the parentheses and using the commutation properties $\fixF_n R_1 = R_1 \fixF_n$ and $\fixC_n R_2^{-1} = R_2^{-1} \fixC_n$ the latter equation becomes:
\begin{multline*}
g_n(\chi_1 \biggrouplaw \chi_2) = \fixF_n \fixX_1 + R_1 \fixF_n \fixX_2 + R_1 R_2 \fixC_n \bodyX_2 + R_1 \fixC_n \bodyX_1 + \fixd_n - R_1 R_2 \fixd_n.
 \end{multline*}
Right-hand side of \eqref{eq::morphism_fix} can be computed from \eqref{eq::gR}  and \eqref{eq::gfix}:
\begin{multline*}
g_n^\fixX(\chi_1) + g_n^R(\chi_1) * g_n^\fixX(\chi_2) =
\fixF_n \fixX_1 + R_1 \fixC_n \bodyX_1 + \fixd_n - R_1 \fixd_n  \\
  + R_1 \left[ \fixF_n \fixX_2 + R_2 \fixC_n \bodyX_2 + \fixd_n - R_2 \fixd_n  \right],
 \end{multline*}
where we recover all terms of $g_n^\fixX(\chi_1\biggrouplaw \chi_2)$ just  above, after distributing the factor $R_1$ and noticing cancellation of $R_1 \fixd_n $ terms, which proves \eqref{eq::morphism_fix}.
Left-hand side of \eqref{eq::morphism_body} can be computed using \eqref{group:comp} then \eqref{eq::gbody}:
\begin{multline*}
g_n^\bodyX(\chi_1\biggrouplaw \chi_2)  =
\bodyF_n (\bodyX_2 + R_2^{-1} \bodyX_1) + R_2^{-1} R_1^{-1} \bodyC_n (\fixX_1 + R_1 \fixX_2)
+ R_2^{-1} R_1^{-1} \fixu_n - \fixu_n.
\end{multline*}
Expanding the parentheses and using the commutation properties $\bodyF_n R_2^{-1} = R_2^{-1} \bodyF_n$ and $\bodyC_n R_1 = R_1 \bodyC_n$ the latter equation becomes:
\begin{multline*}
g_n^\bodyX(\chi_1\biggrouplaw \chi_2)  =
\bodyF_n \bodyX_2 + R_2^{-1} \bodyF_n \bodyX_1 +   R_2^{-1} R_1^{-1} \bodyC_n \fixX_1 + R_2^{-1} \bodyC_n \fixX_2
+ R_2^{-1} R_1^{-1} \fixu_n -\fixu_n.
\end{multline*}
Right-hand side of \eqref{eq::morphism_body} can be computed from \eqref{eq::gR}  and \eqref{eq::gbody}:
\begin{multline*}
g_n^\bodyX(\chi_2) + g_n^R(\chi_2)^{-1} * g_n^\bodyX(\chi_1) =
\bodyF_n \bodyX_2 + R_2^{-1} \bodyC_n \fixX_2 + R_2^{-1} \fixu_n \\ - \fixu_n
+ R_2^{-1} \left[ \bodyF_n \bodyX_1 + R_1^{-1} \bodyC_n \fixX_1 + R_1^{-1} \fixu_n - \fixu_n\right],
\end{multline*}
where we recover all terms of $g_n^\bodyX(\chi_1\biggrouplaw \chi_2)$ after distributing the factor $R_2^{-1}$ and noticing cancellation of $R_2^{-1}\fixu_n $ terms, which proves \eqref{eq::morphism_body}, and thus \eqref{eq::morphism} and hence Theorem \ref{thm::group-affine}.
 $\blacksquare$
 
\vspace{.2cm}

\subsection{Natural frame dynamics as  group affine dynamics}\label{nature:sec}

The last step of our analysis of two-frames systems regards natural frame dynamics of Def. \ref{def::frame-shift}. A counterpart of the result of the previous section is derived, then two cases of specific interest are highlighted.

\begin{thm} \label{thm::unbiased}
Natural frame dynamics of Def. \ref{def::frame-shift}, satisfies the group affine property \eqref{groupaffine:prop} with respect to the TFG law \eqref{group:comp}.
\end{thm}

\paragraph{proof}
First step is writing \eqref{def::frame-shift} as $s_n=s_n' \circ s_n''$ with:
\begin{equation*}
s_n'\begin{pmatrix}
R \\ \fixX \\ \bodyX
\end{pmatrix} =
\begin{pmatrix}
O_n \\ 0 \\ 0
\end{pmatrix}
\biggrouplaw
\begin{pmatrix}
R \\ \fixX \\ \bodyX
\end{pmatrix}
\biggrouplaw
\begin{pmatrix}
\Omega_n \\ 0 \\ 0
\end{pmatrix}, \quad
s_n'' \begin{pmatrix}
R \\ \fixX \\ \bodyX
\end{pmatrix} =
\begin{pmatrix}
R \\ O_n^{-1} \fixX \\ \Omega_n \bodyX
\end{pmatrix}
\end{equation*}
This identity can be checked by direct computation based on \eqref{group:comp}.
The operator $s_n'(\chi)$ is mixed-invariant and always satisfies the group-affine property (see   Corollary 19 of \cite{barrau2019linear}), so $s_n$ is group-affine if $s_n''$ is, as a composition of two group-affine operators.
As a second step note that $s_n''$ matches the definition of  vector dynamics, with the maps $\fixX \mapsto O_n^{-1} * \fixX$ and $\bodyX \mapsto \Omega_n * \bodyX$ playing the roles of $\fixF_n$ and $\bodyF_n$. If they commute with the action of $G$ as required by Def. \ref{def::vector_discrete_dynamics}, Theorem \ref{thm::group-affine} then ensures $s_n''$  is group-affine and so is  $s_n$. All we have to do is thus show the map $\fixX \mapsto O_n^{-1} \fixX$ commutes with the action of $G$ if the map $\fixX \mapsto O_n \fixX$ does. Let us assume the latter is true. Then for any $R \in G, \fixX \in \fixV$ we have $O_n^{-1}   R * \fixX = O_n^{-1}   (R  O_n) * (O_n^{-1} * \fixX)=O_n^{-1}   (O_n  R) * (O_n^{-1} * \fixX) = R  O_n^{-1} * \fixX$, proving the last step, and thus Theorem \ref{thm::unbiased}.
 $\blacksquare$
 
\vspace{.2cm}

The difficulty here is to figure out when the commutation property required by the definition of natural frame shifts holds. In practice, the following specific cases are meaningful.
\begin{thm}\label{prop::no-mu}
Frame dynamics of the shape Eq. \eqref{eq::frame-shift} are  \emph{natural} frame dynamics as soon as we have (a) or (b) or (c):
\begin{description}
\item[(a)] The state variable boils down to $(R,\bodyX)$,
and  frame dynamics read  $s_n^R(R, \fixX, \bodyX) =  O_n \grouplaw R  $,
\item[(b)] The state variable boils down to $(R,\fixX)$,
and  frame dynamics read  $s_n^R(R, \fixX, \bodyX) =  R \grouplaw~\Omega_n$,
\item[(c)] Frame dynamics read  $s_n^R(R, \fixX, \bodyX) =  R$.
\end{description}
As a result, in each of these cases frame dynamics of the shape Eq. \eqref{eq::frame-shift}  satisfy the group-affine property \eqref{groupaffine:prop} with respect to the TFG law \eqref{group:comp}.
\end{thm}
\paragraph{proof}
Let us check the commutation hypotheses of Def. \ref{def::frame-shift}. Consider, e.g., (a).  The map  $\fixX \mapsto O_n * \fixX$ is then a trivial action, hence commutes. Besides, 
$\bodyX \mapsto \Omega_n * \bodyX$ commutes with action of $G$ as $\Omega_n=Id$. Thus Theorem \ref{thm::unbiased} applies.
 $\blacksquare$

\vspace{.2cm}

Albeit extremely degenerate with respect to the entire TFG theory,   Case  {(b)} of Thm. \ref{prop::no-mu}  actually covers \emph{all} previously discovered group affine dynamics. The TFG structure with $G=SO(d)$ then boils down to the group $SE_k(d)$.

In particular, coming back to example of Sec. \ref{expl::nav_IMU}, we recover immediately from   Thm. \ref{thm::group-affine} and Thm.  \ref{prop::no-mu} case (b) that in the absence of IMU biases (which are encoded by $\bodyX_n$), the equations of inertial navigation are group affine, which is one of the main discoveries of  \cite{barrau2017invariant}. 
Moreover, if we ignore  gyroscope bias  $b^\omega$,  but we want to estimate an accelerometer bias, we let   $\bodyX_n=b^a_n$, and we get    frame dynamics of the shape  \eqref{eq::frame-shift} with   $ O_n  =\Matrix{I}_3, \Omega_n= \expm{m}{\Delta t{ \Matrix{[    \omega_n]_\times}}}$.  
Case (c) of Thm. \ref{prop::no-mu} shows that frame dynamics become   group affine as soon as $\omega_n=0$ yielding $\Omega_n=\Matrix{I}_3$, i.e., attitude $R$ is constant  but possibly unknown.   
We  thus recover from Thm. \ref{thm::group-affine} and Thm.  \ref{prop::no-mu} case (c) that inertial navigation dynamics  with accelerometer bias become group affine   whenever the craft is moving straight forward, which is quite  common when navigating. 

Thm.  \ref{prop::no-mu} seems to suggest that whenever    $\fixX$ and $\bodyX$   coexist in the state, see case (c), we must have $R_n=R_{n-1}$ for  frame dynamics to be group affine. This is not always true. Notably, our second major result on frame dynamics is as follows.  
\begin{thm}\label{prop::abelian}
If   $G$ is an abelian group, i.e., for all $R_1,R_2\in G$ we have $R_1R_2=R_2R_1$, then all frame dynamics of the shape \eqref{eq::frame-shift} are \emph{natural} frames dynamics. As a result, they satisfy the group-affine property \eqref{groupaffine:prop} with respect to the TFG law \eqref{group:comp}.
\end{thm}\paragraph{proof}They  then satisfy commutation hypotheses of Def. \ref{def::frame-shift} as group elements commute with $O_n$ and $\Omega_n$.  $\blacksquare$

\vspace{.2cm}

This result is of major importance as the group $G=SO(2)$ is abelian, and it models the heading of wheeled robots in 2D, so that a wide class of systems of practical interest are two-frames systems based on an abelian group $G$.

 \subsection{Two-frames systems as linear observed systems on the TFG}
 
We now have the machinery to meet our end objective: cast two-frames systems into the framework of invariant filtering.
\begin{dfn}[from \cite{barrau2019linear}]
\label{def::LOSG}
A \emph{linear observed system on group} is a dynamical system $(\chi_n)_{n\geq 0}$ observed through measurements $\fixY_n$ or $\bodyY_n$ following equations of the form:
\begin{align}\boxed{
\underbrace{
\begin{matrix}
\chi_{n} & = & \phi_n(\chi_{n-1}) \\
\fixY_n & = & \chi_n \star b
\end{matrix}}_{\emph{Left action case}}
\qquad \text{or} \qquad
\underbrace{
\begin{matrix}
\chi_{n} & = &\phi_n(\chi_{n-1}) \\
\bodyY_n &= &\chi_n^{-1} \star b
\end{matrix}
}_{\emph{Right action case}}} \label{eq:LOSOG}
\end{align}
where 
  $\star$ is an   {action}  on the output vector space $\outSpace_n$  and $\phi_n$ satisfies the  group affine property   \eqref{groupaffine:prop}.
\end{dfn}
The terminology for the right-hand case comes from the map $\left( \chi, b \right) \mapsto \chi^{-1} \star ~b$ being a right action, see  \cite{Olve93a}.  

Linear observed systems on groups are the systems brought up and studied by the  invariant filtering theory. They encompass    \cite{barrau2017invariant} where the studied groups are matrix Lie groups and  the action on the output space is then a mere matrix-vector multiplication, and where the group affine property  \eqref{groupaffine:prop} is given in continuous time (see Supplementary material below). 
In discrete time   Def. \ref{def::LOSG} appears in \cite{barrau2019linear}, see also \cite{barrau2018invariant}.
Gathering our results we get:
\begin{thm}\label{thm:aff}Using the  TFG law \eqref{group:comp}  and    action \eqref{star}, natural two-frames observed systems of Def. \ref{def::two-frames-system} fit the Def. \ref{def::LOSG} of  linear observed systems \eqref{eq:LOSOG} on the group $\tFrames$,  with $\phi_n=s_n\circ f_n$.
\end{thm}
The introduction of the TFG thus allows for     the discovery of new systems  that actually  fit into the invariant filtering framework. Concrete examples are provided in Sect. \ref{sect::expl}.

\section{Observer design for two-frames systems}
\label{sect::design}

The TFG structure and action enabled us to write natural two-frames as in   Def. \ref{def::LOSG}. As soon as this can be done,  invariant filtering theory \cite{barrau2017invariant, barrau2018invariant, barrau2019linear}  automatically brings observers that inherit some key  properties of linear observers. 

\subsection{Linear observers on the two-frames group}
 The following Prop. \ref{prop::innov}, \ref{prop::update5}, \ref{prop::propagation5} are  already known from \cite{barrau2019linear}, but proofs are given  so that the paper is self-contained.

For systems  that formally write as \eqref{eq:LOSOG}, one may introduce ``linear observers on groups'', see \cite{barrau2019linear}, by mimicking linear observers, and then readily inherit a number of properties of the linear case. To start things off,  we   build alternative innovation terms using the group action on the output  as:
\begin{equation}
\label{eq::innov}
\bodyZ_n = \hat{\chi}_{n|n-1}^{-1} \star \fixY_n - b \quad or \quad  \fixZ_n = \hat{\chi}_{n|n-1} \star \bodyY_n-  b.
\end{equation} 
Those terms measure a prediction error between the predicted output and the measured output, in a way that differs from the classical linear innovation $ \fixY-\hat\chi \star b$ or $\bodyY_n-\hat \chi_n^{-1} \star b$.

Observers are then readily defined as a copy of the dynamics followed by a ``multiplicative'' update, that accounts for the innovation (prediction error). 
\begin{dfn}[Linear observer on group \cite{barrau2019linear}]\label{inv:ob:d}
A linear observer  on  group $\tFrames$  for   system  \eqref{eq:LOSOG} of  Def. \ref{def::LOSG}  consists of estimates $\hat{\chi}_{n|n-1}$ and $\hat{\chi}_{n|n}$ defined through a succession of propagation   (copy of the dynamics) and update steps:
\begin{gather*}
 \hat{\chi}_{n|n-1} = \phi_n(\hat{\chi}_{n-1|n-1}) \qquad \emph{[Propagation step]}\\
\left\{
\begin{aligned}
\hat{\chi}_{n|n} & = \hat{\chi}_{n|n-1} \biggrouplaw  L_n(\bodyZ_n) 
 & \emph{[Update step: left action case]}
\\
\hat{\chi}_{n|n} & =  {L}_n(\fixZ_n)  \biggrouplaw \hat{\chi}_{n|n-1}
 & \emph{[Update step:  right action case]}
\end{aligned}
\right.
\end{gather*}
with $L_n: \outSpace \mapsto \tFrames$  any mapping and $\bodyZ_n,\fixZ_n$ given by \eqref{eq::innov}.
\end{dfn}

Linear observers on groups are designed to ensure striking properties of a specific type of error variables called ``invariant error variables" and playing a   central role in the theory of invariant filtering and equivariant observers:
\begin{align}\boxed{
\bodyE_{n|n} = \hat{\chi}_{n|n}^{-1} \biggrouplaw \chi_n,
\qquad
\fixE_{n|n} = \chi_{n} \biggrouplaw \hat{\chi}_{n|n}^{-1},}\label{errors:eqq}
\end{align}
and similarly   we let $\bodyE_{n|n-1} = \hat{\chi}_{n|n-1}^{-1} \biggrouplaw \chi_n$ and $
\fixE_{n|n-1} = \chi_{n} \biggrouplaw \hat{\chi}_{n|n-1}^{-1}$.   $\bodyE$ and $\fixE$ are called respectively  left- and right-invariant error variables. In previous work they were denoted by $\eta_{n|n}^L$ and $\eta_{n|n}^R$, but   here we use $\bodyE_{n|n}$ and $\fixE_{n|n}$  to alleviate  notation.

\begin{dfn}The error evolution is said  state-trajectory independent (or autonomous) when $e_{n|n}$ (resp. $E_{n|n}$)  depends only of  $e_{n-1|n-1}$ (resp. $E_{n-1|n-1}$) and $n$, i.e., it does not explictly depend on the estimated state variables $\hat \chi$.\label{SIEP:def}
\end{dfn} 
State-trajectory independence of the error is   arguably the most important result of invariant filtering, in that it ensures the mapping $L_n$ can be tuned independently from the actual (unknown) trajectory followed by the state.  
In this regard, the group-theoretic approach allows for an extension of the properties of the linear case to the nonlinear case.  Let us now study this property at each step of the observer's construction. 
\begin{prop}
\label{prop::innov}
Innovation  $\fixZ_n$ (resp. $\bodyZ_n$)  is a function of the invariant error $\fixE_{n|n-1}$ (resp. $\bodyE_{n|n-1}$)  only. We have indeed:
\begin{equation}\label{innov:abs}
\bodyZ_n = \bodyE_{n|n-1} \star b-b,  \qquad \fixZ_n = \fixE_{n|n-1}^{-1} \star b-b.
\end{equation}
\end{prop} 
\paragraph{proof}
Having cast the outputs as   actions in \eqref{eq:LOSOG} allows writing  $\hat{\chi}_{n|n-1}^{-1} \star \fixY_n=\hat{\chi}_{n|n-1}^{-1}\star (\chi_n  \star b)=(\hat{\chi}_{n|n-1}^{-1}\biggrouplaw  \chi_n )\star b=E_{n|n-1}\star b$ so $\bodyZ_n =  E_{n|n-1} \star b-b$, and similarly for $z_n$.
 $\blacksquare$

\vspace{.2cm}

The latter feature is a pivotal property that linear observed systems on groups share with conventional linear observers. This allows proving in turn:
\begin{prop}
\label{prop::update5}Error evolution at update step is state-trajectory independent. Indeed, we have
\begin{equation}\label{update:abs5}\bodyE_{n|n}  =  L_n(\bodyZ_n)^{-1}\biggrouplaw \bodyE_{n|n-1},\quad 
 \fixE_{n|n}   =\fixE_{n[n-1}\biggrouplaw L_n(\fixZ_n)^{-1}.
\end{equation}
\end{prop} 
\paragraph{proof}
 We obviously have from Def. \ref{inv:ob:d} that $
\bodyE_{n|n}  = \hat{\chi}_{n|n}^{-1} \biggrouplaw \chi_n=L_n(\bodyZ_n)^{-1}  \biggrouplaw\hat{\chi}_{n|n-1}^{-1}  \biggrouplaw \chi_n=L_n(\bodyZ_n)^{-1}\biggrouplaw\bodyE_{n|n-1}$ and similarly 
$\fixE_{n|n}  = \chi_{n} \biggrouplaw \hat{\chi}_{n|n}^{-1}= \chi_{n} \biggrouplaw \hat{\chi}_{n|n-1}^{-1}\biggrouplaw L_n(\fixZ_n)^{-1}=\fixE_{n|n-1}\biggrouplaw L_n(\fixZ_n)^{-1}$. Using \eqref{innov:abs} completes the result. 
 $\blacksquare$

\vspace{.2cm}

Finally, state-trajectory independence at propagation step is  \emph{equivalent} to group affine dynamics, see \cite{barrau2017invariant,barrau2019linear}. This implies: 
\begin{prop}
\label{prop::propagation5}If $ \phi_n$ satisfies the group affine property \eqref{groupaffine:prop} then error evolution at propagation step is state-trajectory independent. Indeed, we have
\begin{equation}\label{propagation:abs5}\begin{aligned}\bodyE_{n|n-1}  &= \phi_n(Id)^{-1} \biggrouplaw~\phi_n(\bodyE_{n-1|n-1} ),\\
 \fixE_{n|n-1}   &= \phi_n(\fixE_{n-1|n-1} )\biggrouplaw~ \phi_n(Id)^{-1}.\end{aligned}
\end{equation}
\end{prop} 
\paragraph{proof}Let's prove the first equality. The second is proved similarly, see \cite{barrau2019linear}.
{
First, Eq. \eqref{groupaffine:prop} gives $\phi_n(\hat  \chi_{n-1|n-1}^{-1}) \phi_n(Id)^{-1} \phi_n(\hat  \chi_{n-1|n-1}) = \phi_n(\hat  \chi_{n-1|n-1}^{-1} \hat  \chi_{n-1|n-1})  = \phi_n(Id) $, that we re-write as $\phi_n(\hat  \chi_{n-1|n-1})^{-1} = \phi_n(Id)^{-1} \phi_n(\hat  \chi_{n-1|n-1}^{-1}) \phi_n(Id)^{-1} $. We use this identity in order to propagate the error: $\bodyE_{n|n-1} = \hat  \chi_{n|n-1}^{-1} \chi_{n}$ =$ \phi_n(\hat  \chi_{n-1|n-1})^{-1} \phi_n(\chi_{n-1}) = \phi_n(Id)^{-1} \phi_n(\hat \chi_{n-1|n-1}^{-1}) \phi_n(Id)^{-1}\phi_n(\chi_{n-1})$ and using \eqref{groupaffine:prop} again we obtain: $$\bodyE_{n|n-1} = \phi_n(Id)^{-1} \phi_n(\hat \chi_{n-1|n-1}^{-1}\chi_{n-1}) = 
\phi_n(Id)^{-1} \phi_n(\bodyE_{n-1|n-1}).$$
}
 $\blacksquare$

\vspace{.2cm}

Gathering the last results we see linear observers on groups automatically yield state-trajectory independent error evolution for linear systems on groups at all steps.  Having picked $\biggrouplaw$ as the TFG law \eqref{group:comp} and $\star$ as action  \eqref{star}, and recalling   Thm. \ref{thm:aff},   Def. \ref{inv:ob:d} thus allows us to  meet our second   objective:  to manage to build state-trajectory independent observers for two-frames systems.
\begin{thm}\label{thm:error}
The evolution of error \eqref{errors:eqq} is state-trajectory independent both at propagation and update steps for the observer of Def. \ref{inv:ob:d} applied to \emph{any} natural two-frames system.
\end{thm}

\subsection{Invariant extended Kalman filter  design }\label{IEKF22:sec}
So far, we have shown one may derive observers for the class of two-frames systems of Section  \ref{2frame:sec} that possess properties akin to the linear case, since they fit into the linear observed systems on groups  (or more simply invariant filtering) framework. Theses properties facilitate gain design and convergence analysis and open the door to observers possessing strong mathematical  guarantees.  The   simplest yet very  efficient approach to design meaningful gains is then to follow the extended Kalman filter (EKF)  methodology, which consists in linearizing the error equation and tune the gains on the linearized error system using the linear Kalman filter, i.e., least squares techniques.  Indeed, the invariant extended Kalman filter  (IEKF)   \cite{barrau2015non, barrau2017invariant,barrau2018invariant}  consists of a linear observer on the group, as in Def. \ref{inv:ob:d}, where  update terms are chosen to be: 
\begin{align}
\text{\normalfont{Obs. \eqref{eq::output_fixed}}} \Rightarrow L_n(\bodyZ_n) &= \exp_{\tFrames}(\Matrix{K_n} \bodyZ_n) \label{eq::gains1}\\\text{\normalfont{Obs. \eqref{eq::output_body}}} \Rightarrow  L_n(\fixZ_n) &= \exp_{\tFrames}(\Matrix{K_n} \fixZ_n) \label{eq::gains2}
\end{align}
with $\exp_{\tFrames}(\cdot)\in \tFrames$  
the exponential map  of the TFG    and where the gains   $\Matrix{K}_n$ are tuned using the Kalman filter applied to the linearized error system. Deriving IEKFs for two-frames systems thus requires $i)$ computing the exponential map of the TFG, and $ii)$  linearizing  the error equations   obtained previously,   yielding Jacobian matrices.  One of the main contributions of the present paper is to have derived generic formulas usable for any two-frames systems.  Explicit formulas for the exponential map are given in Appendix \ref {expo:sec}. Error equations   \eqref{update:abs5}, \eqref{propagation:abs5} in terms of two-frames variables are provided in Appendix \ref{error:ec:sec}.  The main idea to linearize them  is to define  linearized errors $\bodyXi_{n|n}, \fixXi_{n|n}$, see \cite{arxiv-08,barrau2017invariant,barrau2018invariant}, via
\begin{align}\label{exp:gain}
\bodyE_{n|n} = \exp_{\tFrames}(\bodyXi_{n|n}), \qquad e_{n|n} = \exp_{\tFrames}(\fixXi_{n|n}).
\end{align}
This allows linearizing the error system   in $\bodyXi_{n|n}$ and $\fixXi_{n|n}$ as
\begin{align}
\xi_{n|n-1} & = \Matrix{A}^s_n \Matrix{A}^v_n \xi_{n-1|n-1} \label{eq::first-order1},\\
\xi_{n|n} & = \left( \Matrix{I}  - \Matrix{K}_n \Matrix{H}_n \right) \xi_{n|n-1},\label{eq::first-order2}
\end{align}
with $\Matrix{A}^s_n ,\Matrix{A}^v_n,\Matrix{H}_n$ Jacobian matrices corresponding respectively to  frame dynamics, vector dynamics, and output map, and where $\Matrix{K}_n$ is the Kalman gain. Proofs  and explicit generic formulas for the Jacobians are provided in Appendix     \ref{jacob:sec}.

Eq. \eqref{eq::first-order1}  is the linearization   of the propagation of the error \eqref{propagation:abs5}, while  \eqref{eq::first-order2}  is the linearization of update  \eqref{update:abs5} under IEKF gain design  \eqref{eq::gains1}, \eqref{eq::gains2}. The IEKF does not directly  use  \eqref{eq::first-order1}-\eqref{eq::first-order2}, though. Instead, the IEKF methodology associates a ``noisy'' system to the problem,  based on realistic sensor noise characteristics and then completes  \eqref{eq::first-order1}-\eqref{eq::first-order2} with the linearized noise terms to tune the gain $\Matrix{K}_n$. As a desirable byproduct, it  allows the Riccati's covariance matrix $\Matrix{P_{n|n}}$ to convey the correct extent of statistical uncertainty about the state, and the IEKF equations then provide  \emph{first-order optimal tuning}   for the probablistic   problem. The noisy system, and the resulting  computation of noise matrices $\Matrix{\hat{Q}_n},\Matrix{\hat{N}_n}$,  have been moved to   Appendix  \ref{noise:sec}. By letting the underlying Lie group in the IEKF methodology   of  \cite{barrau2017invariant} be the TFG,  we readily obtain: 
\begin{dfn}[TFG-IEKF]\label{def:TFGIEKF}The propagation step of TFG-IEKF is a copy of the dynamics. For fixed-frame [resp. body-frame] observations of the form   \eqref{eq::output_fixed}   hence  \eqref{output11} [resp. \eqref{eq::output_body}   hence  \eqref{output12}], update is defined as the left-action [resp. right-action] case of Def. \ref{inv:ob:d}. $L^R,L^\bodyX,L^\fixX$ are extracted from  \eqref{eq::gains1} [resp. \eqref{eq::gains2}] using the exponential of the TFG,   where the  gain is tuned through the following Riccati equation:
\begin{equation}
\label{eq::riccati}
\begin{aligned}
&\Matrix{P_{n|n-1}} =\Matrix{A}^s_n \Matrix{A}^v_n \Matrix{P_{n-1|n-1}}(\Matrix{A}^s_n \Matrix{A}^v_n)^T+ \Matrix{\hat{Q}_n},\\
& \Matrix{S_{n}} =\Matrix{H}_n \Matrix{P_{n|n-1} } \Matrix{H}_n^T+ \Matrix{\hat{N}_n} ,\quad \Matrix{K_n}= \Matrix{P_{n|n-1}}\Matrix{H}_n^T \Matrix{S_{n}}^{-1},\\
& \Matrix{P_{n|n}} =\left( \Matrix{I} - \Matrix{K}_n \Matrix{H}_n \right) \Matrix{P_{n|n-1}},
\end{aligned}
\end{equation}
with $\Matrix{P_{0|0}}$ the prior covariance of error $\bodyE_0$ [resp. $\fixE_0$]. 
\end{dfn}
In Appendix  \ref{noise:sec}, a formula allows for retrieving  $\Matrix{P_{0|0}}$ from the  initial covariance matrix in the original variables $(R,\fixX,\bodyX)$.

\subsection{Properties of the TFG-IEKF}

Various properties of the IEKF are  recovered. Let us start with one of the most striking properties of invariant filtering. 
\begin{thm}[log-linear property]
\label{thm::log-linear}
Natural two-frames dynamics fully possess the log-linear property of the error of \cite{barrau2017invariant}, as they were shown to be group affine. This means deterministic nonlinear error propagation \eqref{propagation:abs5}  is \emph{ exactly equivalent} to 
\eqref{eq::first-order1}, via the  nonlinear correspondance \eqref{exp:gain}. 
\end{thm}
Invariant Kalman filtering targets state independent Jacobians. 
\begin{prop} Each part involved in the definition of a natural two-frames system brings a state-independent Jacobian.\begin{itemize}\item The Jacobian matrix $\Matrix{H}_n$ is state-trajectory independent as soon as the output is natural, \item  the Jacobian matrix $\Matrix{A}^v_n$ is state-trajectory independent as soon as the vector dynamics is natural,\item   the Jacobian matrix $\Matrix{A}^s_n$ is state-trajectory independent as soon as  the frame dynamics is natural.\end{itemize}
\end{prop} 
By contrast,  noise matrices $\Matrix{\hat{Q}_n},\Matrix{\hat{N}_n}$ often depend on the state. However, this is not an impediment to convergence properties. 
\begin{prop}\label{prop:con} For natural two-frames observed systems, local convergence of TFG-IEKF as an observer is guaranteed  under a list of observability assumptions about the {true} state trajectory displayed in  \cite{barrau2017invariant}.
\end{prop}
 We recall in passing that one of the strengths of the IEKF framework is to enable one to prove convergence properties under standard observability assumptions on the followed trajectory, and not under uncheckable observability assumptions concerning the estimated trajectory, see  \cite{barrau2017invariant}.
 
 \vspace{.2cm}
 
 It would be misleading, though, to believe the   group affine property is required to design and use  IEKFs in practice. 
\begin{rem}Any system of the form \eqref{eq:LOSOG} gives rise to an IEKF, worth testing, even if $\phi_n$ is not group affine. Besides,  if  frame dynamics is {not} group affine but vector dynamics is, then \emph{only} Jacobian $\Matrix{A}_n^s$ is state-dependent, and such an IEKF is   likely to outperform the standard EKF, as illustrated in Sect. \ref{IMU:exp}, although convergence properties are no longer proved. 
\end{rem}

 {
As an aside,  let us see how the insight of \cite{barrau2019extended} combined with the present properties  leads to a powerful result having relevant practical consequences. For terrestrial vehicles  equipped with a 3D  IMU, actual trajectories are often virtually planar and the orientation $R_n$ boils down to rotations around the vertical, which commute, leading to an  abelian subgroup of $SO(3)$. To cast such system into the IEKF framework even when some body-frame vectors such as a lever arm are estimated, it is thus tempting to apply Thm. \ref{prop::abelian}, readily yielding Thm. \ref{thm:error}.  Precisely,  Thm. \ref{prop::abelian} combined with the other results ensures  the following:
\begin{crl}
\label{permanent:crl2}
If $ \forall n\in\mathbb N ~O_n,\Omega_n,R_n,\hat R_{n|n}$  all commute with each other,  the error has state independent evolution.
\end{crl}
The hypothesis  holds as long as $O_n,\Omega_n,R_n,\hat R_{n|n}$ are  rotations around the vertical axis, which seems true during the phases where the motion is planar. There is a catch, though, as   this means   that the observer   estimates $\hat R_{n|n}$ remain  in the commuting subspace, which is not obvious. But the main result of \cite{barrau2019extended} is that if at some point  $\hat R_{n|n}$ falls into the subgroup of rotations around the vertical, and if the  covariance matrix  is consistent with that information, then the IEKF's   estimates  $\hat R_{n|n}$ remain  in this subgroup indeed, so that Corollary \ref{permanent:crl2} applies.  This means 3D terrestrial vehicle navigation offers an almost theoretically  perfect  application of the (TFG-) IEKF. 
}

\section{Applications}
\label{sect::expl}
The TFG has allowed us to cast two-frames observed systems into the framework of invariant filtering, and to readily inherit  strong properties. However,   to implement the observers in practice, we need  to derive explicit formulas in the original two-frames state variables $\hat R,\hat\fixX,\hat\bodyX$. As a byproduct, this provides the reader with a more concrete picture of the  derived class of   observers.  This is the first step of this section devoted to applications. Then, we consider three nontrivial systems of practical interest. The first two   examples are  novel and have never been shown to fit into the theory of invariant filtering. Applying the theory we derive  the form of observers   having autonomous error equations, without writing down all their properties, owing to space limitation. The third example shows the methodology  allows for significant improvement over state-of-the-art IEKF   for navigation.

\subsection{Explicit formulas for implementation}\label{explicit:sec}

Let us first  transpose the abstract formulas of Def. \ref{inv:ob:d}  by using the definition of the TFG law \eqref{group:comp} and   action \eqref{star}, starting with the innovation. Let $\fixY_n$  and $\bodyY_n $ be natural outputs in, respectively, the fixed and body frame as in Def. \ref{def::2frame_obs}. The innovation variable $\bodyZ_n$ in \eqref{eq::innov} associated to fixed-frame output $\fixY_n$, i.e., Obs. \eqref{eq::output_fixed}, writes:
\begin{equation}
\label{eq::innovBody}\boxed{
\bodyZ_n = \hat{R}_{n|n-1}^{-1}  \action \left( \fixY_n - \Matrix{H}^\fixX \hat{\fixX}_{n|n-1} \right)- \Matrix{H}^{\bodyX} \hat{\bodyX}_{n|n-1} - \bodyb_n}
\end{equation}
while the innovation variable $\fixZ_n$ in \eqref{eq::innov}  associated to body-frame output $\bodyY_n$,  i.e., Obs. \eqref{eq::output_body},  writes:
\begin{equation}
\label{eq::innovFix}\boxed{
\fixZ_n = \hat{R}_{n|n-1} \action  \left( \bodyY_n + \Matrix{H}^{\bodyX}\hat{\bodyX}_{n|n-1} \right) + \Matrix{H}^\fixX \hat{\fixX}_{n|n-1} - \fixb_n }
\end{equation}
We recognize   that we merely have $ \bodyZ_n = \hat{R}_{n|n-1}^{-1} *  (\fixY_n -\hat \fixY_n )$, where $ \hat \fixY_n =  \Matrix{H}^\fixX \hat{\fixX}_{n|n-1} - \hat{R}_{n|n-1} *  \left[\Matrix{H}^{\bodyX} \hat{\bodyX}_{n|n-1} +   \bodyb_n \right]
$ denotes the predicted output. Similarly $\fixZ_n$ is the usual output error $
 \bodyY_n - \hat{R}_{n|n-1}^{-1} *  \fixb_n + \hat{R}_{n|n-1}^{-1}  * \Matrix{H}^\fixX \hat{\fixX}_{n|n-1} + \Matrix{H}^{\bodyX} \hat{\bodyX}_{n|n-1}
$, moved to the opposite reference frame by application of $\hat{R}_{n|n-1}$'s action. This modification is characteristic of invariant filtering.

Now, to transpose the observer formulas of Def. \ref{inv:ob:d}  into the original variables, let us denote the estimates  at time $n$ by:
\begin{equation*}
\hat{\chi}_{n^-|n-1}, \qquad \hat{\chi}_{n|n-1}, \qquad \hat{\chi}_{n|n},
\end{equation*}
respectively after applying the vector dynamics (Def. \ref{def::vector_discrete_dynamics}), after applying the frame dynamics (Def. \ref{def::frame-shift}) and after taking into account the observation of Def. \ref{def::2frame_obs}.  Moreover we split the correction term $L_n(\cdot)\in\tFrames$ as $\bigl(L_n^R(\cdot),L_n^\fixX(\cdot),L_n^\bodyX(\cdot)\bigr)$. 

We  find using   \eqref{group:comp} and Def. \ref{inv:ob:d} that left-invariant observers for System \eqref{eq::two-frame-dynamics} of Def. \ref{def::two-frames-system} with fixed-frame observation $\fixY_n$  write:
\begin{equation}
\label{eq::invObsFix}\boxed{
\text{\normalfont{Obs. \eqref{eq::output_fixed}}} \Rightarrow  \left\{
    \begin{array}{lll}
         \hat{\chi}_{n^-|n-1}  = f_n \left( \hat{\chi}_{n-1|n-1} \right) \\
 \hat{\chi}_{n|n-1}  = s_n \left( \hat{\chi}_{n^-|n-1} \right) \\
 \hat{\chi}_{n|n}   =
\begin{pmatrix}
\hat{R}_{n|n-1} L_n^R(\bodyZ_n) \\
\hat{\fixX}_{n|n-1} + \hat{R}_{n|n-1} \action L_n^\fixX(\bodyZ_n) \\
L_n^\bodyX(\bodyZ_n) + L_n^R(\bodyZ_n)^{-1} \action\hat{\bodyX}_{n|n-1}
\end{pmatrix}
    \end{array}
\right.}
\end{equation}
where $L_n^R: \mathcal{Y} \mapsto G$, $ L_n^\fixX: \mathcal{Y} \mapsto V$ , $L_n^\bodyX: \mathcal{Y} \mapsto B$ can be any functions, and $\bodyZ_n$ is the innovation variable from Eq. \eqref{eq::innovBody}.

When confronted with body-frame observations, right-invariant observers for System \eqref{eq::two-frame-dynamics}   with  observation $\bodyY_n$ write:
\begin{equation}
\label{eq::invObsBody}\boxed{
\text{\normalfont{Obs. \eqref{eq::output_body}}} \Rightarrow  \left\{
    \begin{array}{lll}
  \hat{\chi}_{n^-|n-1}  = f_n \left( \hat{\chi}_{n-1|n-1} \right) \\
  \hat{\chi}_{n|n-1}  = s_n \left( \hat{\chi}_{n^-|n-1} \right) \\
  \hat{\chi}_{n|n} =
\begin{pmatrix}
L_n^R(\fixZ_n)  \hat{R}_{n|n-1} \\
L_n^\fixX(\fixZ_n) + L_n^R(\fixZ_n)\action \hat{\fixX}_{n|n-1} \\
\hat{\bodyX}_{n|n-1} + \hat{R}_{n|n-1}^{-1}\action  L_n^\bodyX(\fixZ_n) 
\end{pmatrix}
   \end{array}
\right.}
\end{equation}
where $L_n^R: \mathcal{Y} \mapsto G$, $L_n^X: \mathcal{Y} \mapsto V$ , $L_n^\bodyX: \mathcal{Y} \mapsto B$ can be any functions, and $\fixZ_n$ is the innovation variable from Eq. \eqref{eq::innovFix}.


We see the error variables \eqref{errors:eqq} provide alternative definitions of ``discrepancies'' between the   true state $(R,\fixX,\bodyX)$ and the estimated state $(\hat{R},\hat{\fixX},\hat{\bodyX})$ of a two-frames system indeed. In the case where  observations are performed in the fixed frame \eqref{eq::output_fixed}, one should use  the left-invariant error $\bodyE_{n|n}$ in \eqref{errors:eqq} which writes using the TFG law   \eqref{group:comp} as
\begin{equation}
\label{eq::error_variables_body}\text{\normalfont{Obs. \eqref{eq::output_fixed}}} \Rightarrow\boxed{
\bodyE_{n|n} =
\begin{pmatrix}
\hat{R}_{n|n}^{-1}\grouplaw R_n \\ \hat{R}_{n|n}^{-1}\action (\fixX_n-\hat{\fixX}_{n|n}) \\ \bodyX_n -( R_n^{-1}\grouplaw\hat{R}_{n|n})\action  \hat{\bodyX}_{n|n} 
\end{pmatrix}.}
\end{equation}
 If observations are performed instead in the body frame \eqref{eq::output_body}  one should use the right-invariant error $\fixE_{n|n} $ in \eqref{errors:eqq}, i.e.,
\begin{equation}
\label{eq::error_variables_fixed}\text{\normalfont{Obs. \eqref{eq::output_body}}} \Rightarrow\boxed{
\fixE_{n|n} = 
\begin{pmatrix}
R_n\grouplaw \hat{R}_{n|n}^{-1} \\ \fixX_n-(R_n \grouplaw \hat{R}_{n|n}^{-1})\action  \hat{\fixX}_{n|n} \\ \hat{R}_{n|n}\action (\bodyX_n-\hat{\bodyX}_{n|n})
\end{pmatrix}.}
\end{equation}
The indexes ${n|n}$ can be replaced with ${n^-|n-1}$ (resp. ${n|n-1}$) everywhere to define the error variables after vector dynamics but before frame dynamics (resp. after frame dynamics but before update). Regarding $\fixX_n$ and $\bodyX_n$,  we see the error variables we consider much differ from the classical linear difference $\fixX_n-\hat{\fixX}_{n|n}$ (resp. $\bodyX_n-\hat{\bodyX}_{n|n}$).  

All the (autonomous) error equations then governing the evolution of $\bodyE_{n|n}$ and $\fixE_{n|n}$ are provided in Appendix \ref{error:ec:sec}.

\subsection{Methodology to attack examples}
\label{sect::method}

When facing a novel example, the user shall
first cast the navigation system into the two-frames systems equations and endow the state with the TFG structure. Derivation of a
TFG-IEKF is then automatic  following the methodology, and  always worth being tested, but
is of course especially relevant when Thm. 7 holds. This is guaranteed when  1)   commutation properties required by definitions of natural output and   vector dynamics (Defs.    \ref{def::2frame_obs} and  \ref{def::vector_discrete_dynamics}) hold, which is often checked using Prop. \ref{easy:prop} in practice, 2)  frame dynamics is natural. In practice, it suffices that frame dynamics  has  the shape   \eqref{eq::frame-shift} and that one of the conditions of Thm. \ref{prop::no-mu}, or those  of  Thm.   \ref{prop::abelian}, hold.  

\subsection{Odometer-GNSS navigation  with unknown lever arm}
\label{sect::lever_arm}
The  problem of navigating with unknown GNSS antenna lever arm is both very  relevant in practice and not addressed until now by the theory of invariant filtering.

Herein, we consider the classical 2D model of a non-holonomic car, see e.g., \cite{barrau2017invariant}. The position of the car in 2D  is described  by   the middle point  of the rear wheels axle $\fixX_n \in \RR^2$ , and its orientation (heading)  denoted by $\theta_n \in \RR$ and parameterized by the planar rotation matrix $\Matrix{R_n} = \Rot{\theta_n}$ of angle $\theta_n$ where $ \Rot{\theta}:= \begin{pmatrix}\cos\theta & -\sin\theta\\\sin \theta&\cos\theta\end{pmatrix}$.
The car is equipped with a GNSS antenna located at unknown position $\bodyX_n \in \RR^2$ in the car frame with respect to point $x_n$,  which provides the world (fixed) frame position measurements   
\begin{equation}
\label{eq::lever-arm}
\fixY_n = x_n + \Matrix{R_n} \bodyX_n \in \RR^2.
\end{equation}
In the schematic diagram below, the triangle is the car and the square is the position measured by the GNSS.

\tikzset{every picture/.style={line width=0.75pt}} 

\begin{center}
\begin{tikzpicture}[x=0.75pt,y=0.75pt,yscale=-1,xscale=1]

\draw  [color={rgb, 255:red, 189; green, 16; blue, 224 }  ,draw opacity=1 ][line width=1.5]  (249,138.97) -- (227.13,177.22) -- (205.68,147.07) -- cycle ;
\draw [color={rgb, 255:red, 65; green, 117; blue, 5 }  ,draw opacity=1 ]   (232.7,150.56) -- (262.93,89.12) ;
\draw [shift={(263.81,87.32)}, rotate = 476.2] [color={rgb, 255:red, 65; green, 117; blue, 5 }  ,draw opacity=1 ][line width=0.75]    (10.93,-3.29) .. controls (6.95,-1.4) and (3.31,-0.3) .. (0,0) .. controls (3.31,0.3) and (6.95,1.4) .. (10.93,3.29)   ;

\draw  [color={rgb, 255:red, 74; green, 144; blue, 226 }  ,draw opacity=1 ] (267.48,84.72) -- (266.56,91.2) -- (260.14,89.93) -- (261.06,83.45) -- cycle ;
\draw    (252.91,136.18) -- (311.47,94.21) ;

\draw    (257.93,150.9) -- (330.67,151.33) ;

\draw    (332,145.33) .. controls (332.64,127.96) and (327.7,112.46) .. (317.17,98.21) ;
\draw [shift={(316,96.67)}, rotate = 412.31] [color={rgb, 255:red, 0; green, 0; blue, 0 }  ][line width=0.75]    (10.93,-3.29) .. controls (6.95,-1.4) and (3.31,-0.3) .. (0,0) .. controls (3.31,0.3) and (6.95,1.4) .. (10.93,3.29)   ;

\draw (202.33,171) node [scale=1,color={rgb, 255:red, 189; green, 16; blue, 224 }  ,opacity=1 ]  {$\fixX_n$};
\draw (346,115) node   {$\theta_n $};
\draw (239.33,113.33) node [color={rgb, 255:red, 65; green, 117; blue, 5 }  ,opacity=1 ]  {$\bodyX_n$};
\draw (278.67,74.67) node [color={rgb, 255:red, 74; green, 144; blue, 226 }  ,opacity=1 ]  {$\fixY_n$};
\end{tikzpicture}
\end{center}
Differential odometers  measure  linear velocity and angular rate that may be compounded over a time step $dt$ into position and angular shifts $\body\bodyu_n, \omega_n$, while the lever arm $\bodyX_n$ between the reference point and the GNSS antenna remains constant,  albeit unknown. The dynamics in discrete time write  \cite{barrau2017invariant}:
\begin{equation}
\label{eq::odo}
\begin{aligned}
\Matrix{R_n} & = \Matrix{R_{n-1} \Omega_n} ,  \quad 
\fixX_n = \fixX_{n-1} + \Matrix{R_{n-1}}  \bodyu_n,   \quad 
\bodyX_n = \bodyX_{n-1},
\end{aligned}
\end{equation}
with $\Matrix{\Omega_n} = \Rot{ \omega_n}$.
This system has been already studied for invariant filtering in \cite{bonnabel2008symmetry, barrau2017invariant}, but without the lever arm $\bodyX_n$ which makes observer design more difficult. It is pedagogical to first illustrate the results very concretely, without referring to the theory above. Consider the following observer shape:
\begin{equation}
\label{eq::car-observer-prop}
\begin{gathered}
\Matrix{\hat{R}_{n|n-1}} = \Matrix{\hat{R}_{n-1|n-1} \Omega_n} , \\
\fixX_{n|n-1} = \fixX_{n-1|n-1} + \Matrix{\hat{R}_{n-1|n-1}}  \bodyu_n,   \quad 
\bodyX_{n|n-1} = \bodyX_{n-1|n-1},
\end{gathered}
\end{equation}
\begin{equation}
\label{eq::lever-arm-update}
\begin{aligned}
\Matrix{\hat{R}}_{n|n}  & = \Matrix{\hat{R}}_{n|n-1} \Matrix{L_n^R(\bodyZ_n)} ,  \\ 
\hat{\fixX}_{n|n-1} & = \hat{\fixX}_{n|n-1} + \Matrix{\hat{R}}_{n|n-1} L_n^\fixX ( \bodyZ_n ),    \\
\hat{\bodyX}_{n|n-1} & =  L_n^\bodyX ( \bodyZ_n )+\Matrix{L_n^R(\bodyZ_n)}^T \hat{\bodyX}_{n|n-1} , \\
\text{with } \bodyZ_n & = \Matrix{\hat{R}}_{n|n-1}^T \left( \fixY_n - \hat{\fixX}_{n|n-1} \right)- \hat{\bodyX}_{n|n-1},
\end{aligned}
\end{equation}
where $L_n(\bodyZ)=\left(\Matrix{L_n^R (\bodyZ)}, L_n^\fixX(\bodyZ), L_n^\bodyX(\bodyZ)  \right) \in SO(2) \times \RR^2\times\RR^2$ can be any function. Now, let the error variable be defined as:
\begin{equation}
\label{eq::car-error}
\bodyE_{n|n} =
\begin{pmatrix}
\Matrix{\hat{R}_{n|n}^{-1} R_n} \\
\Matrix{\hat{R}_{n|n}^{-1}} (\fixX_n-\hat{\fixX}_{n|n}) \\
\bodyX_n - \Matrix{R_n^{-1}\hat{R}_{n|n}}  \hat{\bodyX}_{n|n} 
\end{pmatrix}.
\end{equation}
We invite the reader to first  check ``manually'' Theorem \ref{prop::lever-arm}.
\begin{thm}
\label{prop::lever-arm}
Observer \eqref{eq::car-observer-prop}, \eqref{eq::lever-arm-update} ensures the error variable \eqref{eq::car-error}  has   state-independent  evolution.
\end{thm}
\paragraph{proof}   Using the more concrete variable $\theta$ via the relation  $\Matrix{R}:=\Rot{\theta}$ the   error  \eqref{eq::car-error}  re-writes:
\begin{equation}
\label{eq::lever-arm-error}
\bodyE_{n|n}=\begin{pmatrix} \bodyE_{n|n}^\theta\\\bodyE_{n|n}^\fixX\\\bodyE_{n|n}^\bodyX
\end{pmatrix}=\begin{pmatrix}
\theta_n - \hat{\theta}_{n|n} \\
\Rot{\hat{\theta}_{n|n}}^T \left( \fixX_n - \hat{\fixX}_{n|n} \right) \\
\bodyX_n - \Rot{\hat{\theta}_{n|n}-\theta_n} \hat{\bodyX}_{n|n}
\end{pmatrix}.
\end{equation}
We readily see the innovation is a function of the error as $ \bodyZ_n   = \Matrix{\hat{R}}_{n|n-1}^T \left( \fixX_n - \hat{\fixX}_{n|n-1} \right)+\Matrix{\hat{R}}_{n|n-1}^T\Matrix{{R}}_{n}{\bodyX}_{n}- \hat{\bodyX}_{n|n-1}=\bodyE_{n|n-1}^\fixX+\Rot{\bodyE_{n|n-1}^\theta}\bodyE_{n|n-1}^\bodyX$.  Let us study for instance the evolution of $\bodyE_{n|n-1}^\bodyX$ under update \eqref{eq::lever-arm-update}. Let us write $\Matrix{L_n^R}:=\Rot{l^\theta(\bodyZ_n  )}$. We find at the update step that  $\hat\theta,\hat\bodyX$ are transformed as $\hat\theta\to\hat\theta+l^\theta$ and  $\hat  \bodyX\to L^\bodyX+\Rot{-l^\theta}\hat\bodyX$ so that error evolves as $E^\bodyX\to \bodyX_n - \Rot{\hat{\theta}+l^\theta-\theta} (L^\bodyX+\Rot{-l^\theta}\hat\bodyX)=E^\bodyX- \Rot{l^\theta-E^\theta} L^\bodyX$ which is a function of the error at previous step indeed. 
 $\blacksquare$

\vspace{.2cm}

Of course, checking Theorem \ref{prop::lever-arm} is the ``easy'' part while finding the observer and error variable is the hard part necessitating the proposed theory. Let us see how the theory formally  applies indeed. Letting $\action$ be standard product, we readily recognize    that:
{
\begin{itemize} 
\item \eqref{eq::odo} formally writes as vector dynamics given by \eqref{eq::vector_dynamics}   with frame dynamics \eqref{eq::frame-shift}, where  $G=SO(2)$  and  $\fixF_n=\bodyF_n=\Matrix{O}_n=\Matrix{I_2}$, $\bodyC_n=\fixC_n=\0{2,2}$, \item Observation \eqref{eq::lever-arm} formally writes as the natural output \emph{in the fixed frame}   \eqref{eq::output_fixed} with $\Matrix{H}^\fixX_n = \Matrix{H}^\bodyX_n = \Matrix{I_2}, \bodyb_n=0_2$. \end{itemize}
Let us apply the method of Sect. \ref{sect::method}: 1)  commutation properties required by definitions of natural output and natural vector dynamics (Defs.    \ref{def::2frame_obs} and  \ref{def::vector_discrete_dynamics})  directly stem from Prop. \ref{easy:prop}; 2) frame dynamics has the shape \eqref{eq::frame-shift} and   $G=SO(2)$ is abelian, thus Thm. \ref{prop::abelian} guarantees that  frame dynamics is natural.
} Besides, the left-invariant error  \eqref{eq::error_variables_body} specifies as  \eqref{eq::car-error}.
We have all the conditions of Theorem \ref{thm:aff} and know the methodology of previous sections will lead to an observer ensuring state-independent  error evolution, as can be seen from Thm. \ref{thm:error}.

The present nontrivial example was shown to fit into our theory.  Beyond,  our approach suggests a novel way to treat lever arms. Looking at  \eqref{eq::lever-arm-update}
we see that despite a simplified car model and an abelian group $G=SO(2)$, we woud probably not have managed  to come up with an observer guaranteeing state-independent  error evolution without  the present theory. {A last important remark is that this system still fits into the theory if an unknown scale factor affects odometry: $G=SO(2)$  is simply replaced with the (abelian) group of scaled rotations.}

\subsection{Inertial 3D SLAM with  moving objects tracking}

Simultaneous Localization and Mapping with Moving Objects Tracking (SLAMMOT) is the subject of a rich and vast literature, see e.g. the landmark paper  \cite{wang2007simultaneous}. In \cite{bonnabel2012symmetries} a group structure was discovered to make standard SLAM dynamics   left-invariant,  later called $SE_k(d)$ in \cite{barrau2015ekf,barrau2015non} and $SLAM_n(3)$ in \cite{mahony2017geometric} (which designs a geometric observer for  SLAM with robot and features having known velocity). For the SLAM problem, TFG boils down to $SE_k(d)$ and the present theory allows recovering those results. A question one could legitimately ask is then as follows. \emph{Are features allowed to move while preserving group affine dynamics?} The present theory provides  (positive) answers.

Let us consider a robot equipped with an IMU  evolving in a 3D    environment containing   static unknown features with positions $l^k_n \in \RR^3$ and moving features with unknown position  $q_n^i \in \RR^3$ and unknown velocity $c_n^i \in \RR^3$. Attitude, velocity and position of the robot are denoted by $\Matrix{R}_n \in SO(3)$, $v_n \in \RR^3$ and $p_n \in \RR^3$, while preintegrated inertial factors \cite{forster2017manifold} are denoted by $\Matrix{\Omega}_n \in SO(3), a_n^v  \in \RR^3, a_n^p \in \RR^3$. Dynamics write:
\begin{equation}
\label{eq::dyn_slam}
\begin{aligned}
\Matrix{R}_n & = \Matrix{R}_{n-1} \Matrix{\Omega}_n, \\
v_n & = g + \Matrix{R}_{n-1} a_n^v, \\
p_n & = p_{n-1} + dt ~ v_{n-1} + \Matrix{R}_{n-1} a_n^p,
\end{aligned}
\qquad
\begin{aligned}
l_n^k & = l_{n-1}^k, \\
q_n^i & = q_{n-1}^i + dt ~ c_n^i, \\
c_n^i & = c_{n-1}^i.
\end{aligned}
\end{equation}
A wide range of observations measured in the frame of the vehicle can be considered, such as:
\begin{enumerate}
\item Position of static feature points: $\Matrix{R}_n^T(l^k_n-p_n)$,
\item Position of moving features: $\Matrix{R}_n^T(q_n^i-p_n)$, \label{item::target} 
\item Velocity of moving features: $\Matrix{R}_n^T(c_n^i-v_n)$,
\item Position of some known landmarks $r_n^m$: $\Matrix{R}_n^T(r_n^m-p_n)$, \label{item::landmark} 
\item Magnetic field $\beta$ (known): $\Matrix{R}_n^T\beta$,
\item Magnetic field $\beta_n$ (to be estimated): $\Matrix{R}_n^T\beta_n$  \label{item::magnetic}.
\end{enumerate}
Recalling $\Matrix{R}_n^T=\Matrix{R}_n^{-1}$, we immediately see all   of them are natural outputs in the body frame as in Def. \ref{def::2frame_obs}, i.e.,   Obs. \eqref{eq::output_body}, which evidences the broad scope of the present theory. Consider, e.g., position of moving features $q_n^i$   in body frame:
\begin{equation}
\label{eq::obs_slam}
\begin{gathered}
\bodyY_n^i = \Matrix{R}_n^T(q_n^i-p_n)
\end{gathered}
\end{equation}
Here, we recognize that:
\begin{itemize}
\item  The group of frame changes is $G=SO(3)$, acting term by term on the fixed-frame multi-vector $\fixX_n=(v_n,p_n,l_n^k,q_n^i,c_n^i)\in \fixV$ as in Eq. \eqref{tbt:eq},
\item   \eqref{eq::dyn_slam} formally writes as the combination of  \eqref{eq::vector_dynamics} and   \eqref{eq::frame-shift}, with no $\bodyV $ (hence no $ \bodyX_n,\bodyu_n$), $O_n=\Matrix{I}_3$   and where
$$
\fixF_n =
\begin{pmatrix}
\Matrix{I}_3 & \0{3} & \0{3} & \0{3} & \0{3} \\
dt \Matrix{I}_3 & \Matrix{I}_3 & \0{3} & \0{3} & \0{3} \\
\0{3} & \0{3} & \Matrix{I}_3 & \0{3} & \0{3} \\
\0{3} & \0{3} & \0{3} & \Matrix{I}_3 &dt \Matrix{I}_3 \\
\0{3} & \0{3} & \0{3} & \0{3}& \Matrix{I}_3 \\
\end{pmatrix}, \quad   \fixu_n =
\begin{pmatrix}
a_n^v \\ a_n^p \\ 0_{3} \\ 0_{3} \\ 0_{3}  
\end{pmatrix},
$$
\item 
Observation \eqref{eq::obs_slam} {writes as} a natural output \emph{in the body frame} \eqref{eq::output_body} with:
 $
\Matrix{H}^\fixX =
\begin{pmatrix}
\0{3}  & \Matrix{I}_3  & \0{3} &-\Matrix{I}_3& \0{3} 
\end{pmatrix} 
$. 
The minus sign   in  matrix $\Matrix{H}^\fixX$ looks switched due to Def.  \ref{eq::output_body}. 
\end{itemize}Let's apply the method of Sect. \ref{sect::method}:   1) 
 commutations required by Defs.    \ref{def::2frame_obs} and  \ref{def::vector_discrete_dynamics} of   $\Matrix{H}^\fixX $ and $\fixF_n$  with {$R$ acting as a term-by-term rotation stem from Prop. \ref{easy:prop},}
2) frame dynamics has the shape \eqref{eq::frame-shift} and  we are in the case (b) of Thm. \ref{prop::no-mu} so  it is natural. We have all conditions of Theorem \ref{thm:aff} and know the methodology of previous sections will lead to an observer ensuring autonomous error evolution, as can be seen from Thm. \ref{thm:error}.  With  the observations in the body frame \eqref{eq::output_body},   observers are defined by \eqref{eq::innovFix}, \eqref{eq::invObsBody}.
\begin{thm}
\label{prop::moving}
System \eqref{eq::dyn_slam}, \eqref{eq::obs_slam}, where \eqref{eq::obs_slam} could be replaced with any item of the list above,   is a linear observed system on the group $\tFrames=SO(3)^+_{(2+K+2I),0}$ with body-frame observations. A  family of invariant observers is defined as follows. The propagation step for the estimated state  $(\hat{R}_n, \hat{v}_n, \hat{p}_n, \hat{l}_n^k, \hat{q}_n^i, \hat{c}_n^i)$ is a copy of \eqref{eq::dyn_slam}.
Denoting $z_n=(z_n^{1},\cdots,z_n^J)$,   update   writes:
\begin{equation}
\label{eq::upd_inv_slam}
\begin{aligned}
\Matrix{\hat{R}}_{n|n}  & = \Matrix{L_n^R(\fixZ_n)} \Matrix{\hat{R}}_{n|n-1} \\
\hat{\square}_{n|n}  & = L_n^\square(\fixZ_n) + \Matrix{L_n^R(\fixZ_n)} (\hat{\square}_{n|n-1}) \\
\text{with } \fixZ_n^i & = 
\Matrix{\hat{R}_{n|n-1}} \bodyY_n^i - \hat{q}_{n|n-1}^i + \hat{p}_{n|n-1},
\end{aligned}
\end{equation}
where $\square$ must be replaced  with $v,p,l^k, q^i, c^i$, and $L_n(\fixZ)=\left(\Matrix{L_n^R (\fixZ)}, L_n^\fixX(\fixZ)  \right) \in SO(3) \times V$ can be any function. The innovation vector $\fixZ_n$ must be adapted if Eq. \eqref{eq::obs_slam} is replaced by one or various  alternative observations proposed above. These observers ensure  the   right-invariant error \eqref{eq::error_variables_fixed} has   state-trajectory independent evolution.
\end{thm}
The reader can  alternatively check this result injecting Eq. \eqref{eq::dyn_slam}, \eqref{eq::obs_slam}, \eqref{eq::upd_inv_slam} into Eq. \eqref{eq::error_variables_fixed}. A TFG-IEKF is readily   built  using Def.   \ref{def:TFGIEKF}, i.e., tuning the observer via  \eqref{eq::gains2}, \eqref{eq::riccati}. Jacobians $\Matrix{A}^v_n,\Matrix{A}^s_n,\Matrix{H}_n$ are retrieved from Prop. \ref{prop:lin2} and  noise matrices from \eqref{Q:eq} (both in the Appendix).  It inherits the group affine and error log-linearity properties, and   potential consistency properties  \cite{barrau2015ekf,brossard2018exploiting}.

 In onboard radar tracking contexts, one may refine the moving objects model. Indeed, constant velocity assumption is simplistic, and does not suit targets that perform manoeuvers.  A   celebrated model in the tracking literature and industry is the Singer model \cite{singer1970estimating}, that assumes the acceleration of the target is a Gauss-Markov process with auto-correlation time $\tau=1/\gamma$. This leads to modifying \eqref{eq::dyn_slam} by letting 
 $
c_n^i  = c_{n-1}^i + dt . a_{n-1}^i, $ $ 
a_n^i  =  a_{n-1}^i-dt.\gamma . a_{n-1}^i+w_n^i.
$ with $w_n^i$ a process white noise.
\begin{prop}
\label{prop::moving2}
The SLAM equations with features moving according to a Singer tracking model with noise turned off define a deterministic linear observed system on group  as commutation also stems from Prop. \ref{easy:prop}.
\end{prop}
More generally, the state can  be augmented with     vectors of the fixed frame, as long as their dynamics commute with $R$. 

\subsection{Inertial navigation with IMU biases}\label{IMU:exp}
Let us  come back to the prototypical example of Section \ref{expl::nav_IMU}. 
Inertial navigation without biases has been shown to possess the group affine property and   autonomous errors thanks to the introduction of the group $SE_2(3)$ in  \cite{barrau2017invariant}, which is here generalized by the TFG. To treat the case where IMU biases need be estimated, various publications have built on \cite{barrau2017invariant} for models including gyro and accelerometer bias by using the imperfect IEKF of \cite{barrau2015non}, i.e., simply using as group structure the Cartesian product of $SE_2(3)$ (for attitude, velocity and position) and $\RR^3 \times \RR^3$ for gyroscope and accelerometer bias, see \cite{barrau2018invariant,Hartley-RSS-18,hartley2019contact,wu_invariant-ekf_2017,heo_consistent_2018,caruso_2018}.

However the   TFG structure advocated herein provides an alternative approach and the following experiments prove it should be preferred even  if dynamics \eqref{eq::nav-flat} are not  group affine for the TFG. The state space can be cast into a two-frames state space with $G=SO(3), \fixV=(\RR^3)^2, \bodyV = (\RR^3)^2$.

Many two-frames natural outputs could be considered for this system, let us stick with the setting of \cite{barrau2017invariant} by choosing:
\begin{equation}
\begin{gathered}
\bodyY_n^m = \Matrix{R}_n^T(r^m-p_n) \end{gathered}
\end{equation}
where $r^m$ for $m \in [1, M]$ are landmarks of known position and $Y_n^m$ is the position of feature point $m$ at time step $n$ observed by the IMU carrier in its own reference frame. It corresponds to \eqref{eq::output_body}  with $ \Matrix{H}_m^\fixX =
\begin{pmatrix}
\0{3}  & \Matrix{I}_3  
\end{pmatrix}, ~
\Matrix{H}_m^\bodyX = \0{3,6},~b^m=r^m$. 

Let us detail the  TFG-IEKF with body-frame observations, using   Def.   \ref{def:TFGIEKF} and the Appendix. Propagation step    is a  copy of   dynamics \eqref{eq::nav-flat}. Using  \eqref{eq::gains2} where we let $\Matrix{K_n} \fixZ_n=(\Matrix{K_n^R} \fixZ_n,\Matrix{K_n^v} \fixZ_n,\Matrix{K_n^p} \fixZ_n,\Matrix{K_n^{b^\omega}} \fixZ_n,\Matrix{K_n^{b^a}} \fixZ_n)$, then extracting  $L^R \in SO(3)$, $L^\bodyX=(L^v,L^p),$ $L^\fixX=(L^{b^\omega},L^{b^a})$ from the exponential map definition  \eqref{exp:formula},  and applying  the observer update formula \eqref{eq::invObsBody} where we recall $\action$ denotes the  term-by-term rotation of Ex. \ref{expl::term-by-term}, the  update writes:

\begin{equation}
\begin{aligned}
\Matrix{\hat{R}_{n|n}} & = \expm{SO(3)}{K_n^R \fixZ_n} \Matrix{\hat{R}_{n|n-1}} \\
\hat{v}_{n|n} & = \Matrix{\nu_3(K_n^v \fixZ_n)} \Matrix{K_n^v} \fixZ_n+ \expm{SO(3)}{K_n^R \fixZ_n} \hat{v}_{n|n-1} \\
\hat{p}_{n|n} & = \Matrix{\nu_3(K_n^p \fixZ_n)}\Matrix{K_n^p} \fixZ_n + \expm{SO(3)}{K_n^R \fixZ_n} \hat{p}_{n|n-1} \\
\hat{b}_{n|n}^\omega & = \hat{b}_{n|n-1}^\omega + \Matrix{\hat{R}_{n|n-1}^T}  \Matrix{\nu_3(-K_n^R \fixZ_n)} \Matrix{K_n^{b^\omega}} \fixZ_n \\
\hat{b}_{n|n}^a & = \hat{b}_{n|n-1}^a + \Matrix{\hat{R}_{n|n-1}^T}  \Matrix{\nu_3(-K_n^R \fixZ_n)} \Matrix{K_n^{b^a}} \fixZ_n
\end{aligned}
\end{equation}
where $\Matrix{\nu_3(\cdot)}$ is taken from Prop. \ref{prop::exp-rot} of Appendix, and matrix
$
\Matrix{K_n}  
$
is obtained from the Riccati equation \eqref{eq::riccati}. Matrices $\Matrix{A}_n^v, \Matrix{H_n}$ are retrieved from Prop. \ref{prop:lin2},  referring to \eqref{TFS:inert} and Example \ref{ex4} proving $\repu{r^m}{Y}=\Matrix{(r^m)_\times}$, while $\Matrix{\hat Q_n},\Matrix{\hat N_n}$  are read on \eqref{Q:eq}, where in this case $\Matrix{Ad}_R=\Matrix{R}$. It yields
\begin{gather*}
\Matrix{A^v_n} =\left(
\begin{array}{c|cc|cc}
 \Matrix{I}_3 & \0{3,3} & \0{3,3} & \0{3,3} & \0{3,3} \\\hline 
 \Delta t \Matrix{(g_n)_\times} & \Matrix{I_3} & \0{3,3} & \0{3,3} & \Delta t\Matrix{I_3} \\
 \0{3,3} & \Delta t \Matrix{I_3} & \Matrix{I_3} & \0{3,3} & \0{3,3} \\\hline 
 \0{3,3} & \0{3,3} & \0{3,3} & \Matrix{I_3} & \0{3,3} \\
 \0{3,3} & \0{3,3} & \0{3,3} & \0{3,3} & \Matrix{I_3}  
\end{array}\right)
, \\
\Matrix{H}_n = \begin{pmatrix}
 \Matrix{(r^m)_\times} & \0{3,3} &  -\Matrix{I_3} & \0{3,3} & \0{3,3}
\end{pmatrix}.
\end{gather*}
They are both state-trajectory independent as could be anticipated from the theory. 
Matrix $\Matrix{A}_n^s$ must be derived manually.
\begin{prop} Jacobian for frame dynamics writes
\begin{align}
& \Matrix{A}_n^s = \begin{pmatrix}
\Matrix{I}_3 & \0{3,3} & \0{3,3} & \Matrix{M_1} & \0{3,3} \\
\0{3,3}  & \Matrix{I}_3 & \0{3,3}  & \Matrix{(\hat{v}_{n^-|n-1})_\times M_1} & \0{3,3} \\
\0{3,3} & \0{3,3} & \Matrix{I}_3 & \Matrix{(\hat{p}_{n^-|n-1})_\times M_1} & \0{3,3} \\
\0{3,3} & \0{3,3} & \0{3,3} & \Matrix{M_2} & \0{3,3} \\
\0{3,3} & \0{3,3} & \0{3,3} & \0{3,3} & \Matrix{M_2}  
    \end{pmatrix},\label{Fsbias}
\end{align}with $
  \Matrix{M_1}  = \Delta t\Matrix{\hat{R}_{n|n-1} \Matrix{\bar J}   \hat{R}_{n-|n-1}^T},$   $
   \Matrix{M_2}  = \Matrix{\hat{R}_{n|n-1} \hat{R}_{n-|n-1}^T}$,   $\Matrix{\bar J}  = 
 \Matrix{I_3} - \frac{1-\cos(||\mu||)}{||\mu||^2} \Matrix{(\mu)_\times} - \frac{\sin(||\mu||)-||\mu||}{||\mu||^3} \Matrix{(\mu)_\times}^2$,  $\mu:= \omega+\hat b^\omega$.\label{prop:nav:IEKF}
\end{prop}
A detailed proof is provided in the supplementary material.    Let's sketch it here. 
\paragraph{proof}
We   study to the first order the  effect of   $\Matrix{R_n}=\Matrix{R_{n-1}} \expm{m}{\Delta t{[    \omega_n+b^\omega_{n-1}]_\times}}$   and $\Matrix{\hat R_{n|n-1}} =\Matrix{\hat R_{n-|n-1}} \expm{m}{\Delta t{ \Matrix{[    \omega_n+\hat b^\omega_{n-|n-1}]_\times}}}$ on \eqref{eq::error_variables_fixed}, which is  linearized   as $  \Matrix{R_n\grouplaw \hat{R}^{-1}}\approx Id+\repu{\xi^R}{}$, $ \xi^\fixX\approx \fixX -   \hat{\fixX}-\repu{\xi^R}{\fixV} \hat{\fixX}$, $\xi^\bodyX=  \Matrix{ \hat{R}}\action (\bodyX-\hat{\bodyX})$, see also the Appendix. To do so we use the right-Jacobian formula \cite{Chirikjian}   $\expm{m}{\Delta t{ \Matrix{[    \omega+b^\omega]_\times}}}\approx\expm{m}{\Delta t{ \Matrix{[    \omega+\hat b^\omega]_\times}}} \expm{m}{\Delta t{ \Matrix{[  \bar J( b^\omega-\hat b^\omega)]_\times}}}$, leading to $\xi_{n|n-1}^R=\xi_{n-|n-1}^R+\Matrix{M_1}\xi_{n-|n-1}^{b^\omega}$, which yields the first row of $ \Matrix{A}_n^s $, and similarly for the other rows. $\blacksquare$

\begin{figure}[h]\centering
\includegraphics[width=.6\columnwidth]{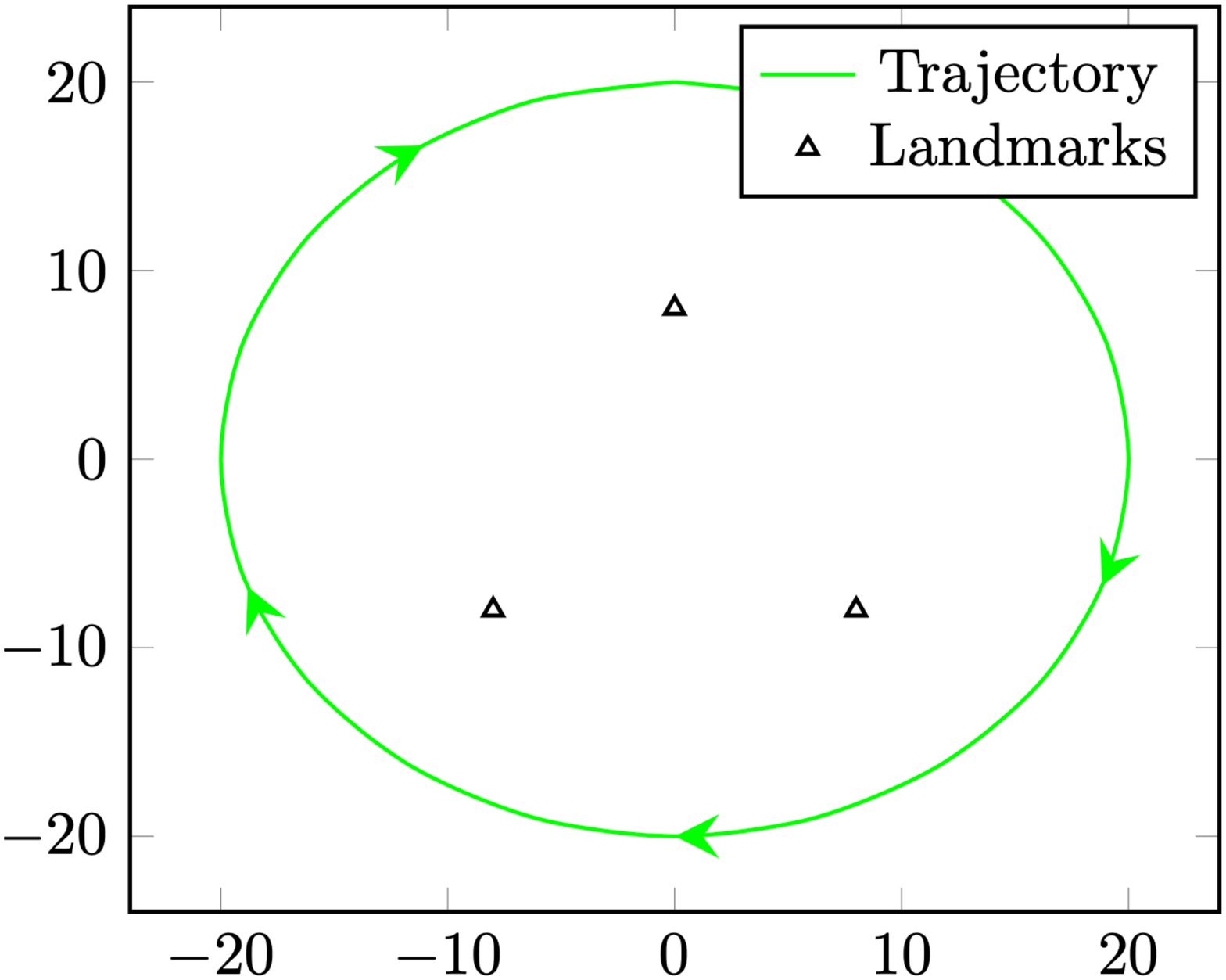}
\caption{Experimental setting of Sec. \ref{simuss:sec}. Aircraft follows a loop  while observing one landmark of known position for $20 \, s$, then two more landmarks with known position for $60 \, s$.}\label{fig::setting}
\end{figure}

\begin{figure}[!htb] 
\hspace*{-1cm}  
\includegraphics[width=1.1\columnwidth]{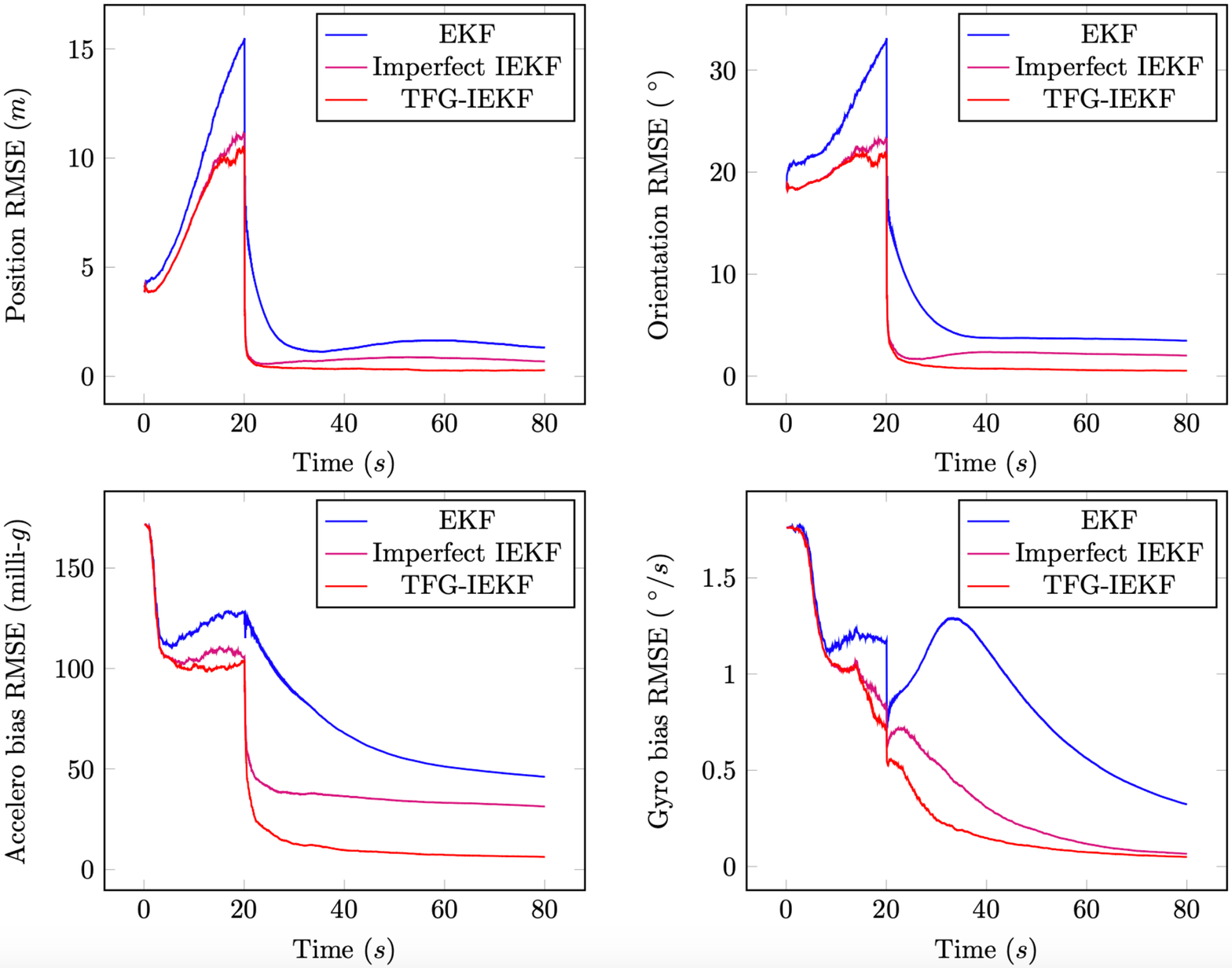}
\caption{Initial orientation error is aleatory with  a $30^\circ$ standard deviation, observations  and IMU signals    are noisy, and 100 Monte-Carlo runs are averaged. One known feature point is first visible  then two additional feature points make the full state observable at $t=20 \, s$. We see the IEKF based on the TFG structure outperforms EKF, as well as  the ``Imperfect IEKF'' advocated in quite a number of recent publications. We note the benefits  of the methodology are substantial for all variables, including gyro biases, although   their presence prevents frame dynamics to be natural and hence   group affine.}
\label{fig::results}
\end{figure}

\subsection{Simulation results}\label{simuss:sec}
The (TFG-) IEKF of Sect. \ref{IMU:exp} is tested on the setting described by Fig. \ref{fig::setting}, and compared to both conventional EKF, and ``imperfect IEKF''   as in previous work \cite{barrau2015non,Hartley-RSS-18,hartley2019contact,wu_invariant-ekf_2017,heo_consistent_2018,heo2018consistent, barrau2018invariant,brossard2017unscented,van2020invariant,cohen2020navigation}. The name imperfect IEKF was  coined in \cite{barrau2015non}, and this   filter has   led to a high-performance commercial product, see \cite{barrau2018invariant}.

Initial  errors on biases and orientation are Gaussian  with respective standard deviations $1 \, \text{deg}/s$, $0.1 \, g$, $30^\circ$. Results averaged on 100 Monte-Carlo runs are presented on Fig. \ref{fig::results}. We see conventional multiplicative EKF based on $SO(3)\times \RR^{12}$ is outperformed by IEKF based on   $SE_2(3) \times \RR^6$, i.e., imperfect IEKF. The latter is in turn outperformed by TFG-IEKF based on  $\tFrames=SO(3)^+_{2,2}$.

This example is of major importance   as most recent successes of IEKF are related to inertial navigation problems where gyroscope and accelerometer bias came with a simple additive group structure. We claim this is  probably not the most efficient approach and  TFG structure should be used from now on when implementing IEKFs. The example shows indeed that even when full state independence is not achieved, using the TFG structure for body-frame vectors is beneficial in practice. Similarly the treatment of GNSS or camera lever arms should also follow the TFG approach and more generally it should be tested for all systems that involve estimating a change of frame along with variables defined in the two frames.  The observed improvements may stem from the drastic reduction of the dependency of the Jacobians and error evolution on the trajectory.

\section{Conclusion}

In this paper a novel class of systems was introduced, and a novel group structure was shown to endow it with strong properties regarding observer design. The obtained versatile and constructive framework unifies a large body of successes of the IEKF to date, and also allows for additional vectors to be estimated   such as IMU biases, lever arms, moving landmarks when doing SLAM,  as long as some commutation relations hold.

We focused on state independence of error, but the four properties of the IEKF displayed in Introduction apply to natural two-frames systems. Re-deriving them all in detail  and studying their consequences obviously goes beyond the scope of this paper, e.g., consistency of   TFG-IEKF SLAM with moving object tracking, see  \cite{barrau2015ekf}, and is left as a perspective.

\section{Appendix}

\subsection{Exponential map of the TFG}\label{expo:sec}
 
A complete theory is provided in the  supplementary material.  Herein, we provide a numerically efficient formula when the   actions of $G$   are the term-by-term rotations of Example \ref{expl::term-by-term}.  
\begin{prop}[Exponential map for rotation TFGs]
\label{prop::exp-rot}
Assume $G$ is $SO(2)$ or $SO(3)$ and its action is a multi-vector rotation as in Example \ref{expl::term-by-term}. Then we have:
\begin{align}\label{exp:formula}{
\exp_{G_{N_1,N_2}^+}
\begin{pmatrix}
\xi^R \\ \xi^\fixX_1 \\ \vdots \\ \xi^\fixX_{\nbVectFix} \\ \xi^\bodyX_1 \\ \vdots \\ \xi^\bodyX_{\nbVectBody}
\end{pmatrix}
=
\begin{pmatrix}
\expm{SO(\dimEmbedG)}{\xi^R} \\
\Matrix{\nu_\dimEmbedG(\xi^R)} \xi^\fixX_1 \\ \vdots \\ \Matrix{\nu_\dimEmbedG(\xi^R)} \xi^\fixX_{\nbVectFix} \\
\Matrix{\nu_\dimEmbedG(-\xi^R)} \xi^\bodyX_1 \\ \vdots \\ \Matrix{\nu_\dimEmbedG(-\xi^R)}  \xi^\bodyX_{\nbVectBody}
\end{pmatrix}}
\end{align}
Where $\Matrix{\nu_\dimEmbedG}$ is given for $\dimEmbedG=3$ and $\dimEmbedG=2$ by:
\begin{align*}
\Matrix{\nu_3}(\xi) & = \Matrix{I}_3 + \frac{1-\cos(||\xi||)}{||\xi||^2} \Matrix{(\xi)}_{\times }+\frac{||\xi||-\sin(||\xi||)}{||\xi||^3}\Matrix{(\xi)}_{\times }^2, \\
\nuMat{\xi} & =\frac{\sin ||\xi||}{||\xi||} \Matrix{I}+ \frac{1-\cos||\xi||}{||\xi||} \Matrix{J}\quad \text{with}\quad \Matrix{J}:=\Matrix{\rho}(\pi/2),
\end{align*}
where $\Matrix{\rho}(\theta)$ denotes the 2$\times$2 rotation matrix of angle $\theta$. 
\end{prop} 
The proof is based on an embedding into a matrix Lie group where group composition boils down to matrix multiplication (see Appendix \ref{jacob:sec} for notation):
\begin{equation}
 (R, \fixX, \bodyX) \mapsto
 \begin{pmatrix}
 \rep{  R  }{\fixV} & \fixX & \0{\dimFixV, \dimBodyV} & \0{1,\dimBodyV} \\
 \0{1, \dimFixV} & 1 & \0{1, \dimBodyV} & 0 \\
 \0{\dimBodyV,\dimFixV} & \0{\dimBodyV, 1} &  \rep{  R  }{\bodyV} & \rep{  R  }{\bodyV} \bodyX \\
 \0{1, \dimFixV} & 0 & \0{1, \dimBodyV} & 1 \\ 
 \end{pmatrix} 
\end{equation}
It may be checked that by multiplying such two matrices we recover the TFG group law  \eqref{group:comp}.   
\color{black}

\subsection{Error equations in the original variables}\label{error:ec:sec}

To tune the gains of an IEKF, one must linearize the error equations. A first step is to translate the various abstract error-related formulas on the TFG in terms of  the original variables. 

\textbf{\underline{Notation}}: From that point onwards, we may omit $\action$ and denote   $R \action x$ as $Rx$ to alleviate calculation and formulas.

Let us first transpose the result of Prop. \ref{prop::innov} using the original variables.  By using the TFG action \eqref{star}, Eq. \eqref{innov:abs}  
proves the innovation  \eqref{eq::innovBody} or \eqref{eq::innovFix} computed from a natural output is   given by \eqref{eq::innovBodyError} or \eqref{eq::innovFixError} below:
\begin{align}
\label{eq::innovBodyError}
&\text{\normalfont{Obs. \eqref{eq::output_fixed}}} \Rightarrow\bodyZ_n = \Matrix{H}^\fixX \bodyE_n^x + \Matrix{H}^\bodyX \bodyE_n^R \bodyE_n^{\bodyX} + \bodyE_n^R  \bodyb_n-\bodyb_n,
\\
\label{eq::innovFixError}
&\text{\normalfont{Obs. \eqref{eq::output_body}}}\Rightarrow  \fixZ_n = - \Matrix{H}^\bodyX \fixE_n^\bodyX -\Matrix{H}^\fixX \left(\fixE_n^R \right)^{-1} \fixE_n^x + \left( \fixE_n^R \right)^{-1}  \fixb_n-\fixb_n.
\end{align}
This result is of major importance regarding observer design, and should be put in contrast  with innovation terms (a.k.a. prediction errors) used for general nonlinear observers and the EKF:
$
z_n = h(x_n) - h(\hat{\fixX}_{n|n-1})
$
where the innovation is not a function of the error except in the linear case where innovation has the form $z_n = H e_{n|n-1}$, with $e_{n|n-1} = x_n-\hat{\fixX}_{n|n-1}$.

\begin{rem}
In the \emph{particular case} where $\bodyV=\RR^3$ and $\bodyX \in \bodyV$ denotes a bias  or a lever arm in the body frame,   \eqref{eq::error_variables_body} and \eqref{eq::error_variables_fixed} advocate two different errors:  $\bodyX-R^{-1}\hat R\hat \bodyX$ or $\hat{R}(\bodyX-\hat{\bodyX})$. The former was already proposed in \cite{andrle2015attitude} using physical arguments, see also \cite{chang2020se}. But while the authors of \cite{andrle2015attitude} consider this error variable is intrinsically better,    Prop \ref{prop::update_aux}  shows  the innovation based on it is better  only when observations are performed in the fixed frame, a counterintuitive fact.  
\end{rem}

By using the TFG law, \eqref{update:abs5} readily proves the following:
\begin{prop}[Error evolution at update]
\label{prop::update_aux}
For fixed-frame observations and   observer \eqref{eq::invObsFix} we have:
\begin{equation}
\label{eq::error-update-body}
\begin{aligned}
\bodyE^R_{n|n} & = L_n^R(\bodyZ_n)^{-1} \bodyE_{n|n-1} \\
\bodyE^\fixX_{n|n} & = L_n^R(\bodyZ_n)^{-1} \action\left( \bodyE^\fixX_{n|n-1} -  L_n^\bodyX(\bodyZ_n) \right) \\
\bodyE^\bodyX_{n|n} & = \bodyE_{n|n-1}^\bodyX - \left( \bodyE_{n|n-1}^R \right)^{-1}  L_n^R(\bodyZ_n) \action L_n^\bodyX(\bodyZ_n),
\end{aligned}
\end{equation}
with $\bodyZ_n$ as in \eqref{eq::innovBody}. On the other hand,  for body-frame observations  and  observer \eqref{eq::invObsBody}  with $  \fixZ_n$ as in \eqref{eq::innovFix} we have: 
\begin{equation}
\label{eq::error-update-fix}
\begin{aligned}
\fixE^R_{n|n} & = \fixE_{n|n-1} L_n^R(\fixZ_n)^{-1} \\
\fixE^\fixX_{n|n} & = \fixE^\fixX_{n|n-1} - \fixE^R_{n|n-1} L_n^R(\fixZ_n)^{-1}\action L_n^\fixX(\fixZ_n) \\
\fixE^\bodyX_{n|n} & =   L_n^R(\fixZ_n)\action\left( \fixE_{n|n-1}^\bodyX - L_n^X(\fixZ_n)  \right).
\end{aligned}
\end{equation} 
\end{prop}
Recalling \eqref{eq::innovBodyError}, \eqref{eq::innovFixError},  we   see that 
    \eqref{eq::error-update-body},  \eqref{eq::error-update-fix} illustrate  our point, which is that state variables  $\chi_n$, $\hat \chi_{n|n}$ , $\hat \chi_{n|n-1}$  vanish as in the linear case, leading to  state independant error evolution at the update step.  Then, we proved at Theorem \ref{thm::group-affine} that vector dynamics is group affine. Thus, by letting $\phi_n=f_n$, the result \eqref{propagation:abs5} holds true. Using the TFG law  \eqref{group:comp}, this readily proves: 
\begin{prop}[Error   via natural vector dynamics]\label{vect:prop}
The    left-invariant error through \eqref{eq::vector_dynamics}  satisfies:
\begin{equation}
\label{eq::error-vector-body}
\begin{aligned}
\bodyE^R_{n^-|n-1} & = \bodyE^R_{n|n-1} \\
\bodyE^\fixX_{n^-|n-1} & = \fixF_n \bodyE^\fixX_{n|n-1} +  \fixC_n \bodyE^R_{n|n-1}\bodyE^\bodyX_{n|n-1} + \bodyE^R_{n|n-1} \bodyu_n - \bodyu_n \\
\bodyE^\bodyX_{n^-|n-1} & = \bodyF_n \bodyE_{n|n-1}^\bodyX + \bodyd_n - (\bodyE_{n|n-1}^R)^{-1} \bodyd_n  + \bodyC_n (\bodyE_{n|n-1}^R)^{-1}  \bodyE_{n|n-1}^\fixX
\end{aligned}
\end{equation}
and the right-invariant error  through \eqref{eq::vector_dynamics}  satisfies:
\begin{equation}
\label{eq::error-vector-fix}
\begin{aligned}
e^R_{n^-|n-1} & = e^R_{n|n-1} \\
e^\fixX_{n^-|n-1} & =
\fixF_n e^\fixX_{n|n-1} + \fixC_n e_{n|n-1}^R e_{n|n-1}^\bodyX + \fixd_n-e_{n|n-1}^R \fixd_n \\
e^\bodyX_{n^-|n-1} & = \bodyF_n e_{n|n-1}^\bodyX + \bodyC_n (e_{n|n-1}^R)^{-1}  e_{n|n-1}^\fixX  + (e_{n|n-1}^R)^{-1} \fixu_n - \fixu_n.
\end{aligned}
\end{equation}
\end{prop}
The formulas illustrate more concretely what was already known from Thm. 
\ref{thm::group-affine} and Prop. \ref {prop::propagation5}, that is,  state independence of the error evolution  systematically occurs at natural vector propagation step.  

Finally, error evolution during natural frame dynamics reads as follows in the  original variables:
\begin{prop}[Error    via natural frame dynamics]
The    left-invariant error through \eqref{eq::frame-shift}  satisfies:
\begin{equation}
\label{eq::error-frame-body}
\begin{aligned}
\bodyE^R_{n|n-1} & = \Omega_n^{-1} \bodyE^R_{n^-|n-1} \Omega_n \\
\bodyE^\fixX_{n|n-1} & = \Omega_n^{-1} O_n^{-1} \bodyE^\fixX_{n^-|n-1} \\
\bodyE^\bodyX_{n|n-1} & = \bodyE^\bodyX_{n^-|n-1}
\end{aligned}
\end{equation}
and the right-invariant error  through \eqref{eq::frame-shift}   satisfies:
\begin{equation}
\label{eq::error-frame-fix}
\begin{aligned}
e^R_{n|n-1} & = O_n e^R_{n^-|n-1} O_n^{-1} \\
e^\fixX_{n|n-1} & = e^\fixX_{n^-|n-1} \\
e^\bodyX_{n|n-1} & = O_n \Omega_n e^\fixX_{n^-|n-1} \\
\end{aligned}
\end{equation}
\end{prop}
 
Although this was known from Thm. \ref{thm:error}, explicit  formulas \eqref{eq::innovBodyError}, \eqref{eq::innovFixError},  
    \eqref{eq::error-update-body},  \eqref{eq::error-update-fix}, \eqref{eq::error-vector-body}, \eqref{eq::error-vector-fix} and then \eqref{eq::error-frame-body}, \eqref{eq::error-frame-fix} concretely illustrate the state-trajectory independence of the error. Note that all the preceding error equations can also be derived directly, as a good exercise, see supplementary material (Section \ref{direct::sec}) below where it is  done.

\subsection{Linearization of error equations and IEKF Jacobians}\label{jacob:sec}
Linearizing on groups  requires a few technical ingredients. It is useful to be familiar with $SO(3)$ and think of $G$ as $SO(3)$.

\vspace{.2cm}

\emph{Matrix $\rep{R}{V}$}: In the definition of a group action Def. \ref{def::action} we assumed    the mapping $x \mapsto \elem{R} \action x$ is linear on vector space $V$. 
In particular, if a basis of $V$ has been chosen, it can be described by a matrix we denote by $\rep{R}{V}$ verifying $(\rep{R}{V} )x = \elem{R} \action x$ for all $x\in V$. For the   term-by-term action \eqref{tbt:eq}  of the group $SO(d)$, writing $N$-tuples of vectors as stacked vectors of $W=\RR^{d   N}$, we obviously have 
$
\rep{R}{W} = \text{diag}  \left( \Matrix{R}, \dots, \Matrix{R} \right)$. Note $\rep{R}{W} $ also denotes a (other) matrix verifying $(\rep{R}{B} )\bodyX = \elem{R} \action \bodyX$ for all $\bodyX\in B$. Albeit a different matrix,   confusion is hardly possible and using the same notation is consistent with our choice to use $\action$ in both cases hitherto.

\vspace{.2cm}

\emph{First-order expansion in the $R$ element}: Let $\xi=(\xi^R,\xi^\fixX,\xi^\bodyX)$ in the Lie algebra of the TFG, that is,   $\xi^R\in\mathfrak g$ the Lie algebra of $G$ and $(\xi^\fixX,\xi^\bodyX)\in V\times B$, see supplementary material below. By denoting $R=\exp_{G}(\xi^R)$ we may define - see e.g.,  \cite{Hall} - a linear map $\repu{\xi^R}{V}$ on $V$ through equality  $\rep{R}{V}~ = \expm{m}{\repu{\xi^R}{V}}$, where $\expm{m}{\Matrix{M}}=\Matrix{I}+\Matrix{M}+\frac{1}{2}\Matrix{M}^2+\cdots$ denotes the  matrix exponential. This yields $\rep{(\exp_{G}(\xi^R))}{V} = \Matrix{I}_q + \repu{\xi^R}{V} + \circ(\xi^R)$ and an other linear map shall be defined on $B$, using the same notation.

\vspace{.2cm}

\emph{Linearization w.r.t. $\xi^R$}: As   $\repu{\xi^R}{V} x$ is also linear  w.r.t. $\xi^R$, we define $q\times d$ matrix $\dg{x}$ via $\dg{x}\xi^R := -\repu{\xi^R}{V} x$, so:
\begin{equation}
\rep{\exp_{G}(\xi^R)}{V} x \approx x + \repu{\xi^R}{V}~ x  
=x - \dg{x}\xi^R.\label{exp:lin}
\end{equation} 
\begin{expl}\label{ex4}
Let $\Matrix{R}\in SO(3)$  and $x\in V=\RR^3$. In this case we merely have $R*x=\Matrix{R}x$ and thus find $\rep{R}{V}=\Matrix{R}$. Besides, $\repu{\xi}{V}=\Matrix{(\xi)_\times}$ where $\Matrix{(\xi)_\times}$ denotes the skew-symmetric matrix associated with $\xi$ so the notation is coherent. This ensures in turn that  $\dg{x} = \Matrix{(x)_\times}$.  In the same way, coming back to Ex. \ref{expl::term-by-term}, we have    $\repu{\xi}{V}~ x=((\xi)_\times x^1,\dots,(\xi)_\times x_N)=-\dg{x}\xi$ so that     $\dg{x}=  -\left( \Matrix{(x^1)_\times}, \dots, \Matrix{(x^N)_\times} \right)^T\in \RR^{Nd \times d}$.
\end{expl}

\vspace{.2cm}

\emph{Adjoint  $Ad$ on Lie algebra $ \mathfrak g$}: is defined through the relation: $R\grouplaw\exp_{G}(\xi^R)
\grouplaw R^{-1}=\exp_{G}(Ad_R\xi^R)$ for $R\in G$.

\vspace{.2cm}

The following ``Rosetta stone'' allows for translation of error equations on the TFG, such as  formulas \eqref{eq::innovBodyError} up to \eqref{eq::error-frame-fix},  into linearized vector error equations. The sign $\approx$ means a quantity may be readily replaced by its linearized counterpart.

\vspace{.2cm}

\emph{TFG linearization ``Rosetta stone''}: The  exponential on $\tFrames$ (see supplementary material) ensures $\bodyE^R=\exp_G(\xi^R)$, and   $\bodyE^\fixX\approx\xi^\fixX$ and  $\bodyE^\bodyX\approx\xi^\bodyX$, and naturally $\Omega\action \bodyE^\fixX\approx (\rep{\Omega}{V})\xi^\fixX$.  From \eqref{exp:lin} we have  $\rep{\bodyE^R}{V} \bodyu\approx\bodyu+\repu{\xi^R}{B}\bodyu =\bodyu- \dg{\bodyu}\xi^R$ and $\rep{\bodyE^R}{V}\bodyE^\fixX\approx\xi^\fixX$, as $\repu{\xi^R}{B}\xi^\fixX$ is second order, and similarly on $B$. As $(\bodyE^R)^{-1}=\exp_G(-\xi^R)$ we have  $\rep{(\bodyE^R)^{-1}}{V} \bodyu\approx \bodyu+ \dg{\bodyu}\xi^R$. The counterparts for the right-invariant error  $\fixE$   are identical.

\begin{prop}\label{prop:lin1}
For fixed-frame observations i.e., \emph{left-invariant error} $\bodyE_{n|n}$, the linearized error system writes \eqref{eq::first-order1}-\eqref{eq::first-order2}, and splitting $\bodyXi  $ as $ (\xi^R,\xi^\fixX,\xi^\bodyX)$, Jacobians read:
\begin{equation}
\label{eq::Fv-body}
\begin{aligned}
&\Matrix{A}^v_n =
\begin{pmatrix}
 \Matrix{I}_\dimG & \0{\dimG, \dimFixV} & \0{\dimG, \dimBodyV} \\
 -\dg{\bodyu_n} & \Matrix{F}_n &  \Matrix{C}_n \\
 -\dg{\bodyd_n} &\Matrix{\Gamma}_n & \Matrix{\Phi}_n
\end{pmatrix}
,~~\Matrix{H}_n = \begin{pmatrix}
-\dg{\bodyb_n} & \Matrix{H}^\fixX & \Matrix{H}^\bodyX
\end{pmatrix}\\&\Matrix{A}_n^s =\begin{pmatrix}
\Matrix{Ad_{\Omega_n}^{-1}} & \0{\dimG, \dimFixV} & \0{\dimG, \dimBodyV} \\
\0{\dimFixV, \dimG}  &  \rep{(\Omega_n^{-1}  O^{-1})}{\fixV} & \0{\dimFixV, \dimBodyV}  \\ \0{\dimG,\dimBodyV }& \0{ \dimFixV,\dimBodyV} & \Matrix{I}_{\dimBodyV}
\end{pmatrix},\end{aligned}\end{equation}where we recall that in the (I)EKF theory, $\Matrix{H}_n$ always relates the linearized innovation to the linearized error. 
\end{prop}
\paragraph{proof}
 Using the ``Rosetta stone'' to substitute $\bodyE$ with  $\xi$  in  \eqref{eq::error-vector-body}  yields $\xi^{ \bodyE}_{n^-|n-1}\approx\Matrix{A}^v_n\xi^{ \bodyE}_{n-1|n-1}$ and similarly with \eqref{eq::error-frame-body} to get $\Matrix{A}^s_n$.  Doing similarly in \eqref{eq::innovBodyError} proves $\bodyZ\approx\Matrix{H}_n  \bodyXi$. Albeit possible, there is no need to analyze \eqref{eq::error-update-body} to get \eqref{eq::first-order2}. This may be done more concisely at a higher level, as using \eqref{eq::gains1}, the definition  \eqref{exp:gain} and $\bodyZ\approx\Matrix{H}_n  \bodyXi$,   update  \eqref{update:abs5} rewrites  $\exp_{\tFrames}(\xi_{n|n}^{\bodyE})\approx\exp_{\tFrames}(-\Matrix{KH}\xi_{n|n-1}^{\bodyE} )\exp_{\tFrames}(\xi_{n|n-1}^{\bodyE})\approx\exp_{\tFrames}(-\Matrix{KH}\xi_{n|n-1}^{\bodyE}+\xi_{n|n-1}^{\bodyE})$ using the BCH formula and then we take the logarithm. 
 $\blacksquare$

\vspace{.2cm}

\begin{prop}\label{prop:lin2}
For body-frame observations, i.e., \emph{right-invariant error} $\fixE_{n|n}$,  we   get  the following Jacobians:
\begin{equation}
\label{eq::Fv-fix}
\begin{aligned}
&\Matrix{A}^v_n =
\begin{pmatrix}
 \Matrix{I}_\dimG & \0{\dimG, \dimFixV} & \0{\dimG, \dimBodyV} \\
 \dg{\fixd_n} & \Matrix{F}_n & \Matrix{C}_n \\
 \dg{\fixu_n} & \Matrix{\Gamma}_n & \Matrix{\Phi}_n
\end{pmatrix}
,~\Matrix{A}^s_n =
\begin{pmatrix}
\Matrix{Ad_{O_n}} & \0{\dimG, \dimFixV} & \0{\dimG, \dimBodyV} \\  \0{\dimFixV,\dimG} & \Matrix{I}_{\dimFixV} & \0{\dimFixV, \dimBodyV} \\
\0{\dimBodyV, \dimG} & \0{\dimBodyV, \dimFixV} & \rep{O_n   \Omega_n  }{\bodyV}
\end{pmatrix}\\&
\Matrix{H}_n = \begin{pmatrix}
 \dg{\fixb_n} &- \Matrix{H}^\fixX & -\Matrix{H}^\bodyX
\end{pmatrix} .\end{aligned}\end{equation}
\end{prop}Proof similarly stems from applying the TFG linearization ``Rosetta stone'' to the error equations of Appendix   \ref{error:ec:sec}. Note that in the case of generic frame dynamics, $\Matrix{A}^s_n$ may be more difficult to obtain, as in the example of Sec. \ref{expl::nav_IMU}. But Jacobians $\Matrix{A}_n^v,\Matrix{H}_n$ can then still be retrieved from the formulas above. 

\subsection{Computation of noise covariance matrices}\label{noise:sec}
Following the (I)EKF methodology \cite{barrau2017invariant}, we associate a noisy system with the dynamics. In practice, noises should reflect the magnitude of the sensors' uncertainty. 
\begin{dfn}
\label{defn::noise}
Consider   system \eqref{eq::two-frame-dynamics} where $s_n$ is generic,  with noise turned on, i.e.,  noisy system $\chi_{n }=u_n(\chi_{n-1})$, $\fixY_n  = \fixh(\chi_n) + V_n$ or $\bodyY_n  = \bodyh(\chi_n) + V_n$ with $u_n(R , \fixX , \bodyX)$ defined as
\begin{align*}
\begin{pmatrix}
s_n^R(R, \fixX, \bodyX)\exp_G(w_n^R) \\
[\fixF_n \fixX + \fixd_n] + R \action [\fixC_n \bodyX + \bodyu_n]+ \Matrix{\dNoiseFix} \left( \chi_n \right) w_n^\fixX\\
[\bodyF_n \bodyX + \bodyd_n] + R^{-1}\action  [\bodyC_n \fixX + \fixu_n] +  \Matrix{\dNoiseBody} \left( \chi_n \right) w_n^\bodyX
\end{pmatrix} 
\end{align*}
where $w_n^R,w_n^\fixX,w_n^\bodyX$ are process noises with covariance matrices $\Matrix{Q^R}, \Matrix{Q^\fixX},  \Matrix{Q^\bodyX}$;   $\Matrix{\dNoiseFix} \left( \chi_n \right),\Matrix{\dNoiseBody} \left( \chi_n \right) $ are state-dependent matrices; $V_n$ is an observation noise with covariance matrix  $ \Matrix{N}_n$.  
\end{dfn}
The IEKF is then fed with the following noise parameters. 
\begin{prop}
\label{prop::noise} Denote $\Matrix{\hat{\dNoiseFix}} = \Matrix{\dNoiseFix} \left( \hat\chi_{n } \right)$ and $\Matrix{\hat{\dNoiseBody}} = \Matrix{\dNoiseBody} \left(\hat \chi_{n } \right)$ we have regarding the left-invariant error  $\bodyE$:
\begin{equation}\label{QL:eq}
\begin{aligned}
\Matrix{\hat{Q}} =
\begin{pmatrix}
\Matrix{Q}^R & \0{\dimG,\dimFixV} & \Matrix{Q}^R {\dg{ \hat{\bodyX}_n}}^T \\
\0{\dimFixV,\dimG} & \Matrix{\hat{Q}}^\fixX & \0{\dimFixV,\dimBodyV} \\
\dg{ \hat{\bodyX}_n} \Matrix{Q}^R & \0{\dimBodyV,\dimFixV}  & \Matrix{\hat{Q}}^\bodyX
\end{pmatrix}
\\
\Matrix{\hat{N}} = \left(\rep{\hat{R}_n}{\outSpace} \right)^{-1} \Matrix{N} \left(\rep{\hat{R}_n}{\outSpace} \right)^{-T}\hspace{1.3 cm} & \\
 \left \lbrace
\begin{aligned}
 \Matrix{\hat{Q}}^\fixX & = \left(\rep{\hat{R_n}}{\fixV} \right)^{-1} \Matrix{\hat{\dNoiseFix}} \Matrix{Q}^\fixX {\Matrix{\hat{\dNoiseFix}}}^T \left(\rep{\hat{R}_n}{\fixV} \right)^{-T} \\
 \Matrix{\hat{Q}}^\bodyX  & = \dg{ \hat{\bodyX}_n} \Matrix{Q}^R {\dg{ \hat{\bodyX}_n}}^T + \Matrix{\hat{\dNoiseBody}} \Matrix{Q}^\bodyX \Matrix{\hat{\dNoiseBody}}^T
\end{aligned} \right.\end{aligned}
\end{equation}
and regarding the right-invariant error $\fixE$:
\begin{equation}\label{Q:eq}
\begin{aligned}
\Matrix{\hat{Q}} =
\begin{pmatrix}
\Ad{\hat{R}_n} \Matrix{Q}^R  \Ad{\hat{R}_n}^T &  \Matrix{Q}^R \left(\dg{\hat{x}_n} \Ad{\hat{R}_n} \right)^T  & \0{\dimG, \dimBodyV} \\
 \dg{\hat{x}_n} \Ad{\hat{R}_n} \Matrix{Q}^R & \Matrix{\hat{Q}}^\fixX & \0{\dimFixV, \dimBodyV} \\
\0{\dimBodyV, \dimG} & \0{\dimBodyV, \dimFixV} &  \Matrix{\hat{Q}}^\bodyX 
\end{pmatrix}
\\
\Matrix{\hat{N}}  = \left( \rep{\hat{R}_n}{\outSpace} \right) \Matrix{N} \left(\rep{\hat{R}_n}{\outSpace} \right)^T\hspace{3.5 cm} &  \\
\left \lbrace
\begin{aligned}
 \Matrix{\hat{Q}}^\fixX  & = \left(\dg{\hat{x}_n} \Ad{\hat{R}_n} \right) \Matrix{Q}^R \left( \dg{\hat{x}_n} \Ad{\hat{R}_n} \right)^T + \Matrix{\hat{\dNoiseFix}} \Matrix{Q}^\fixX \Matrix{\hat{\dNoiseFix}}^T \\
  \Matrix{\hat{Q}}^\bodyX & = \left( (\rep{\hat{R}_n}{\bodyV}) \Matrix{\hat{\dNoiseBody}} \right) \Matrix{Q}^\bodyX  \left((\rep{\hat{R}_n}{\bodyV}) \Matrix{\hat{\dNoiseBody}} \right)^T.
\end{aligned} \right.
\end{aligned}\end{equation}\end{prop}
The rationale is as follows. At propagation $\xi_{n|n-1}$ becomes $\xi_{n|n}$. If noise is injected  it then  becomes $\xi_{n|n}^{noisy}\approx \xi_{n|n}+\Matrix{B}w_n$. Owing to  noise being centered and independent we have $\mathbb E\bigl(\xi_{n|n}^{noisy}(\xi_{n|n}^{noisy})^T\bigr)\approx\mathbb E\bigl(\xi_{n|n}\xi_{n|n}^T\bigr)+\Matrix{B}\mathbb E\bigl(w_nw_n^T\bigr)\Matrix{B}^T=\mathbb E\bigl(\xi_{n|n}\xi_{n|n}^T\bigr)+
\Matrix{\hat Q}$ with $\Matrix{\hat Q}:=\Matrix{B}\Matrix{Q}\Matrix{B}^T$. We can now turn to the proof. 
\paragraph{proof}For instance, let us compute $\Matrix{B}$ for error $\fixE$. 

$\boldsymbol{\cdot}$ $ (\fixE_n^R)^{noisy}:= R_n^{noisy}\hat{R}_n^{-1}=R_n\exp(w_n^R)\hat{R}_n^{-1}=R_n\hat{R}_n^{-1}\exp(Ad_{\hat{R}_n}w_n^R)=\fixE_n^R\exp_G(Ad_{\hat{R}_n}w_n^R)$ using the  BCH formula yields $(\xi_{n}^R)^{noisy}\approx\xi_{n}^R+Ad_{\hat{R}_n}w_n^R$.

$\boldsymbol{\cdot}$ $ (\fixE_n^\fixX)^{noisy}:=\fixX_{n}^{noisy}-R_n^{noisy}\hat{R}_n^{-1}\action\hat\fixX_n=\fixX_{n}^{noisy}- (\fixE_n^R)^{noisy}\action\hat\fixX_n\approx\fixX_{n}^{noisy}-\hat\fixX_n- \repu{(\xi_n^R)^{noisy}}{V}\hat\fixX_n$ using the ``Rosetta stone''. Besides the stone shows $\xi^\fixX\approx\fixE_n^\fixX \approx \fixX_{n}-\hat\fixX_n- \repu{\xi_n^R}{V}\hat\fixX_n$. Moreover $\fixX_{n}^{noisy}=\fixX_{n}+\Matrix{\hat{\dNoiseFix}}w_n^\fixX$ and $(\xi_{n}^R)^{noisy}$ was derived, so that $(\xi_n^\fixX)^{noisy}\approx \fixE_n^\fixX+\Matrix{\hat{\dNoiseFix}}w_n^\fixX- \repu{Ad_{\hat{R}_n}w_n^R}{V}\hat\fixX_n \approx  \fixE_n^\fixX+\Matrix{\hat{\dNoiseFix}}w_n^\fixX+\dg{\hat{x}_n} \Ad{\hat{R}_n}w_n^R$. 

$\boldsymbol{\cdot}$  $ \xi_n^\bodyX\approx (\fixE_n^\bodyX)^{noisy}:=\hat R_n\action (\bodyX_{n}^{noisy}-\hat\bodyX_n)=\hat R_n\action (\bodyX_{n} +\Matrix{\hat{\dNoiseBody}} w_n^\bodyX-\hat\bodyX_n)=\fixE_n^\bodyX+(\rep{\hat R_n}{B})w_n^\bodyX\approx \xi_n^\bodyX+(\rep{\hat R_n}{B})w_n^\bodyX$. 

 $\blacksquare$

\vspace{.2cm}

\begin{rem}If   initial covariance matrix  is known for the more classical error variable $\left( \Matrix{\hat{R}^{-1} R}, \fixX - \hat{\fixX}, \bodyX- \hat{\bodyX} \right)$ (with value $\Matrix{\bar{P}}$), then $\Matrix{P_{0|0}}$ in Def. \ref{def:TFGIEKF} is given by
 $
\Matrix{P_{0|0}} = \Matrix{L \bar{P} L^T},
$ 
with $\Matrix{L}=\begin{pmatrix}
\Matrix{I_3} &  \0{3}    & \0{3} \\
    \0{3}    & \rep{\hat{R}_n^{-1}}{\fixV} & \0{3} \\
    -\dg{\hat{\bodyX}} & \0{3} & \Matrix{I_3}
\end{pmatrix}$ for error $\bodyE$. Similarly if  $\Matrix{\bar{P}}$ denotes the covariance of $\left( \Matrix{R_n\hat{R}^{-1} }, \fixX- \hat{\fixX}, \bodyX - \hat{\bodyX}\right)$ then $\Matrix{L}=
\begin{pmatrix}
  \Matrix{I_3}     &        \0{3}            &        \0{3}           \\
\dg{\hat{\bodyX}} & \Matrix{I_3} &        \0{3}           \\
    \0{3}          &        \0{3}            & \rep{\hat{R}_n}{\fixV}  
\end{pmatrix}$ for error $\fixE$. \end {rem}

\bibliographystyle{IEEEtran}

 \clearpage

 \section{Supplementary Material to ``The Geometry of Navigation Problems''}\label{supp:sec}

\subsection{Exponential map of the TFG} 

Reading the present section requires having read the Appendix of the paper before, notably   Appendix \ref{jacob:sec}  that introduces a number of notations. 
 
We first recall that if a Lie group admits an embedding in a matrix group, then the exponential map coincides with the matrix exponential. 
Now, let us apply this  method to find the exponential of $\tFrames$. This has the merit to be a systematic method to derive the exponential of a Lie group, as long as an embedding into a matrix Lie group has first been found. 

\color{black}
\begin{lem}The Lie algebra of the TFG can be parameterized by $\mathfrak{g} \times \fixV \times \bodyV$.
\end{lem}
Indeed, although we are not dealing with a matrix group, any element of the Lie algebra can be considered as having three components, the first one being an element of the Lie algebra of $G$ and the two others being the vector spaces themselves. In practice, $G$ is a matrix group, so that its Lie algebra may be interpreted as what is customarily done with matrix groups in robotics. Now that we have a Lie algebra, we can introduce the exponential. 
\color{black}
\begin{prop_n}[Exponential map of the TFG]
\label{prop::exp} 
A general formula for the Lie exponential map is then obtained as $\exp_{\tFrames}(\xi^R,\xi^\fixX,\xi^\bodyX)=(R,\fixX,\bodyX)$, where $R,\fixX,\bodyX$ are extracted from:
\begin{equation}\label{exp1}
\begin{pmatrix}
\rep{  R  }{\fixV} &  \fixX \\
 0 & 1
\end{pmatrix}
=
\expm{m}{
\begin{matrix}
\repu{  \xi^R }{\fixV} & \xi^\fixX \\
 0 & 0
\end{matrix}}.
\end{equation}
\begin{equation}\label{exp2}
\begin{pmatrix}
\rep{  R  }{\bodyV} & \rep{  R  }{\bodyV}  \bodyX \\
 0 & 1
\end{pmatrix}
=
\expm{m}{
\begin{matrix}
\repu{  \xi^R  }{\bodyV} & \xi^\bodyX \\
 0 & 0
\end{matrix}}.
\end{equation}
\end{prop_n}
\paragraph{proof} 
The proof is based on an embedding into a matrix Lie group where group composition boils down to matrix multiplication, which requires changing the frame of $\bodyX$:
\begin{equation}
 (R, \fixX, \bodyX) \mapsto
 \begin{pmatrix}
 \rep{  R  }{\fixV} & \fixX & \0{\dimFixV, \dimBodyV} & \0{1,\dimBodyV} \\
 \0{1, \dimFixV} & 1 & \0{1, \dimBodyV} & 0 \\
 \0{\dimBodyV,\dimFixV} & \0{\dimBodyV, 1} &  \rep{  R  }{\bodyV} & \rep{  R  }{\bodyV} \bodyX \\
 \0{1, \dimFixV} & 0 & \0{1, \dimBodyV} & 1 \\ 
 \end{pmatrix}\label{embed:eq}
\end{equation}
It may be checked that by multiplying such two matrices we recover the TFG group law \eqref{group:comp}. 
Lie exponential map coincides with $\expm{m}{\cdot}$  for matrix groups. The block structure allows us to split the exponential into two matrix exponentials.  $\blacksquare$

\subsection{Proof of Proposition \ref{prop::exp-rot}}

Let us reproduce this  result from the paper herein for simplicity, before proving it. 
\begin{prop_n}[Exponential map for rotation TFGs]
Assume $G$ is $SO(2)$ or $SO(3)$ and its action is a multi-vector rotation as in Example 1. Then we have:
\begin{align}\label{exp:formula2}{
\exp_{G_{N_1,N_2}^+}
\begin{pmatrix}
\xi^R \\ \xi^\fixX_1 \\ \vdots \\ \xi^\fixX_{\nbVectFix} \\ \xi^\bodyX_1 \\ \vdots \\ \xi^\bodyX_{\nbVectBody}
\end{pmatrix}
=
\begin{pmatrix}
\expm{SO(\dimEmbedG)}{\xi^R} \\
\Matrix{\nu_\dimEmbedG(\xi^R)} \xi^\fixX_1 \\ \vdots \\ \Matrix{\nu_\dimEmbedG(\xi^R)} \xi^\fixX_{\nbVectFix} \\
\Matrix{\nu_\dimEmbedG(-\xi^R)} \xi^\bodyX_1 \\ \vdots \\ \Matrix{\nu_\dimEmbedG(-\xi^R)}  \xi^\bodyX_{\nbVectBody}
\end{pmatrix}}
\end{align}
Where $\Matrix{\nu_\dimEmbedG}$ is given for $\dimEmbedG=3$ and $\dimEmbedG=2$ by:
\begin{align*}
\Matrix{\nu_3}(\xi) & = \Matrix{I}_3 + \frac{1-\cos(||\xi||)}{||\xi||^2} \Matrix{(\xi)}_{\times }+\frac{||\xi||-\sin(||\xi||)}{||\xi||^3}\Matrix{(\xi)}_{\times }^2, \\
\nuMat{\xi} & =\frac{\sin ||\xi||}{||\xi||} \Matrix{I}+ \frac{1-\cos||\xi||}{||\xi||} \Matrix{J}\quad \text{with}\quad \Matrix{J}:=\Matrix{\rho}(\pi/2),
\end{align*}
where $\Matrix{\rho}(\theta)$ denotes the 2$\times$2 rotation matrix of angle $\theta$. 
\end{prop_n}
\paragraph{proof}It is known that  in the case where $\Matrix{R}\in SO(d)$ with $d=2$ or $d=3$ that \eqref{exp1} yields  elements of the form  $\Matrix{R}=\expm{SO(\dimEmbedG)}{\xi^R}$ and $x=
\Matrix{\nu_\dimEmbedG(\xi^R)} \xi^\fixX$, with $\Matrix{\nu_\dimEmbedG(\xi^R)}$ as in the proposition,  see  
\cite{barrau2015ekf,barrau2017invariant,barrau2015non}. This is also illustrated by the fact that  $SE(3)$  coincides with the TFG group $\tFrames=SO(3)^+_{1,0}$.

Looking at \eqref{exp2}, this may in turn be used to prove the part that concerns $\bodyX$. Indeed in \eqref{exp2} the exponential of the right member is similar to those in $\eqref{exp1}$, so we may use our first result to write$$\bodyX=\Matrix{R}^{-1}\Matrix{\nu_\dimEmbedG(\xi^R)}  \xi^\bodyX.$$  To prove our result, we thus would like to show that  $\Matrix{R}^{-1}\Matrix{\nu_\dimEmbedG(\xi^R)}  =\Matrix{\nu_\dimEmbedG(-\xi^R)}  $, where $\Matrix{R}=\expm{SO(\dimEmbedG)}{\xi^R}$. 

First, using the matrix exponential definition in \eqref{exp1} proves $\Matrix{R}=\mathbf{I}+\repu{  \xi^R }{\fixV}+\frac{1}{2!}\repu{  \xi^R }{\fixV}^2+\frac{1}{3!}\repu{  \xi^R) }{\fixV}^3+\dots$ and  that $\Matrix{\nu_\dimEmbedG(\xi^R)} =\mathbf{I}+\frac{1}{2!}\repu{  \xi^R }{\fixV}+\frac{1}{3!}\repu{  \xi^R) }{\fixV}^2+\dots$. Let us use those series expansions to show our desired equality $\Matrix{R}\Matrix{\nu_\dimEmbedG(-\xi^R)}=\Matrix{\nu_\dimEmbedG(\xi^R)} $. In the latter expression, we are confronted with the Cauchy product of two series where it is clear from the expansion of $ \Matrix{\nu}$ that  $\Matrix{\xi^R}\Matrix{\nu_\dimEmbedG(\xi^R)}=\Matrix{R}-\mathbf{I}$. As all matrices commute, proving the desired equality amounts to proving in the scalar case that for any scalar $z$ we have $\frac{e^z-1}{z}=e^z\frac{e^{-z}-1}{-z}$, an obviously true equality whose  series expansions coincide with those in the desired equality.  $\blacksquare$

Note that embedding \eqref{embed:eq} can be used for theoretical analysis but is not recommended for implementation as the dynamics of $\bodyX_n$ is usually very simple ($\bodyX_n$ is often constant) while the dynamics of $  \rep{  R_n  }{\bodyV} \bodyX_n$ may  be more convoluted. However, the method of the matrix group embedding is general and systematic  which may be quite reassuring to the practitioner when prototyping a filter  (albeit usually cumbersome in terms of numerical computations). \\

\emph{Linear approximation of the exponential}: In the ``Rosetta stone'' of the paper's appendix we wrote that from the definition of the exponential on $\tFrames$, we have $\bodyE^R=\exp_G(\xi^R)$, and   $\bodyE^\fixX\approx\xi^\fixX$ and  $\bodyE^\bodyX\approx\xi^\bodyX$. Let us justify it in the light of the developments above. By definition $(\bodyE^R,\bodyE^\fixX,\bodyE^\bodyX)=\exp_{\tFrames}(\xi^R,\xi^\fixX,\xi^\bodyX)$. The formula  $\bodyE^R=\exp_G(\xi^R)$ stems from \eqref{exp1} by writing the matrix exponential as a series and doing block multiplication as $\expm{m}{\Matrix{M}}=\Matrix{I}+\Matrix{M}+\frac{1}{2}\Matrix{M}^2+\cdots$. A first order expansion proves $\bodyE^\fixX\approx\xi^\fixX$ . Using \eqref{exp2} and the first order expansion $\rep{\exp_{G}(\xi^R)}{V}   \approx I_q+ \repu{\xi^R}{B}$, which is formula  \eqref{exp:lin}  in the article, we similarly find $\bodyE^\bodyX\approx\xi^\bodyX$ up to the first order.

\subsection{Error  linearization of generic frame dynamics}

Reading the present section requires having read the appendix of the paper before.

In the paper, we gave all the formulas for the Jacobians of natural vector and frame dynamics. In the present section, we first give a few guidelines for the linearization of generic frame dynamics, see Definition \ref{biased:def}.  This is then applied to fully prove Proposition \ref{prop:nav:IEKF}, the proof being only sketched in the paper. 

\subsubsection{Guidelines for linearization of generic frame dynamics}

It is harmless to assume $G$ to be a matrix Lie group (and even $G=SO(3)$ if one whishes to have a concrete picture in mind). Let us recall a few useful facts  beyond the TFG linearization ``Rosetta stone'', that prove useful to linearize general frame dynamics. 

\begin{itemize}
\item As an application of the Baker-Campbell-Hausdorff (BCH) formula, we have
\begin{align*}
 \exp_G(\alpha)\exp_G(\beta) =\exp_G(\alpha+\beta+\dots)
\end{align*}
where ``$\dots$'' represent second order terms of magnitude $O\bigl(||\alpha||^2,||\alpha||||\beta||,||\beta||^2\bigr)$. This is useful when both terms are small. 
\item The latter formula is not useful to expand $\exp_G(\alpha+\beta)$ when, e.g., $\beta$ is small but $\alpha$ is not. However, one may then   use   the right-Jacobian formula \cite{Chirikjian}. Up to second order terms in $\beta$, we have   $\expm{SO(3)}{\alpha+\beta}\approx\expm{SO(3)}{\alpha}\expm{SO(3)}{\exp(J\beta)}$ with the Jacobian defined by ${J}  = 
 {I_3} - \frac{1-\cos(||\alpha||)}{||\alpha||^2} {(\alpha)_\times} - \frac{\sin(||\alpha||)-||\alpha||}{||\alpha||^3} {(\alpha)_\times}^2$. 
 \item One shall not forget that if $G=SO(3)$ we have $\exp_G(\alpha)=\expm{m}{(\alpha)_\times}$ where $\expm{m}{}$ denotes the usual matrix exponential, so that $\expm{m}{(\alpha)_\times}=I_3+(\alpha)_\times+O(\||\alpha||^2)$.
 \item Adjoint: if $R\in SO(3)=G$ we have $R\exp_G(\xi)R^T=\exp_G(Ad_R\xi)= \exp_G(R\xi)$. 
\end{itemize}

Let us use  the points above to derive the 
Jacobian associated to frame dynamics for inertial navigation with IMU biases. \\

\subsubsection{Proof of Proposition \ref{prop:nav:IEKF} }
Let us reproduce  the result here for simplicity before proving it. 
\begin{prop_n} Jacobian for frame dynamics writes
\begin{align}
& \Matrix{A}_n^s = \begin{pmatrix}
\Matrix{I}_3 & \0{3,3} & \0{3,3} & \Matrix{M_1} & \0{3,3} \\
\0{3,3}  & \Matrix{I}_3 & \0{3,3}  & \Matrix{(\hat{v}_{n^-|n-1})_\times M_1} & \0{3,3} \\
\0{3,3} & \0{3,3} & \Matrix{I}_3 & \Matrix{(\hat{p}_{n^-|n-1})_\times M_1} & \0{3,3} \\
\0{3,3} & \0{3,3} & \0{3,3} & \Matrix{M_2} & \0{3,3} \\
\0{3,3} & \0{3,3} & \0{3,3} & \0{3,3} & \Matrix{M_2}  
    \end{pmatrix}, \label{Fsbias2}
\end{align}with $
  \Matrix{M_1}  = \Delta t\Matrix{\hat{R}_{n|n-1} \Matrix{ J}   \hat{R}_{n-|n-1}^T},$   $
   \Matrix{M_2}  = \Matrix{\hat{R}_{n|n-1} \hat{R}_{n-|n-1}^T}$,   $ \Matrix{ J}  = 
 \Matrix{I_3} - \frac{1-\cos(||\mu||)}{||\mu||^2} \Matrix{(\mu)_\times} - \frac{\sin(||\mu||)-||\mu||}{||\mu||^3} \Matrix{(\mu)_\times}^2$,  $\mu:= \omega+\hat b^\omega$. 
\end{prop_n}
\paragraph{proof}
We   study to the   effect of propagation through frame dynamics  $\Matrix{R_n}=\Matrix{R_{n-1}} \expm{m}{\Delta t{[    \omega_n+b^\omega_{n-1}]_\times}}$   and $\Matrix{\hat R_{n|n-1}} =\Matrix{\hat R_{n-|n-1}} \expm{m}{\Delta t{ \Matrix{[    \omega_n+\hat b^\omega_{n-|n-1}]_\times}}}$ on the right-invariant error \eqref{eq::error_variables_fixed}  up to the first order.  

$\boldsymbol{-}$ After frame propagation, error $\fixE^R$ is equal to
\begin{align}
\Matrix{R_n}\Matrix{\hat R_{n|n-1}}^{-1}=&\Matrix{R_{n-1}} \expm{m}{\Delta t{[    \omega_n+b^\omega_{n-1}]_\times}}\\&\quad\expm{m}{- \Delta t{\Matrix{[    \omega_n+\hat b^\omega_{n-|n-1}]_\times}}}\Matrix{\hat R^T_{n-|n-1}}
\end{align}
Let us drop all the indexes for clarity of reading (as well as the bold fonts indicating matrices). 
We see that for small errors  $ \omega+\hat b^\omega$ is a perturbation of $ \omega_n+  b^\omega$ by writing $ \omega_n+  b^\omega = \omega_n+  \hat b^\omega+\delta b^\omega$ with $\delta b^\omega=  b^\omega-\hat b^\omega$.  To expand a matrix exponential to the first order at a point that may be far from identity, we use   the right-Jacobian formula \cite{Chirikjian}. We have   $\exp_m({\Delta t{{[    \omega+b^\omega]_\times}}})=\exp_m({\Delta t{{[    \omega+\hat b^\omega+\delta b^\omega]_\times}}})\approx\exp_m({\Delta t{ {[    \omega+\hat b^\omega]_\times}}}) \exp_m({\Delta t{ {[  J\delta b^\omega]_\times}}})$ with ${J}  = 
 {I_3} - \frac{1-\cos(||\mu||)}{||\mu||^2} {(\mu)_\times} - \frac{\sin(||\mu||)-||\mu||}{||\mu||^3} {(\mu)_\times}^2$ and where $\mu:= \omega+\hat b^\omega$. Thus error $\fixE^R$ after it has been propagated writes
\begin{align*}&{R} \exp_m(\Delta t{[    \omega_n+ \hat b^\omega_{n-1}]_\times}) \exp_m({\Delta t{ {[  J\delta b^\omega]_\times}}})  \exp_m(\Delta t{[    \omega_n+ \hat b^\omega_{n-1}]_\times}) ^T{\hat R}^{T}\end{align*}
We may then write $R=R\hat R^T\hat R=\fixE^R\hat R$.  The indices are such that $R_{n-1}=\fixE^R_{n-|n-1}\hat R_{n-|n-1}$. As we have $\hat R_{n|n-1}=\hat R_{n-|n-1} \exp_m(\Delta t{[    \omega_n+ \hat b^\omega_{n-1}]_\times}) $, we have obtained
\begin{align}\fixE^R_{n|n-1}&=\fixE^R_{n-|n-1}\hat R_{n|n-1} \exp_m({\Delta t{ {[  J\delta b^\omega]_\times}}})\hat R_{n|n-1}^T\\&=\fixE^R_{n-|n-1} \exp_m({\Delta t{ {[ \hat R_{n|n-1} J\delta b^\omega]_\times}}})\label{mmi}
\end{align}
where we have used that $Ad_R=R$.  As both terms are small BCH yields this may be approximated by $  \exp_m({{ {[ \xi^R+ \Delta t\hat R J\delta b^\omega]_\times}}})$, where we recall that $\fixE^R=\exp_m([\xi^R]_\times)$. Now recall the  right-invariant error \eqref{eq::error_variables_fixed}  on the bias, which is a body frame variable,   is defined as  $\xi^\omega:=\hat R(b^\omega-\hat b^\omega)=\hat R\delta b^\omega $.  We may thus write $\hat R_{n|n-1} J\delta b^\omega_{n-|n-1}=\hat R_{n|n-1} J  \hat R_{n-|n-1}^T\hat R_{n-|n-1}\delta b^\omega_{n-|n-1} =(\hat R_{n|n-1} J\hat R_{n-|n-1}^T)\xi^\omega_{n-|n-1}$.   Thus $\xi^R$ propagates as $\xi^R+M_1\xi^\omega$, and this explains the first row in the matrix \eqref{Fsbias2}.

 $\boldsymbol{-}~$ Let us turn to $\xi^\fixX$. The    right-invariant error \eqref{eq::error_variables_fixed}  shows that  $\fixE^\fixX=\fixX-\fixE^R\action\hat \fixX$ and we want to compute its evolution under the frame dynamics, that only affects the $R,\hat R$ elements and thus $\fixE^\fixX$. Let us consider the position $p$ (calculation is identical regarding velocity variable $v$). We have just shown  at \eqref{mmi} the propagated error $\fixE^R$ writes  $    \fixE^R\exp_m[M_1\xi^\omega ]_\times)$  using BCH formula and thus the propated error regarding the $p$ component writes $p-\fixE^R\exp_m[M_1\xi^\omega ]_\times\hat p \approx p-\fixE^R\hat p- [M_1\xi^\omega ]_\times \hat p=\fixE^p+(\hat p)_\times M_1\xi^\omega$. Recalling from the Rosetta stone that $\fixE^p\approx \xi^p$  we find that the frame dynamics turns the linearized position error $\xi^p$ into $\xi^p+M_1\xi^\omega$,  and similarly for the velocity component. This respectively explains the third and second rows of  matrix \eqref{Fsbias2}.

  $\boldsymbol{-}~$ Let us turn to $\xi^\bodyX$, and study the effect of frame propagation on the biases' errors. Recall the right-invariant error \eqref{eq::error_variables_fixed}  shows that  $\fixE^\bodyX=\hat R\action(\bodyX- \hat \bodyX).$ Consider the gyro bias error for instance, that is, $\hat R(b^\omega-\hat b^\omega):=\xi^\omega$. Frame dynamics only change the $  \hat R$ part, through  $\Matrix{\hat R_{n|n-1}} =\Matrix{\hat R_{n-|n-1}} \expm{m}{\Delta t{ \Matrix{[    \omega_n+\hat b^\omega_{n-|n-1}]_\times}}}$. Thus $\xi^\omega$ becomes $\hat R_{n|n-1}(b^\omega-\hat b^\omega)=\hat R_{n|n-1}\hat R_{n-|n-1}^T\hat R_{n-|n-1}(b^\omega- \hat b^\omega)=M_2\xi^\omega$. This  explains the penultimate row of  matrix \eqref{Fsbias2}. The treatment of accelero biases is wholly similar, changing $b^\omega$ into $b^a$.  
$\blacksquare$

\subsection{A direct alternative proof to underline the importance of commutation assumptions}\label{direct::sec}

In the paper, equations  \eqref{eq::error-vector-body} and \eqref{eq::error-vector-fix} stem from the abstract error propagation formula for group affine systems \eqref{propagation:abs5}. Albeit perfectly rigorous, it may be interesting to derive it directly. Indeed, this allows the user to grasp the importance of commutation relations, that are ubiquitous in the calculation. Let us  derive  \eqref{eq::error-vector-body}. We depart from the error:
\begin{equation}
\text{\normalfont{Obs. (3) }} \Rightarrow\boxed{
\bodyE_{n|n} =
\begin{pmatrix}
\hat{R}_{n|n}^{-1}\grouplaw R_n \\ \hat{R}_{n|n}^{-1}\action (\fixX_n-\hat{\fixX}_{n|n}) \\ \bodyX_n -( R_n^{-1}\grouplaw\hat{R}_{n|n})\action  \hat{\bodyX}_{n|n} 
\end{pmatrix}.}\label{EEerror}
\end{equation}
and we want to prove it evolves as 
\begin{equation}
\begin{aligned}
\bodyE^R_{n^-|n-1} & = \bodyE^R_{n|n-1} \\
\bodyE^\fixX_{n^-|n-1} & = \fixF_n \bodyE^\fixX_{n|n-1} +  \fixC_n \bodyE^R_{n|n-1}\bodyE^\bodyX_{n|n-1} + \bodyE^R_{n|n-1} \bodyu_n - \bodyu_n \\
\bodyE^\bodyX_{n^-|n-1} & = \bodyF_n \bodyE_{n|n-1}^\bodyX + \bodyd_n - (\bodyE_{n|n-1}^R)^{-1} \bodyd_n   + \bodyC_n (\bodyE_{n|n-1}^R)^{-1}  \bodyE_{n|n-1}^\fixX
\end{aligned}
\end{equation}
through vector dynamics defined as 
\begin{equation}
f_n \begin{pmatrix}
R \\ \fixX \\ \bodyX
\end{pmatrix} = 
\begin{pmatrix}
R \\
[\fixF_n \fixX + \fixd_n] + R \action [\fixC_n \bodyX + \bodyu_n] \\
[\bodyF_n \bodyX + \bodyd_n] + R^{-1}\action  [\bodyC_n \fixX + \fixu_n] 
\end{pmatrix},
\end{equation}
and where matrices $\fixF_n: \fixV \mapsto \fixV, \quad \bodyF_n: \bodyV \mapsto \bodyV, \quad \fixC_n: \bodyV \mapsto \fixV, \quad \bodyC_n: \fixV \mapsto \bodyV $ \emph{all commute with the action of $G$}  and $\fixd_n, \fixu_n$ (resp. $\bodyd_n$, $\bodyu_n$) are vectors of $V$ (resp. $B$).\\

\paragraph{proof}:

\begin{itemize} \item As $R$ does not evolve, neither does $\bodyE^R$.
\item if  $R,\hat R,x,\hat x,\bodyC,\hat \bodyX$ all evolve through the preceding dynamics, then  $\hat{R}^{-1}  (\fixX-\hat{\fixX})$ becomes (using commutation relations that are paramount, for instance $\hat{R}^{-1}\fixF_n=\fixF_n\hat{R}^{-1}$)\begin{align*}&\hat{R}^{-1}  \bigl([\fixF_n \fixX + \fixd_n] + R \action [\fixC_n \bodyX + \bodyu_n]-[\fixF_n \hat \fixX + \fixd_n]  - \hat R \action [\fixC_n \hat \bodyX + \bodyu_n]\bigr)\\&=\hat{R}^{-1}\fixF_n [\fixX-\hat\fixX]+\hat{R}^{-1}R   [\fixC_n \bodyX + \bodyu_n]-  [\fixC_n\hat \bodyX + \bodyu_n]\\&=\fixF_n\bodyE^x+\bodyE^R \bodyu_n- \bodyu_n+ \hat{R}^{-1}R\fixC_n[\hat \bodyX -R^{-1}\hat R\hat\bodyX]\\&=
\fixF_n\bodyE^x+\bodyE^R \bodyu_n- \bodyu_n+ \bodyE^R\fixC_n[\bodyE^\bodyX]\\&=
\fixF_n\bodyE^x+\bodyE^R \bodyu_n- \bodyu_n+ \fixC_n\bodyE^R\bodyE^\bodyX
\end{align*} 
\item In the same way, the error $\bodyX -( R^{-1}\grouplaw\hat{R}) \hat{\bodyX} $ becomes after vector step
\begin{align*}& 
[\bodyF_n \bodyX + \bodyd_n] + R^{-1}   [\bodyC_n \fixX + \fixu_n]-( R^{-1}\grouplaw\hat{R})[\bodyF_n \hat \bodyX + \bodyd_n] + \hat R^{-1}  [\bodyC_n\hat  \fixX + \fixu_n] \\&=\bodyF_n[\bodyX -( R^{-1}\grouplaw\hat{R})\bodyX]+\bodyd_n-(\bodyE^R)^{-1}\bodyd_n+R^{-1}[\bodyC_n (\hat \fixX -\fixX)]\\&=\bodyF_n\bodyE^\bodyX+\bodyd_n-(\bodyE^R)^{-1}\bodyd_n+\bodyC_n R^{-1}(\hat \fixX -\fixX)\\&=\bodyF_n\bodyE^\bodyX+\bodyd_n-(\bodyE^R)^{-1}\bodyd_n+\bodyC_n (\bodyE^R)^{-1}\bodyE^\fixX 
\end{align*}
where we have used $R^{-1}\grouplaw\hat{R}=(\hat R^{-1} {R})^{-1}=(\bodyE^R)^{-1}$, and the fact that this also proves $ R^{-1}(\hat \fixX -\fixX)= (\bodyE^R)^{-1}\hat R^{-1}(\hat \fixX -\fixX)=(\bodyE^R)^{-1}\bodyE^\fixX$.
\end{itemize}
 \vspace{.3cm}
 
Let us now prove similarly  \eqref{eq::error-vector-fix}. What we want is to prove:
\begin{equation}
\begin{aligned}
e^R_{n^-|n-1} & = e^R_{n|n-1} \\
e^\fixX_{n^-|n-1} & =
\fixF_n e^\fixX_{n|n-1} + \fixC_n e_{n|n-1}^R e_{n|n-1}^\bodyX + \fixd_n-e_{n|n-1}^R \fixd_n \\
e^\bodyX_{n^-|n-1} & = \bodyF_n e_{n|n-1}^\bodyX + \bodyC_n (e_{n|n-1}^R)^{-1}  e_{n|n-1}^\fixX (e_{n|n-1}^R)^{-1} \fixu_n - \fixu_n.
\end{aligned}
\end{equation}Let us depart from error:
\begin{equation}
\text{\normalfont{Obs. (4)}} \Rightarrow\boxed{
\fixE_{n|n} = 
\begin{pmatrix}
R_n\grouplaw \hat{R}_{n|n}^{-1} \\ \fixX_n-(R_n \grouplaw \hat{R}_{n|n}^{-1})\action  \hat{\fixX}_{n|n} \\ \hat{R}_{n|n}\action (\bodyX_n-\hat{\bodyX}_{n|n})
\end{pmatrix},}
\end{equation}and see how it evolves through the vector dynamics. As before $\fixE^R$ does note change at vector dynamics. Let's consider $\fixE^x=\fixX-(R \grouplaw \hat{R}^{-1})\action  \hat{\fixX} $. It evolves as
\begin{align*}
& \fixF_n \fixX+\fixd-R \grouplaw \hat{R}^{-1}[\fixF_n\hat\fixX+\fixd]    + R   [\fixC_n \bodyX + \bodyu_n]-R \grouplaw \hat{R}^{-1}\hat R   [\fixC_n \hat  \bodyX + \bodyu_n]\\&=\fixF_n \fixX+\fixd-\fixF_n R \grouplaw \hat{R}^{-1}\hat\fixX- R \grouplaw \hat{R}^{-1}\fixd  +R\fixC_n (   \bodyX- \hat  \bodyX)\\&=\fixF_n\fixE^x+d-\fixE^R d+\fixC_n R(   \bodyX- \hat  \bodyX)\\&=\fixF_n\fixE^x+d-\fixE^R d+\fixC_n \fixE^R\fixE^\bodyX.
\end{align*}
Finally the last component of the error evolves as:
\begin{align*}
& \hat{R} \bigl(
[\bodyF_n \bodyX -\bodyF_n\hat  \bodyX  + R^{-1}   [\bodyC_n \fixX + \fixu_n]\bigr)-
 \hat{R}[       \hat R^{-1}  [\bodyC_n \hat \fixX + \fixu_n])\\
 &=\bodyF_n \fixE^\bodyX+(\fixE^R)^{-1} [\bodyC_n \fixX + \fixu_n]- [\bodyC_n \hat \fixX + \fixu_n]
 \\
 &=\bodyF_n \fixE^\bodyX+(\fixE^R)^{-1} \fixu_n - \fixu_n+\bodyC_n (\fixE^R)^{-1}[   \fixX- \fixE^R\hat \fixX]
 \\
 &=\bodyF_n \fixE^\bodyX+(\fixE^R)^{-1} \fixu_n - \fixu_n+\bodyC_n (\fixE^R)^{-1} \fixE^\fixX.
\end{align*}

\subsection{Extension of Theorem  \ref{prop::no-mu}  (the rotating Earth case)}

Theorem  \ref{prop::no-mu} case (b) covers all previously discovered group affine dynamics but one, namely inertial nagivation equations in a  rotating frame. Indeed, \cite{bonnabel2020mathematical} shows that after a suitable change of variables (that relies on a nice trick), the equations of inertial navigation on rotating Earth are group affine. Although the obtained system after the trick is group affine, it does not completely fall into one of the cases of   Theorem  \ref{prop::no-mu}, and nor does it fall within the scope of Theorem \ref{prop::abelian} as the group $SO(3)$ is non abelian. However,  a simple extension of Theorem  \ref{prop::no-mu}, case (b), in combination with   Theorem  \ref{thm::group-affine},  allows for covering this result as well.

\begin{thm}
Assume the two-frames  state variable   boils down to $(R,\fixX)$. Consider the following  dynamics 
\begin{equation} 
  \begin{pmatrix}
R_n \\ \fixX_n 
\end{pmatrix} = 
\begin{pmatrix}O_n
R_{n-1}\Omega_n \\
O_n \action [\fixF_n \fixX_{n-1} + \fixd_n]  
\end{pmatrix}, 
\label{eq::vector_dynamics22}\end{equation}
where matrix $\fixF_n: \fixV \mapsto \fixV  $  { commutes with the action of $G$}. Then, this dynamics is group affine  for the TFG structure. 
\end{thm}
\paragraph{proof}We know from invariant filtering theory in discrete time, see \cite{barrau2019linear}, that group affine dynamics is equivalent to (say, left-invariant) autonomous error propagation. Let us assume that $(R,\fixX)$ and $(\hat R,\hat \fixX)$ both evolve through the dynamics \eqref{eq::vector_dynamics22} and let us study the evolution of left-invariant error \eqref{EEerror}.  

We immediately see that $\bodyE^R=\hat R^{-1}R$ propagates as $\Omega^{-1}\bodyE^R \Omega$. Let us now study how the error $\bodyE^\fixX=\hat R^{-1}\action(\fixX-\hat\fixX)$ evolves.   We see $\bodyE^\fixX$ propagates as: \begin{align*}&(\Omega^{-1}\hat{R}^{-1}O^{-1}) \action \bigl(O\action [\fixF \fixX + \fixd_n -\fixF \hat \fixX - \fixd_n]\bigr) \\&=\Omega^{-1}\hat{R}^{-1}\action\fixF (  \fixX -\hat \fixX )\end{align*}
which, using  the commutation assumptions, is equal to $\Omega^{-1}\action \fixF \bodyE^\fixX$. We thus see that $\bodyE^R$  and $\bodyE^\fixX$ propagate autonomously. 
$\blacksquare$
Although the coupling between frame and vector dynamics in \eqref{eq::vector_dynamics22} seems quite specific,  it actually covers (in continuous time) the   inertial navigation with rotating Earth equations, which have been shown to fit into the invariant filtering framework in  \cite{bonnabel2020mathematical}, see also \cite{barrau2021}  for a longer and more detailed version.   It is thus easy to slightly generalize the two-frames theory in this direction to have it actually cover all known group affine dynamics to date. 

\begin{rem}It is worth noting that if we define the alternative state variable $\tilde\fixX_n=O_1^{-1}\dots O_n^{-1}\fixX_n$, we have from \eqref{eq::vector_dynamics22} that $\tilde\fixX_n= O_1^{-1}\dots O_n^{-1}\bigl(O_n \action [\fixF_n \fixX_{n-1} + \fixd_n])=\fixF_n  O_1^{-1}\dots O_{n-1}^{-1}\fixX_{n-1}+\tilde \fixd_n =\fixF_n \tilde\fixX_{n-1} + \tilde\fixd_n$. Hence we recover  natural vector frames dynamics  using the alternative variable $\tilde\fixX_n$.  

In the application studied in \cite{bonnabel2020mathematical} of inertial navigation on rotating Earth, the interpretation is clear. The term $O_n$ encodes rotation of the Earth, and the form of dynamics  \eqref{eq::vector_dynamics22} stems from expressing $\fixX_n$ in a  geographical frame. The presence of $O_n$ in the vector dynamics hence stems from the rotation of the geographic frame with respect to the inertial frame. The alternative variable $\tilde  \fixX_n$ reflects the vectors of the geographic frame corrected with respect to Earth rotation and hence expressed in an inertial frame. It  should  then   come as no surprise that in an inertial frame the vector dynamics be natural.   The apparent  difficulties in \eqref{eq::vector_dynamics22} were related to expressing the state vectors in the geographic frame, which is a rather ``unnatural'' frame, made to match our Earth-fixed system  of coordinates. 
\end{rem}

\subsection{Two-frames systems in continuous time}

Two-frame systems may be introduced in continuous time as well. This may  help  the readers being more familiar with the previous work \cite{barrau2017invariant}   to cast continuous-time navigation problems into the two-frames framework. The theory  and the results  are wholly similar in continuous time. Note that in both the discrete-time and the continuous-time invariant filtering frameworks we consider  observations in discrete time and all that is dealing with outputs in the paper is unchanged. As a result, the present section  only deals with the dynamics.

Let us start with the continuous-time counterpart of Definition 5.
\begin{dfn}[Continuous-time natural vector dynamics]
\label{nat:v}
We define continuous-time natural vector dynamics as $\dotex \chi_t=f_t(\chi_t)$ where $f_t$ is of the form:
\begin{equation}
f_t \begin{pmatrix}
R \\ \fixX \\ \bodyX
\end{pmatrix} = 
\begin{pmatrix}
0\\
[\fixF_t \fixX + \fixd_t] + R \action [\fixC_t \bodyX + \bodyu_t] \\
[\bodyF_t \bodyX + \bodyd_t] + R^{-1} \action  [\bodyC_t \fixX + \fixu_t]  
\end{pmatrix}
\label{eq::vector_dynamics-t}
\end{equation}
and where matrices $\fixF_t: \fixV \mapsto \fixV, \quad \bodyF_t: \bodyV \mapsto \bodyV, \quad \fixC_t: \bodyV \mapsto \fixV, \quad \bodyC_t: \fixV \mapsto \bodyV $ \emph{commute with the action of $G$} and $\fixd_t, \fixu_t$ (resp. $\bodyd_t$, $\bodyu_t$) are vectors of $V$ (resp. $B$).
\end{dfn}

Before turning to natural frame dynamics, let us introduce the counterpart of generic frame dynamics of Definition 8 to see how the separation of frame and vector dynamics work in continuous-time. 

\begin{dfn}[Continuous-time generic frame dynamics]
\label{defB}
Continuous-time generic frame dynamics are defined via a vector field $s_t$ on  $ (G, \fixV, \bodyV)$ of the form:
\begin{equation}
\label{eq::frame-dynamics-t}
s_t \begin{pmatrix}
R \\ \fixX \\ \bodyX
\end{pmatrix}
= \begin{pmatrix}
s_t^R(R, \fixX, \bodyX) \\ 0 \\ 0
\end{pmatrix}.
\end{equation}
\end{dfn}

In continuous time, we obtain a two-frame systems with natural vector dynamics and generic frame dynamics as follows:
\begin{equation}\boxed{
\begin{aligned}
\chi_t &= \left(s_t + f_t \right) \left( \chi_t \right), \\
\fixY_t = \fixh_t(\chi_t)& \quad  \text{or} \quad \bodyY_t = \bodyh_t(\chi_t),
\end{aligned}} \label{eq::two-frame-dynamics-C0}
\end{equation}
where $f_t$ is a continuous-time natural vector mapping \eqref{eq::vector_dynamics-t} of Definition \ref{nat:v}, $s_t$ is a generic frame dynamics \eqref{eq::frame-dynamics-t}   of Definition \ref{defB} and $\fixh_t$ or $\bodyh_t$ are the natural outputs \eqref{eq::output_fixed}, \eqref{eq::output_body} of Definition \ref{def::2frame_obs}.

Let us now turn to the counterpart of Definition \ref{def::frame-shift} of discrete-time natural frame dynamics. Note that  in continuous time   we need the machinery of differentiation on Lie groups. We thus refer the reader to Appendix \ref{jacob:sec} of the paper for the required definitions and notation.

\begin{dfn}[Continuous-time natural frame dynamics]
\label{nat:f}
We define  continuous-time natural frames dynamics  as in Def. \ref{defB}, but with the specificity that  
\begin{equation}
\label{eq::frame-shiftS} 
s_t^R(R, \fixX, \bodyX)=o_tR+R\omega_t
\end{equation}
 with $o_t,\omega_t \in \mathfrak g$   known elements of the Lie algebra of $G$ (inputs).  Moreover it is required that the maps $\fixX \mapsto  \repu{o_t}{V}    \fixX$  and $\bodyX \mapsto \repu{\omega_t}{V} \bodyX$  \emph{commute  with the action} of $\mathfrak g$ on $\fixV$ and  $\bodyV$ respectively.
\end{dfn}

\subsubsection{Example}

Consider the equations of inertial navigation in continuous time, see 
\cite{barrau2017invariant}, with additional sensor biases to be estimated online. They write
\begin{equation}\label{eq::nav-flatS}
\left\lbrace \begin{aligned}
\dotex  \Matrix{R}_t & = \Matrix{R}_{t}  \Matrix{[    \omega_t+b^\omega_{t}]_\times} \\
\dotex v_t & =  g + \Matrix{R}_{t} \left( a_t + b^a_{t} \right) \\
\dotex  p_t &=   v_{t}
\\
\dotex  b^\omega_n  &=0,\quad 
\dotex  b^a_n   = 0
\end{aligned} \right.  
 \end{equation} 
 
 We can define a two-frames state space   where the group $G=SO(3)$ encodes the orientation $R_t$, the vector space $\fixV =  \RR^3 \times \RR^3$ encodes $\fixX_t=(p_t,v_t)$ and  $\bodyV =\RR^3 \times \RR^3$ encodes $\bodyX_t=(b^\omega_t , b^a_t) $. The group $G=SO(3)$ acts through the term-by-term action (1) as: $\Matrix{R} \action \fixX=\Matrix{R} \action (v,p) = (\Matrix{R} v, \Matrix{R} p)$ and  $\Matrix{R} \action \bodyX = \Matrix{R} \action (b^\omega,b^a) = (\Matrix{R} b^\omega , \Matrix{R} b^a)$.
 We   see the vector dynamics part match those of natural vector dynamics of Definition \ref{nat:v} with 
\begin{equation*}
\begin{aligned}
&\fixF_t = 
\begin{pmatrix}
\0{3} & \0{3} \\
  \Id{3} &\0{3}
\end{pmatrix},\quad 
\fixC_t =
\begin{pmatrix}
\0{3} &  \Id{3} \\
\0{3} & \0{3}
\end{pmatrix},~~
\fixd_t = \begin{pmatrix}   g \\ \0{3,1} \end{pmatrix} \\
&\bodyu_t =\begin{pmatrix}   a_n \\ \0{3,1} \end{pmatrix},~
\bodyF_t = 
\begin{pmatrix}
\Id{3} & \0{3} \\
\0{3} &\Id{3}
\end{pmatrix}, ~
\bodyC_t =
\0{6},~
\bodyd_t= 
\fixu_t= \0{6,1} 
\end{aligned}\label{TFS:inertS}
\end{equation*}
as the commutation relations obviously hold from  Proposition 1.

We see the dynamics are generic, owing to  the state variable $b_t^\omega$ appearing in the frame dynamics. Now, let us suppose we neglect the gyroscope and accelerometer biases as in  \cite{barrau2017invariant} and hence remove   $\bodyX_t=(b^\omega_t , b^a_t) $ from the state.  Then the state boils down to $(R,\fixX)$, with $\fixX=(p,v)$. The first line then  corresponds to  \eqref{eq::frame-shiftS} with $o_t=0$ and where the skew-symmetric matrix $\Matrix{[    \omega_t ]_\times}$ is an element of the Lie algebra of $SO(3) $ indeed.  The frame dynamics  then match Def. \ref{nat:f} as the commutation assumptions are obviously satisfied, given there is no $\bodyX$.

\subsubsection{Properties}

The state trajectory independent  error propagation properties are analogous in continuous and in discrete time, as group affine dynamics have been introduced in both contexts.

Indeed the continuous-time counterpart of linear observed systems on groups is introduced in the context of matrix Lie groups in \cite{barrau2017invariant}. The continuous-time dynamics with discrete observations  case boils down to replacing $\chi_n=\phi_n(\chi_{n-1})$  with $\dotex\chi_t=\phi_t(\chi_t)$. The group affine property then  reads $\phi_t(\chi_1   \biggrouplaw \chi_2)=\chi_1\phi_t(\chi_2) +\phi_t(\chi_1)\chi_2-\phi_t(\chi_1)\phi_t(Id)\phi_t(\chi_2)$ and we  recover all the results of the article regarding state-independent propagation in continuous time. 

Interestingly, group affine maps in continuous and discrete time correspond exactly through integration, and the relation between both is strongly related to the notion of preintegration in robotics. The link between the continuous and discrete time is thoroughly discussed in \cite{barrau2019linear}, see also \cite{bonnabel2020mathematical} and \cite{barrau2021}   for applications to preintegration.

\end{document}